\documentclass[article,aps,pra,10pt,twocolumn,superscriptaddress,english,longbibliography]{revtex4-2}

\usepackage{bm,amssymb,amsmath,dsfont,mathrsfs,amstext,latexsym,physics,relsize}
\usepackage{graphicx}

\usepackage[usenames,dvipsnames]{xcolor}
\usepackage[colorlinks=true,citecolor=blue,urlcolor=blue,linkcolor=blue]{hyperref}
\usepackage[all]{hypcap}

\let\Re\relax

\DeclareMathOperator{\Re}{Re}

\newcommand{\figpanel}[2]{\hyperref[#1]{\ref{#1}#2}}

\begin{document}

\title{Driving a Quantum Phase Transition at Arbitrary Rate: Exact solution of the Transverse-Field Ising model}

\author{Andr\'as Grabarits\href{https://orcid.org/0000-0002-0633-7195}{\includegraphics[scale=0.05]{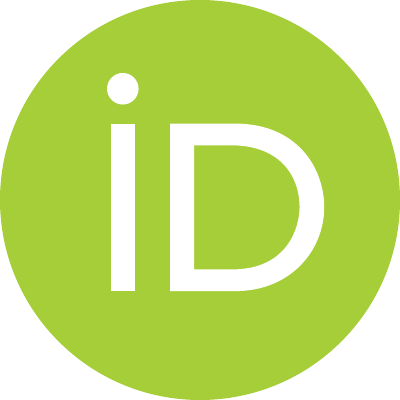}}}
%\email{andras.grabarits@uni.lu}
\affiliation{Department  of  Physics  and  Materials  Science,  University  of  Luxembourg,  L-1511  Luxembourg, Luxembourg}

\author{Federico Balducci\href{https://orcid.org/0000-0002-4798-6386}{\includegraphics[scale=0.05]{orcidid.pdf}}}
%\email{fbalducci@pks.mpg.de}
\affiliation{Max Planck Institute for the Physics of Complex Systems, N\"othnitzer Str. 38, 01187 Dresden, Germany}

\author{Adolfo del Campo\href{https://orcid.org/0000-0003-2219-2851}{\includegraphics[scale=0.05]{orcidid.pdf}}}
\affiliation{Department  of  Physics  and  Materials  Science,  University  of  Luxembourg,  L-1511  Luxembourg, Luxembourg}
\affiliation{Donostia International Physics Center,  E-20018 San Sebasti\'an, Spain}

\date{today}
%###########################################--ABSTRACT--###########################################

\begin{abstract}
We study the crossing of the quantum phase transition in the transverse-field Ising model after modulating the magnetic field at an arbitrary rate, exploring the critical dynamics from the slow to the sudden quench regime. We do so by analyzing the defect density, the complete kink number distribution, and its cumulants upon completion of a linearized quench. 
Our analysis relies on the diagonalization of the model
using the standard Jordan-Wigner and Fourier transformations, 
along with the exact solution of the time evolution in each mode in terms of parabolic cylinder functions. The free-fermion nature of the problem dictates that the kink number distribution is associated with independent and distinguishable Bernoulli variables, each with a success probability $p_k$. We employ a combination of convergent and asymptotic series expansions to characterize $p_k$ without restrictions on the driving rate. When the latter is approximated by the Landau-Zener formula, the kink density is described by the Kibble-Zurek mechanism, and higher-order cumulants follow a universal power-law behavior, as recently predicted theoretically and verified in the laboratory. By contrast, for moderate and sudden driving protocols, the cumulants exhibit a nonuniversal behavior that is not monotonic on the driving rate and changes qualitatively with the cumulant order. The kink number statistics remain sub-Poissonian for any driving rate, as revealed by an analysis of the cumulant rations that vary nonmonotonically from the sudden to the slow-driving regime.  Thanks to the determination of $p_k$ for an arbitrary rate, our study provides a complete analytical understanding of kink number statistics from the slow driving regime to the sudden quench limit.
\end{abstract}

\maketitle

%--------------------------------------------------------------------------------------------------
\section{Introduction}
Critical phenomena are ubiquitous in physical systems and arise in a wide variety of scenarios ranging from cosmology and high-energy physics to condensed matter and quantum technology. By way of example, continuous phase transitions play a crucial role in spontaneous symmetry breaking and are characterized by the divergence of the equilibrium relaxation time, a feature often referred to as critical slowing down. As a result, driving the system across a phase transition from a high-symmetry phase to a low-symmetry phase by modulating a control parameter in finite time leads to the production of topological defects, such as kinks and domain walls in ferromagnetic systems or vortices in superconductors. Continuous quantum phase transitions (QPT) are characterized by the closing of the energy gap between the ground state and the first excited state at the critical point~\cite{Sachdev2011Quantum}. This feature makes the critical dynamics nonadiabatic, leading to excitations in the system after crossing the phase transition. 

Both in the classical and the quantum case, the non-equilibrium dynamics across a phase transition is universal~\cite{Dziarmaga10,DZ14}. 
The so-called Kibble-Zurek mechanism (KZM) predicts the response time of a system driven in a finite (freeze-out) time $\tau$ across its critical point~\cite{Kibble76a,Kibble76b,Zurek85,Zurek96c}. It further identifies the characteristic nonequilibrium correlation length in the broken symmetry phase and, with it, the universal scaling of the density of defects with the quench time $\tau$.
Specifically, the KZM exploits the equilibrium critical scaling  governing the divergence of the relaxation time $\tau_{\rm eq}=\tau_0/\lvert\epsilon\rvert^{z\nu}$ and the correlation length, $\xi_{\rm eq}=\xi_0/\lvert\epsilon\rvert^{\nu}$, as a function of  the linearized control parameter $\epsilon$ across the critical point
\begin{equation}
    \epsilon=\frac{\lambda_c-\lambda}{\lambda_c}=\frac{t}{\tau}\,.
\end{equation}
As a result, the average number of defects scales with the rate as
\begin{equation}
    n\sim \tau^{-\frac{\nu}{z\nu+1}}\,.
\end{equation}
The KZM is a cornerstone in nonequilibrium statistical mechanics, which has applications ranging in cosmology, condensed matter, quantum simulation, and quantum computing. Efforts to verify the universal KZM experimentally have stimulated decades of research in diverse platforms, including ion chains~\cite{delcampo10,Ulm13,Pyka13,Ejtemaee2013,Xu2014,Cui16,Cui20}, colloids~\cite{Keim15}, multiferroics~\cite{Griffin12,Lin14,Du2023}, ultracold quantum gases~\cite{Weiler08,Lamporesi2013,Navon2015,Ko2019,Goo21}. and programmable quantum devices~\cite{Gardas18,Keesling19,Weinberg20,Bando20,King22,King24,Ali2024}. 

Despite its broad applications, the validity of the KZM is restricted to slow quenches near adiabatic dynamics when the density of topological defects is small. In particular, at moderate and fast quenches, the power-law scaling predicted by KZM  breaks down. A crossover to a plateau regime occurs, in which the density of topological defects becomes independent of the quench time but varies universally with the quench depth, i.e., the final value of the control parameter relative to the critical point~\cite{Chesler2015,Huabi2023FastQuench,Xia23prd,Xia2024}. Specifically, in the fast quench limit, the average excitations are predicted to grow with the final quench depth as $n\propto \epsilon^{z\nu}_f$. The critical time for the latter was also shown to vary with the quench depth as $\tau^\mathrm{fast}_c\sim \epsilon^{-z\nu}_f$~\cite{Huabi2023FastQuench}. Similarly, the KZM scaling law is interrupted at slow quenches by the onset of adiabatic dynamics, which can be estimated by matching the nonequilibrium correlation length with the system size, i.e., when the system supports $O(1)$ topological defects \cite{Zurek2005Dynamics,Dziarmaga2005Dynamics}. 

The quest for universal signatures in non-equilibrium critical phenomena beyond the KZM is thus an exciting endeavor motivated by the foundations of physics and its applications. Recent advances in this direction involve the discovery of the universal statistics of topological defects in the broken symmetry phase. Specifically, both the probability distribution of the number of defects~\cite{delCampo2018Universal,GomezRuiz20,Cui20,Bando20,Mayo21,King22}, and several measures of their spatial correlations~\cite{delcampo22,Thudiyangal24}, have been found to be universal. In parallel to these efforts focused on fluctuations beyond KZM, further research has pointed out the universality of critical dynamics in the fast quench regime~\cite{Xia2024}. 

In this work, we study the critical dynamics of an Ising chain driven from the paramagnetic to the ferromagnetic phase by varying the magnetic field in finite time. This dynamics results in kink formation, and we provide a complete characterization of the kink number statistics as a function of the driving rate, from the sudden to the slow driving limit. Using systematic expansions of the exact time evolution, we provide an analytical description of the kink density, kink number distribution, and its cumulants at arbitrary driving rates. Away from the slow quench regime, governed by the Kibble-Zurek mechanism and its generalizations, the critical dynamics as a function of the driving rate ceases to be universal. The cumulants of the kink number distributions no longer follow a monotonic dependence on the quench time, and their behavior varies qualitatively with the cumulant order. The statistics remain sub-Poissonain throughout the complete range of driving rates as revealed by the study of the cumulant rations, which vary nonmonotonically from the slow driving regime to the sudden quench limit.

\section{Exact solution of the driven transverse field Ising model}

In this section, we consider the exact dynamics of a quantum phase transition in the paradigmatic transverse field Ising model (TFIM). 

\subsection{Basics of the Transverse field Ising model}
The transverse field sing model (TFIM) describes the simplest many-body system exhibiting a QPT~\cite{Sachdev2011Quantum,Suzuki2012Quantum}. Its Hamiltonian is given by
\begin{equation}
	\label{eq:H_TFIM}
	\hat{H}(t) = - J \sum_{j=1}^L \left[ \hat{\sigma}^z_j \hat{\sigma}^z_{j+1} + g(t) \hat{\sigma}_j^x \right],
\end{equation}
where $\hat{\sigma}_j^{x,y,z}$ are Pauli matrix operators acting on site $j$. The ferromagnetic coupling $J \equiv1$ fixes the energy scale. The parameter $g(t)$ controls the strength of the transverse fields reaching the critical points at $g_c=\pm1$. For $\lvert g\rvert<1$ the system is in the ferromagnetic phase, in which the spin-spin interactions lead to finite magnetization. The regime $\lvert g\rvert>1$ correspond to the paramagnetic phase, where the transverse fields are strong enough to determine the spin configurations

The transverse field Ising model (TFIM) is an ideal platform to study the non-equilibrium crossing of QPTs~\cite{Zurek2005Dynamics, Dziarmaga2005Dynamics,delcampo12,delCampo2018Universal, Bando20, King22, Balducci2023Large}.
Upon employing the Jordan-Wigner transformation and subsequently a Fourier decomposition, the Hamiltonian~\eqref{eq:H_TFIM} is reduced to that of a set of independent two-level systems (TLSs), labeled by the momenta ~\cite{Suzuki2012Quantum},
\begin{align}
	\label{eq:H_TFIM_momentum}
	\hat{H}(t) &= 2 \sum_{k>0} \hat{\psi}_k^\dagger \left[ (g(t)-\cos k)\tau^z + \sin k \, \tau^x \right] \hat{\psi}_k \nonumber\\
    &= 2 \sum_{k>0} \hat{\psi}_k^\dagger H_k(t) \hat{\psi}_k,
\end{align}
where $\hat{\psi}_k := (\hat{c}_k, \hat{c}_{-k}^\dagger)^T$ is a vector of creation and annihilation operators for fermions of momentum $k=\frac{\pi}{L},\,\frac{3\pi}{L},\,\pi-\frac{\pi}{L}$, and $\tau^{x,y,z}$ are the Pauli matrices acting within the independent TLSs.

We shall characterize the dynamics of the quantum phase transition induced by varying $g(t)$ across the quantum phase point. 
Specifically, we consider a linearized schedule, 
\begin{equation}
g(t)=-g_c\,t/\tau,\quad t\leq 0
\label{Eq:gint}
\end{equation}
from an initial value $g(t_i)\gg1$ to $g(0)=0$.
Here, $\tau$ is the quench time that sets the rate at which the quantum critical point $g_c=1$ is crossed.
The driving results in the formation of defects emerging in pairs of kinks corresponding to the simultaneous excitations of TLSs at momenta $(-k,\,k)$.
The average kink density is given in terms of the expectation value of the kink number operator 
\begin{equation}
    K_L \equiv \frac{1}{2} \sum_{j=1}^{L} \left(1-\sigma^z_j\sigma^z_{j+1}\right),
\end{equation}
as
\begin{equation}
    n_\mathrm{ex}= \langle \Psi(t)|K_L|\Psi(t)\rangle/L.
\end{equation}

We shall further characterize the fluctuations of the kink number associated with the kink number operator $K_L$ after varying $g(t)$ in time.
To this end, we introduce the probability distribution characterizing the eigenvalue statistics of $K_L$ 
\begin{eqnarray}  
 P(N;\tau) = \langle \Psi(t)|\delta(K_L -N)|\Psi(t)\rangle.
 \end{eqnarray}
As shown in Ref.~\cite{delCampo2018Universal}, the kink number distribution is Poisson binomial, associated with independent Bernoulli random variables with success probability  $p_k$; see as well~\cite{Cui20,Bando20,King22}. 
For its analysis, it proves convenient to introduce the characteristic function $\tilde P(\theta;\tau)=\mathbb{E}[e^{i\theta N}]$, and its logarithm, the cumulant generating function, from which specific cumulants can be found via the identity, $\log\tilde P(\theta;\tau)=\sum_{q=1}^\infty\kappa_q(i\theta)^q/q!$. The first three cumulants equal the average defect number, its variance, and the third centered moment, respectively.  

The kink number distribution can be written as the sum of the occupation number in each mode,
$K_L=2\sum_{k>0} \gamma_k^\dagger\gamma_k$,  where  $\gamma_k$ and $\gamma_k$ are the fermionic Bogoliubov operators at  the final time $t=0$ when $g(0)=0$. The multiplicative factor appears due to the fact that excitations come in pairs of $(-k,k)$. Due to the periodic boundary condition, kinks are formed in pairs corresponding to these excitations. However, in real experiments, the number of kinks is more frequently measured.
For this reason, we focus on the statistics of the number of kinks rather than pairs of kinks (see Appendix \ref{app: PDF_CGF_kink_kink_pair} for a detailed discussion of their relation). The average kink density is given by $n_\mathrm{ex}=2\sum_{k>0}p_k/L$~\cite{Dziarmaga2005Dynamics}, where $p_k$ is the excitation probability in each mode.
This feature, along with the fact that the many-body state factorizes into the tensor product of the state in each mode, makes it possible to express  the cumulant generating function as the sum of the individual cumulant generating functions of each independent Bernoulli variable for $k>0$~\cite{Cincio07,delCampo2018Universal}
\begin{eqnarray} 
\log\tilde P(\theta;\tau) &=&\sum_{k>0}\log\left[1+\left(e^{2i\theta}-1\right)p_k\right].
 \end{eqnarray}
From a technical point of view, the description of the nonequilibrium dynamics at an arbitrary rate requires the computation of the excitation probability $p_k$ beyond the Landau-Zener formula restricted to slow quenches, ubiquitously used in the literature~\cite{Dziarmaga2005Dynamics,Polkovnikov05,Damski2006,Cincio07,Cui16,delCampo2018Universal,Cui20,Bando20}. It will thus play a crucial role in our analysis.

\subsection{General solutions for arbitrary pseudo-momentum}

To characterize the dynamics across the quantum phase transition at an arbitrary rate, we consider a linearized schedule in Eq. (\ref{Eq:gint}), 
with a large initial value of the magnetic field (e.g., fixed by $g(-100\tau)=100$, deep in the paramagnetic phase) and finishing at $t=0$, $g(0)=0$, when the Hamiltonian involves exclusively the ferromagnetic interactions and $K_L$ commutes with $H(0)$. Thus, the time of evolution takes negative values $t\in(-\infty,0)$ during the driving protocol.
The time-dependent Schr\"odinger equation governing the time-evolution of the TFIM is also reduced to the independent TLSs and for a given $k$-th mode given by
\begin{align}
    &i\partial_t\lvert \Psi(t)\rangle=H(t)\lvert \Psi(t)\rangle,\\
    &i \partial_t \ket{\psi_k(t)} = H_k(t) \ket{\psi_k(t)},\quad \lvert\Psi(t)\rangle=\bigotimes_k\lvert\psi_k(t)\rangle.
\end{align}

To write down the corresponding differential equation, we choose a basis associated with the fermionic operators in momentum space $\{\hat{c}_k\}$. We represent the time-evolved wavefunction of each momentum mode using the basis of the final Hamiltonian $H_k(0)$ for which $g(0)=0$.

First, we express the excitation probabilities in terms of the time-evolved wave function in this basis. 
The excitation probability is obtained by projecting $\ket{\psi_k(0)}$ onto the excited state of $H_k(0)$, 
\begin{equation}
    H_k(0) = 
    \begin{pmatrix}
		-\cos k   	&\sin k \\
		\sin k		&\cos k
  \end{pmatrix},
\end{equation}
it follows
\begin{equation}
    \ket{\mathrm{GS}_k(0)} = 
    \begin{pmatrix}
        -\cos \frac{k}{2} \\
        \sin \frac{k}{2}
    \end{pmatrix}, \qquad
    \ket{\mathrm{ES}_k(0)} = 
    \begin{pmatrix}
        \sin \frac{k}{2} \\
        \cos \frac{k}{2}
    \end{pmatrix}.
\end{equation}
Thus, the excitation probability reads
\begin{equation}
    \label{eq:def_pk}
    p_k = \big|\braket{\mathrm{ES}_k(0)}{\psi_k(0)}\big|^2 = \left| \psi_{k,1}(0)\sin \frac{k}{2}  + \psi_{k,2}(0)\cos \frac{k}{2} \right|^2,
\end{equation}
where $\psi_{k,1}(t)$ and $\psi_{k,2}(t)$ denote the components of the $k$-th the time evolved state, i.e., $\lvert \psi_k(t)\rangle=(\psi_{k,1}(t), \psi_{k,2}(t))^T$. Next, we provide the exact solutions of the time-dependent Schr\"odigner-equation in this basis, given by
\begin{equation}\label{eq:TLS_eq}
    i \partial_t
    \begin{pmatrix}
        \psi_{k,1} \\
        \psi_{k,2}
    \end{pmatrix}
    = 2J [(g-\cos k) \tau^z + \sin k \, \tau^x]
    \begin{pmatrix}
        \psi_{k,1} \\
        \psi_{k,2}
    \end{pmatrix}.
\end{equation}
The derivations of the exact solutions of these two coupled differential equations can be found in Refs.~\cite{Suzuki2012Quantum}, and we also provide a systematic account of it in App.~\ref{app: LZ_exact} giving the exact solution in terms of the parabolic cylinder function~\cite{NIST-DLMF},
\begin{widetext}
\begin{eqnarray}
\label{eq: parabolic_cylinders}
    \psi_{k,1}(t) &=& \sqrt{\tau} \sin k \, e^{-\frac{\pi}{4} J \tau \sin^2 k} D_{-iJ \tau \sin^2 k-1} \left[ 2\,e^{-3i\pi/4} \sqrt{\tau} \left( \cos k + \frac{t}{\tau}\right) \right], \\
    \psi_{k,2}(t) &=& e^{-\frac{\pi}{4} \tau \sin^2 k +3i\pi/4} D_{-i \tau \sin^2 k} \left[ 2\,e^{-3i\pi/4} \sqrt{\tau} \left( \cos k + \frac{t}{\tau}\right) \right].
\end{eqnarray}

Thus, to get the final excitations, one needs to plug the solutions at $t=0$ into Eq.~\eqref{eq:def_pk},
\begin{eqnarray}\label{eq:psi}
    \psi_{k,1}(0) &=& \sqrt{\tau} \sin k \, e^{-\frac{\pi}{4}  \tau \sin^2 k} D_{-i \tau \sin^2 k-1} \left( 2 e^{-3i\pi/4} \sqrt{\tau} \cos k \right), \\    
    \psi_{k,2}(0) &=& e^{-\frac{\pi}{4} J \tau \sin^2 k+3i\pi/4} D_{-i \tau \sin^2 k} \left( 2\,e^{-3i\pi/4} \sqrt{\tau} \cos k \right).
\end{eqnarray}
\end{widetext}

Even though Eqs.~\eqref{eq:def_pk}--\eqref{eq:psi} provide the exact solutions, the parabolic cylinder functions do not lead to an insightful analytical understanding of the physical properties. For this reason, we expand $\psi_{k,1}$ and $\psi_{k,2}$ for slow annealing protocols compared to the critical scales of fast quenches~\cite{Huabi2023FastQuench}, $\tau\gg1$. In this limit, both the argument \emph{and} the parameter of the parabolic cylinder functions, Eq.~\eqref{eq:psi}, go to infinity with $\tau$. However, there is also a dependence on the momentum $k$ that needs to be taken into account. This constitutes an essential feature of the TFIM that goes beyond the standard Landau-Zener analysis in Refs.~\cite{Damski05,Dziarmaga2005Dynamics,Damski2006}
and has not been investigated before.

The argument and the parameter split the full range momenta into three regions. First, for $k < c \tau^{-1/2}$ as $\tau \to \infty$, where $c$ is a constant, one can expand the parabolic cylinder functions for large argument, as the parameter remains bounded. While the region $k \lesssim \tau^{-1/2}$ shrinks as $\tau \to \infty$, it turns out to dictate most of the physics. Second, for $k>c'\tau^{-1/4}$ as $\tau \to \infty$, where $c'$ is another constant, one can perform a uniform expansion of the parabolic cylinder functions for large argument and large parameter. This region comprises most of the momenta, and will give the most important corrections to the asymptotic scaling. Third, there is the vanishingly small region $c \tau^{-1/2} < k < c'\tau^{-1/4}$, which provides a negligible correction since quantities are regular. For this reason, we will consider together the first and third regions, thus splitting $k \lesssim \tau^{-1/4}$ and $k \gtrsim \tau^{-1/4}$, committing only an exponentially small mistake for the quantities of interest (see below). 

\section{Slow driving protocols}

We next focus on the slow driving limit. We briefly revisit the approximation of the exact solution of the time-dependent Schr\"odinger equation in the slow driving limit, leading to the Landau-Zener approximation. Next, we derive the leading order approximation for the excitation probability replacing the LZ formula above a given momentum threshold. These results reveal new physics beyond the KZM, governing the behavior of the excitations around the fast quench breakdown.

%The comparison between the argument and the parameter splits the full range momenta into two regimes. In particular, one needs to compare the leading order behavior of the absolute values of the parameter, $\sim\tau k^2$ and the argument, $\sim \sqrt\tau$, where it was assumed that $k\ll1$ near the separation scale between the two momentum regimes. As a result, the two expressions become of the same order when $k\sim \tau^{-1/4}$ verifying the assumption on the smallness of $k$. For $k \lesssim \tau^{-1/4}$, one needs to expand the parabolic cylinder functions for large argument, while for $k \gtrsim\tau^{-1/4}$, a uniform asymptotic expansion is required for both large argument and large parameter.

%However, getting closer to the fast quench breakdown with $\tau^\mathrm{fast}\sim O(1)$, important features arise for large enough momenta $k \gtrsim \tau^{-1/4}$. 

%--------------------------------------------------------------------------------------------------
\subsection{Small momenta--Asymptotic expansions for large argument}\label{subsec: small_k_expansion}

In the regime of $k \lesssim \tau^{-1/4}$, the expansion of the parabolic cylinder functions $D_\nu(z)$ at fixed $\nu$ and large $|z|$ strongly depends on the phases of $\arg \nu$ and $\arg z $. At $t=0$, the phase of $z$ is given by $\arg z= -3\pi/4$, allowing for the series expansion~\cite{NIST-DLMF}
\begin{equation}
    D_\nu(z) \approx e^{-z^2/4} z^\nu.
\end{equation}
Plugging this expansion into the expressions of $\psi_{k,1}$ and $\psi_{k,2}$ in Eq.~\eqref{eq:psi},
one obtains up to exponential accuracy the Landau-Zener transition probability (see App.~\ref{app: LZ_exact}),
\begin{equation}
    p_k \approx e^{-2\pi\tau k^2}.
\end{equation}
This result was found originally in Ref.~\cite{Dziarmaga2005Dynamics} and was decisive in helping to establish the validity of the KZ scaling in the quantum regime, along with the Refs.~\cite{Damski05,Zurek2005Dynamics,Polkovnikov05}.

%--------------------------------------------------------------------------------------------------
\subsection{Finite momenta: Uniform asymptotic expansions}

 Consider the region $k \gtrsim \tau^{-1/4}$ when both the argument and the index of the parabolic cylinder functions are large, owing to $\tau \to \infty$. To the best of our knowledge, no result exists on the asymptotics of $D_{-ix}(e^{-3i\pi/4}y)$ when both $x,y \to +\infty$~\cite{NIST-DLMF}. Here, we provide the main steps with the details given in App.~\ref{app:Saddle_expansion_1} and App.~\ref{app:Saddle_expansion_2}.

The parabolic cylinder function for $\mathrm{\Re}\,\nu < 0$ can be put into an integral representation~\cite{NIST-DLMF}
\begin{equation}\label{eq: integral_representation}
    D_{\nu}(z)= \frac{e^{-z^2/4}}{\Gamma(-\nu)} \int_0^\infty dt \, e^{-zt - \frac{1}{2}t^2} t^{-1-\nu},
\end{equation}
allowing for the computation of the resulting correction terms via a saddle-point expansion~\cite{Bleistein86}.
  
%------------------------------------------------
As it turns out,
for the leading order behavior of the excitation probabilities for $k\gtrsim \tau^{-1/4}$ both the leading and the next-to-leading order saddle point corrections are required.
After straightforward but long algebraic steps (see App.~\ref{app:Saddle_expansion_1}) one arrives at
\begin{widetext}
\begin{equation}
    \label{eq:psi1_asymp1}
    \psi_{k,1}(0)\approx\cos\frac{k}{2} e^{-i\pi/4-i\frac{\tau}{2}\left(\cos^2k-2\cos k-1\right)-i\tau\,\sin^2k\log\left(2\sqrt\tau\sin^2\frac{k}{2}\right)}\left(1 + i \frac{4+5\cos k}{48\tau} \tan^2 \frac{k}{2} + \frac{i}{12 \tau \sin^2 k}\right).
\end{equation}
\end{widetext}

%------------------------------------------------

For the lower component of the time-evolved wave function, $\psi_{k,2}(0)$, the exact solution can be written with the help of another integral representation, as in this case $\mathrm{Re}\,\nu=0$~\cite{NIST-DLMF}:
\begin{equation}\label{eq:saddle_inegral_psi2}
\begin{split}
    &D_{\nu}(z)= \sqrt{\frac{2}{\pi}} e^{z^2/4} \int_0^\infty dt \, t^\nu e^{-\frac{1}{2}t^2} \cos \left( zt - \frac{\pi}{2}\nu \right)\\
    &=\sqrt{\frac{2}{\pi}} e^{z^2/4} \int_0^\infty dt \, t^\nu e^{-\frac{1}{2}t^2} \frac{e^{-izt+i\frac{\pi}{2}\nu}+e^{izt-i\frac{\pi}{2}\nu}}{2}.
    \end{split}
\end{equation}
Here, the cosine has been written in terms of the exponentials to make the integral amenable to saddle point expansion. As discussed in App.~\ref{app:Saddle_expansion_2}, after some laborious but straightforward algebraic manipulations, the leading and next-to-leading order saddle point approximations yield 
\begin{eqnarray}
    \label{eq:psi2_asymp1}
    &\psi_{k,2}(0)&\approx\nonumber\\
    &\sin \frac{k}{2} &
    e^{3i\pi/4 -\frac{i}{2} \tau (\cos^2 k - 2\cos k -1) -i \tau \sin^2 k \log 2\sqrt\tau\sin^2\frac{k}{2}  }\nonumber\\
    &\times&\left( 1 + i\frac{5\cos k-4}{48 \tau} \cot^2 \frac{k}{2} \right).
\end{eqnarray}

 The results of Eq.~\eqref{eq:psi1_asymp1} and Eq.~\eqref{eq:psi2_asymp1} can be written in similar forms differing only in the real absolute values and the signs of their phases as
\begin{eqnarray}
    &&\psi_{k,1}(0) \approx\\ 
    &&e^{i\Phi} \cos \frac{k}{2} \left(1 + i \frac{4+5\cos k}{48\tau} \tan^2 \frac{k}{2} + \frac{i}{12 \tau \sin^2 k}\right),\nonumber\\
    &&\psi_{k,2}(0)\approx -e^{i\Phi} \sin \frac{k}{2} \left( 1 + i\frac{5\cos k-4}{48 \tau} \cot^2 \frac{k}{2} \right).
\end{eqnarray}

\begin{figure}
    \centering
    \includegraphics[width=0.9\linewidth]{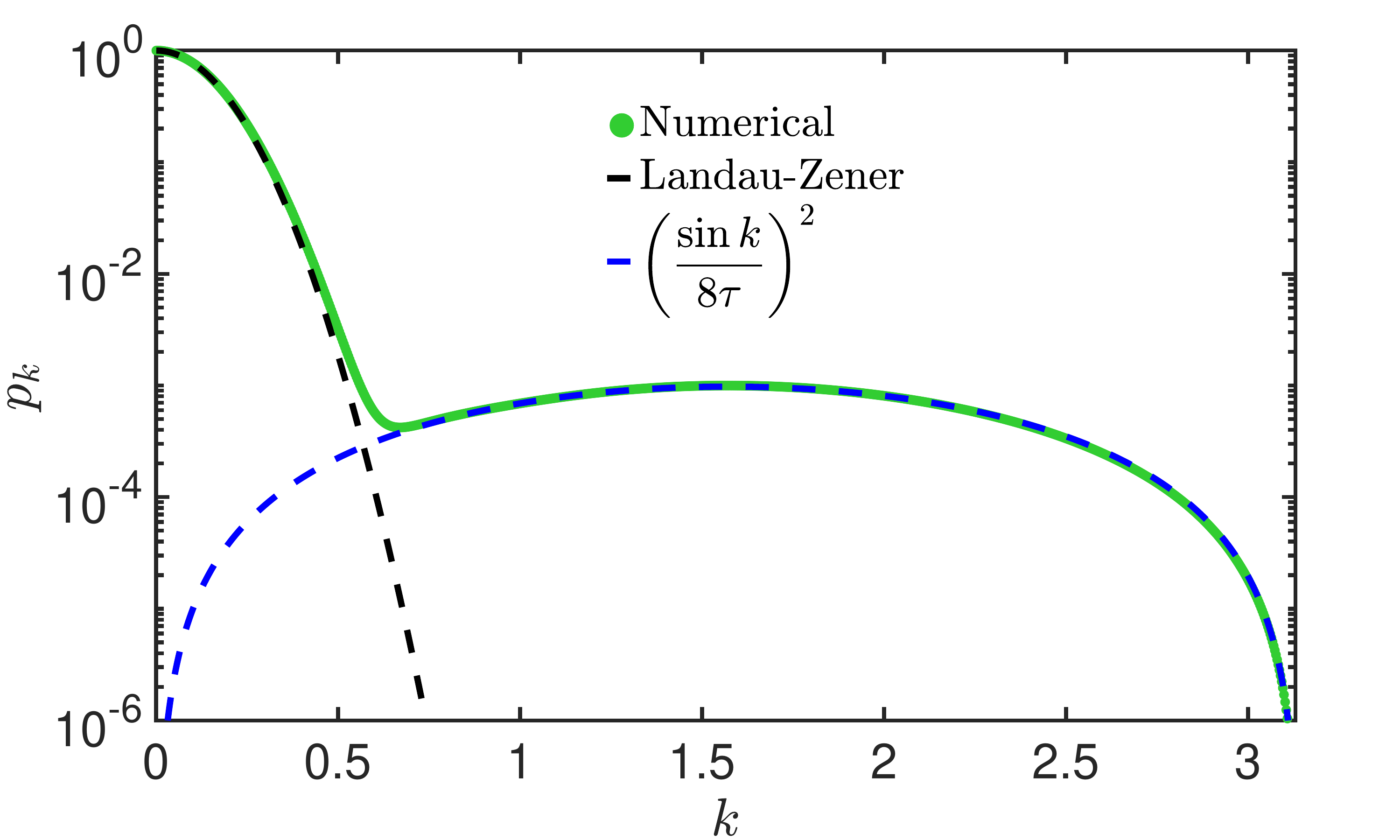}
    \caption{Comparison between the exact and asymptotic excitation probability for $\tau=4$ and $L=4000$. The agreement is very good, except for a very small region around $k \sim \tau^{-1/4}$, where higher orders in the two asymptotic expansions are needed.}
    \label{fig:pk_anneal}
\end{figure}

As a result, the leading order of the excitation probability from $k\gtrsim\tau^{-1/4}$ will originate from the next-to-leading order corrections of the saddle point approximations. These terms give the contribution of
\begin{equation}\label{eq:p_k_klarge}
    p_k =  \left| \sin \frac{k}{2} \, \psi_{k,1}(0) + \cos \frac{k}{2} \, \psi_{k,2}(0) \right|^2
    %= \frac{\sin^2 k}{64 J^2 \tau^2}.
    = \left(\frac{\sin k}{8\tau}\right)^2.
\end{equation}
This correction plays a particularly important role in annealing times near the fast quench breakdown, as discussed in the upcoming section. As demonstrated in Fig.~\ref{fig:pk_anneal}, around the scale of $\tau^{-1/4}$ the structure of $p_k$ exhibits the predicted transition between the LZ formula and the one derived above in Eq.~\eqref{eq:p_k_klarge}. Finally, based on numerical results, the constant prefactor in the separation scale of the momentum is approximately given by $\eta\approx0.7$. Using it, it follows that the excitation probability in each mode is given by
\begin{equation}
    \label{eq:pk_slow_final}
    p_k \simeq 
    \begin{cases}
        e^{-2\pi \tau k^2}                   &\text{if } k <\eta\,\tau^{-1/4}, \\
        \left(\frac{\sin k}{8\tau}\right)^2  &\text{if } k >\eta\,\tau^{-1/4}.
    \end{cases}
\end{equation}

%--------------------------------------------------------------------------------------------------
\subsection{Correction to the KZ scaling of the average kink number}

Using the complete solution of Eq.~\eqref{eq:pk_slow_final} in the expression for the number of defects in the continuum limit, one obtains
\begin{widetext}
\begin{eqnarray}\label{eq: kink_large_tau}
    \kappa_1&=&2\sum_{k>0}\,p_k\approx\frac{L}{\pi}\int_0^\pi\mathrm d
    k\, p_k=\frac{L}{\pi}\int_0^{\eta\,\tau^{-1/4}}\mathrm dk\,e^{-2\pi \tau k^2}
    +\frac{L}{\pi}\int_{\eta\,\tau^{-1/4}}^\pi\mathrm dk\,\left(\frac{\sin k}{8\tau}\right)^2\nonumber\\
    &=&L\,\mathrm{erf}\left(\sqrt{2\pi}\,\eta\,\tau^{1/4}\right)(8\pi^2\tau)^{-1/2}
+L\frac{\sin\left(2\eta\,\tau^{-1/4}\right)-2\eta\,\tau^{-1/4}+2\pi}{256\pi}\tau^{-2}\nonumber\\
    &=&\frac{L}{\sqrt{8\pi^2}}\tau^{-1/2}+\frac{L}{128}\tau^{-2}-\frac{\eta\,L}{192}\frac{\tau^{-11/4}}{\pi}+O(\tau^{-{13/4}}),
\end{eqnarray}
\end{widetext}
showing that the KZ scaling of the defects remains unchanged up to $\tau^{-2}$ corrections. Here, the leading order expansion of the error function was used for large argument, $\mathrm{erf}\left(x\right)=1-\frac{e^{-x^2}}{\sqrt\pi x}+\frac{e^{-x^2}}{2\sqrt\pi x^3}-\dots$, which leads to exponentially suppressed corrections in the first integral. The two additional orders originate from the contributions of the momentum sector, $k\gtrsim\tau^{-1/4}$. This accurately captures the kink number even around the fast quench breakdown, $\tau\approx1$, where the KZ scaling loses its validity.
This is demonstrated in Fig.~\ref{fig:nex_average}(a).

%--------------------------------------------------------------------------------------------------

\subsection{Correction to the cumulant generating function and the cumulants}

\begin{figure}[t]
\centering
\includegraphics[width=.5\textwidth]{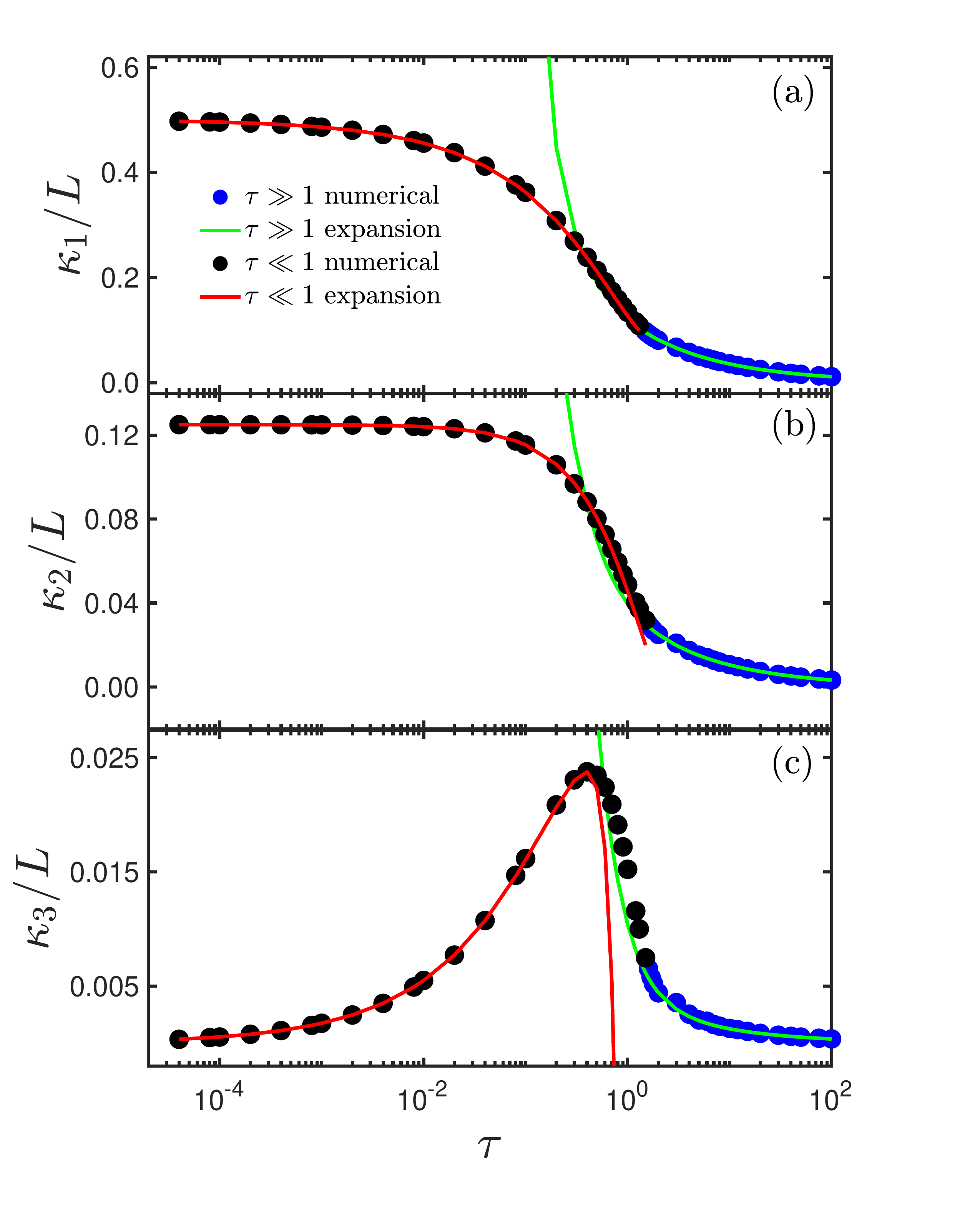}
    \caption{(a) Average kink density as a function of $\tau$ for $L=4000$, reproducing up to high precision the intermediate regime as well by the corrections of the slow driving, Eq.~\eqref{eq: kink_large_tau} and sudden quench limits, Eq.~\eqref{eq: average_tau_small}. (b) A similar agreement is found for the density of the kink number variance. (c) The corrections accurately reproduce the increasing behavior of the third cumulants.}
    \label{fig:nex_average}    
\end{figure}

Using the expression in Eq.~\eqref{eq:pk_slow_final} generalizes the result of the full statistics of excitations of Ref.~\cite{delCampo2018Universal,Cui20,Bando20,King22} to annealing times close to the fast quench breakdown via the corrections in the regime $k\gtrsim \tau^{-1/4}$. In particular, one has to split the summation of the cumulant generating function around the separation scale of $\tau^{-1/4}$,
\begin{widetext}
    \begin{eqnarray}            
    \log\tilde P(\theta;\tau) &\approx& \frac{L}{2\pi}\int_0^\pi\mathrm dk \, \log \left[1+ \left(e^{2i\theta} - 1 \right) p_k \right]\nonumber\\ 
    &\approx& \frac{L}{2\pi}\int_0^{\eta\,\tau^{-1/4}} \mathrm dk \, \log \left[1+ \left(e^{2i\theta} - 1 \right)e^{-2\pi \tau k^2}  \right] 
    + \frac{L}{2\pi}\int_{\eta\,\tau^{-1/4}}^\pi  \mathrm dk \, \log \left[1+ \left(e^{2i\theta} - 1 \right) \frac{\sin^2 k}{64\tau^2} \right]\nonumber\\
    &\approx& -\frac{L}{\sqrt{2\pi^2\tau}} \mathrm{Li}_{3/2} \left(1-e^{2i\theta} \right) + \frac{L}{2\pi}\sum_{n=1}^\infty\frac{1}{n}\left(\frac{1-e^{2i\theta}}{64\tau^2}\right)^{n}\int_{\eta\,\tau^{-1/4}}^\pi\mathrm dk\,(\sin k)^{2n},
    \end{eqnarray}
where $\mathrm{Li}_{3/2}(x)=\sum_{p=1}^\infty\,x^p/p^{3/2}$.
%\end{widetext}

The exact series representation of the second integral is required, as taking the lower limit to $0$ would lead to uncontrolled corrections. The corresponding integrals can be evaluated using the identity
\begin{equation}
    \int_{\eta\,\tau^{-1/4}}^\pi\mathrm dk\,(\sin k)^{2n}=-\frac{B\left[\sin^2\left(\eta\,\tau^{-1/4}\right),n-\frac{1}{2},\frac{1}{2}\right]}{2},
\end{equation}
where $B\left(x,a,b\right)$ is the incomplete beta function.
Exploiting its series representation, $B\left(x,a,b\right)=\sum_{n=0}^\infty\frac{(1-b)_n}{n}\frac{x^{a+n}}{a+n}$, with $(1-b)_n$ denoting the Pochammer symbol, the complete cumulant generating function reads
%\begin{widetext}
\begin{eqnarray}
     \log\tilde P(\theta;\tau)&\approx&-\frac{L}{\sqrt{2\pi^2\tau}} \mathrm{Li}_{3/2} \left(1-e^{2i\theta} \right)
     -L\sum_{n=1}^\infty\frac{1}{4\pi n}\left(\frac{1-e^{2i\theta}}{64\tau^2}\right)^{n}B\left[\sin^2\left(\eta\,\tau^{-1/4}\right),n-\frac{1}{2},\frac{1}{2}\right]\nonumber\\
     &\approx&-\frac{L}{\sqrt{2\pi^2\tau}} \mathrm{Li}_{3/2} \left(1-e^{2i\theta} \right)-L\sum_{n=1}^\infty\frac{1}{4\pi n}\left(\frac{1-e^{2i\theta}}{64\tau^2}\right)^{n}\sum_{m=0}^\infty\frac{(1/2)_m}{m!}\frac{\sin^{2(m+n)-1}\left(\eta\,\tau^{-1/4}\right)}{n+m-1/2}.
     %&\approx N\sum_{n=1}^\infty\frac{1}{n}\left(\frac{1-e^{i\theta}}{64J^2\tau^2}\right)^{n}\frac{\sin^{2n+1} \left((J\tau)^{-1/4}\right)}{2\pi(2n+1)}\,\sum_{m=0}^\infty(-1)^m\binom{-1/2}{m}\frac{(n+1/2)_m}{(n+3/2)_m}\sin^{2m}\left((J\tau)^{-1/2}\right)\\
     %&=N\sum_{n=1,m=0}^\infty\frac{(-1)^m}{n}\binom{-1/2}{m}\left(\frac{1-e^{i\theta}}{64J^2\tau^2}\right)^{n}\frac{\sin^{2(n+m)+1} \left((J\tau)^{-1/4}\right)}{4\pi(n+3/2+m-1)}
     \end{eqnarray}

\end{widetext}
Even though the above expression looks highly complicated, it provides a systematic way to determine the leading order corrections. To this end, we compute the first two orders,
\begin{eqnarray}\label{eq: logP_expansion}
     \log\tilde P(\theta;\tau)&\approx& -\frac{L}{\sqrt{2\pi^2\tau}} \mathrm{Li}_{3/2} \left(1-e^{2i\theta} \right)\nonumber\\
    & &+ \frac{L}{128\pi}\left(\frac{1}{2}\tau^{-2}-\frac{\eta}{3}\tau^{-11/4}\right)(1-e^{2i\theta})\nonumber\\
    & &+ (1-e^{2i\theta})O(\tau^{-13/4}).
\end{eqnarray}
The agreement with the numerically exact results of the cumulant-generating functions with the leading order $\sim\tau^{-2}$ is demonstrated in the App.~\ref{app: Cumulant_Generating_function}.
Note that the derivatives of the cumulant generating function evaluated at $\theta=0$ increase with the cumulant order as $(-i\partial_\theta)^q (1-e^{2i\theta})\Big\vert_{\theta=0}=2^q$, within the first two correction orders.
As a result, the cumulant corrections exhibit an opposite trend than that in the KZ scaling regime and increase exponentially compared to the behavior described by Eq.~\eqref{eq: kink_large_tau},
\begin{equation}\label{eq: cumulant_corrections_slow}
    \delta\kappa_q\approx2^q\frac{L}{128\pi}\left(\frac{1}{2}\tau^{-2}-\frac{\eta}{3}\tau^{-11/4}\right)\propto 2^q\delta\kappa_1.
\end{equation}
We demonstrate these results on the first three cumulants. As shown in Fig.~\ref{fig:nex_average}, the fast quench corrections reproduce the shape of the cumulants around the fast quench breakdown, $\tau\approx 1$, up to high precision.
Due to the identical corrections near the fast quench breakdown, the proportionality between the cumulants disappears. 

\begin{figure*}[t]
    \includegraphics[width=0.8\linewidth]{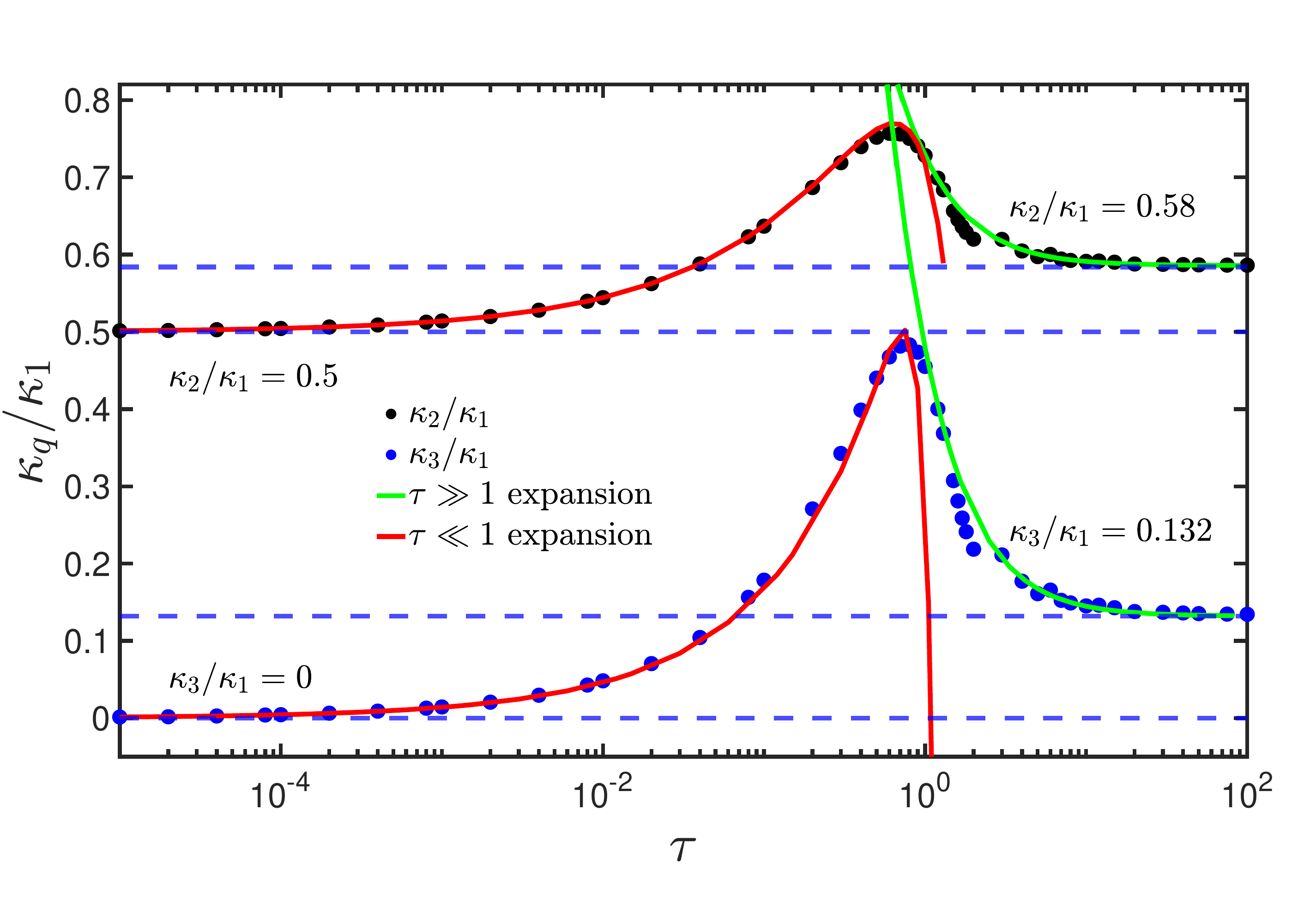}
    \caption{Cumulant ratios as a function of $\tau$, exhibiting a transition around the fast quench breakdown from the sudden quench limit to the KZ scaling constant plateau ($L=4000$). Only in these limits are cumulant rations approximately constant. The values in the KZ limit were predicted in Refs.~\cite{delCampo2018Universal} and experimentally verified in Refs.~\cite{Cui20,Bando20}.}
    \label{fig:Ratios}    
\end{figure*}

We further show in Fig.~\ref{fig:Ratios}, how the cumulant ratios exhibit a transition from the sudden quench limit to the KZ scaling plateau. In the KZ regime, the plateau values match those reported in Refs.~\cite{delCampo2018Universal,Cui20,Bando20,King22}.
 Figure~\ref{fig:Ratios} shows how, as the driving rate is increased, the cumulant ratios grow until reaching a maximum value, after which they decrease monotonically toward the sudden quench limit, in which $\kappa_2/\kappa_1=1/2$ while $\kappa_3$ vanishes.  A detailed characterization of this regime is deferred. Sec.~\ref{sec:quenching}. The statistics remains sub-Poissonian throughout the complete range of annealing times.

Finally, we demonstrate the precision of our results in terms of the kink statistics for annealing times considerably close to the fast quench breakdown (for $\tau=1.3,\,2.5,\,5$). 
The precise matching between the numerical results and the Gaussian limiting distribution with the analytical results of the average and the variance is shown in Fig.~\ref{fig:Stat_slow}. While corrections $\sim\tau^{-2}$  start to deviate for $\tau<2$, the next-to-leading order remains remarkably close to the numerical results even for $\tau\approx 1.3$.

Our results thus span the whole range of quench times from the KZ scaling regime explored in Refs.~\cite{delCampo2018Universal,Cui20,Bando20,King22}, when all cumulants are proportional to each other,  to the sudden quench limit.

\section{Fast quench protocols}
\label{sec:quenching}
In this section, we address the situation $\tau \ll 1$ and provide corrections resulting from finite-time driving protocols with respect to the sudden quench limit ($\tau=0$) for the cumulant generating function and the defect statistics.

\subsection{Excitation probability and kink density near the sudden quench limit}

 Our starting point is again the generalized results for the solution of the Schr\"odinger equation at $t=0$ in Eq.~\eqref{eq: parabolic_cylinders}. However, we will now expand such expressions for small $\tau$. For this, we employ the convergent series representation~\cite{NIST-DLMF}
\begin{widetext}
\begin{equation}\label{eq: D_tau_small}
    D_\nu(z) = \sqrt{\pi} 2^{\nu/2} e^{-z^2/4} \left[ \frac{1}{\Gamma \left( \frac{1-\nu}{2} \right)} \sum_{n=0}^\infty \frac{z^{2n}}{(2n)!} \prod_{k=1}^n (2k-\nu-2) - \frac{\sqrt{2}}{\Gamma \left( -\frac{\nu}{2} \right)} \sum_{n=0}^\infty \frac{z^{2n+1}}{(2n+1)!} \prod_{k=1}^n (2k-\nu-1) \right].
\end{equation}
By substituting Eq.~\eqref{eq: D_tau_small} into Eq.~\eqref{eq:def_pk}, one obtains after some trivial algebraic steps the excitation probability near sudden quenches up to order $O(\tau^2)$, i.e., 
\begin{eqnarray}\label{eq:p_k_fast}
    p_k &=& \cos^2 \frac{k}{2} 
    -\frac{\sqrt{\pi}}{2} \sin^2 k \, \tau^{1/2}
    -\frac{\pi}{2} \sin^2 k \cos k \, \tau 
    -\frac{\sqrt{\pi}}{4} \sin^2 k  [4-(4+\pi-\log 4) \sin^2 k] \tau^{3/2}\nonumber\\
    &-&\frac{1}{12} \sin^2 k \cos k [16 - (3\pi^2+16)\sin^2 k]\tau^2+O(\tau^{5/2})\,.
\end{eqnarray}
The corresponding average kink number reads as
\begin{eqnarray}\label{eq: average_tau_small}
    \kappa_1&\approx&\frac{L}{\pi}\int_0^\pi\mathrm dk\,\cos^2 \frac{k}{2} 
    -\frac{\sqrt{\pi}}{2} \sin^2 k \, \tau^{1/2}
    -\frac{\pi}{2} \sin^2 k \cos k \, \tau 
    -\frac{\sqrt{\pi}}{4} \sin^2 k  [4-(4+\pi-\log 4) \sin^2 k] \tau^{3/2}\nonumber\\
    &=&\frac{L}{2}-\frac{\sqrt\pi\,L}{4}\tau^{1/2}-\frac{\sqrt\pi\,L}{32}(4-3\pi+\log(64))\tau^{3/2}.
    \end{eqnarray}
%\end{widetext}
Combined with the results of the large $\tau$ expansion in Eq.~\eqref{eq: kink_large_tau}, the function of the kink density can be reproduced up to remarkable precision close to the fast quench breakdown $\tau\lesssim 1$, as demonstrated in Fig.~\ref{fig:nex_average}.

\begin{figure}
    \includegraphics[width=0.5\linewidth]{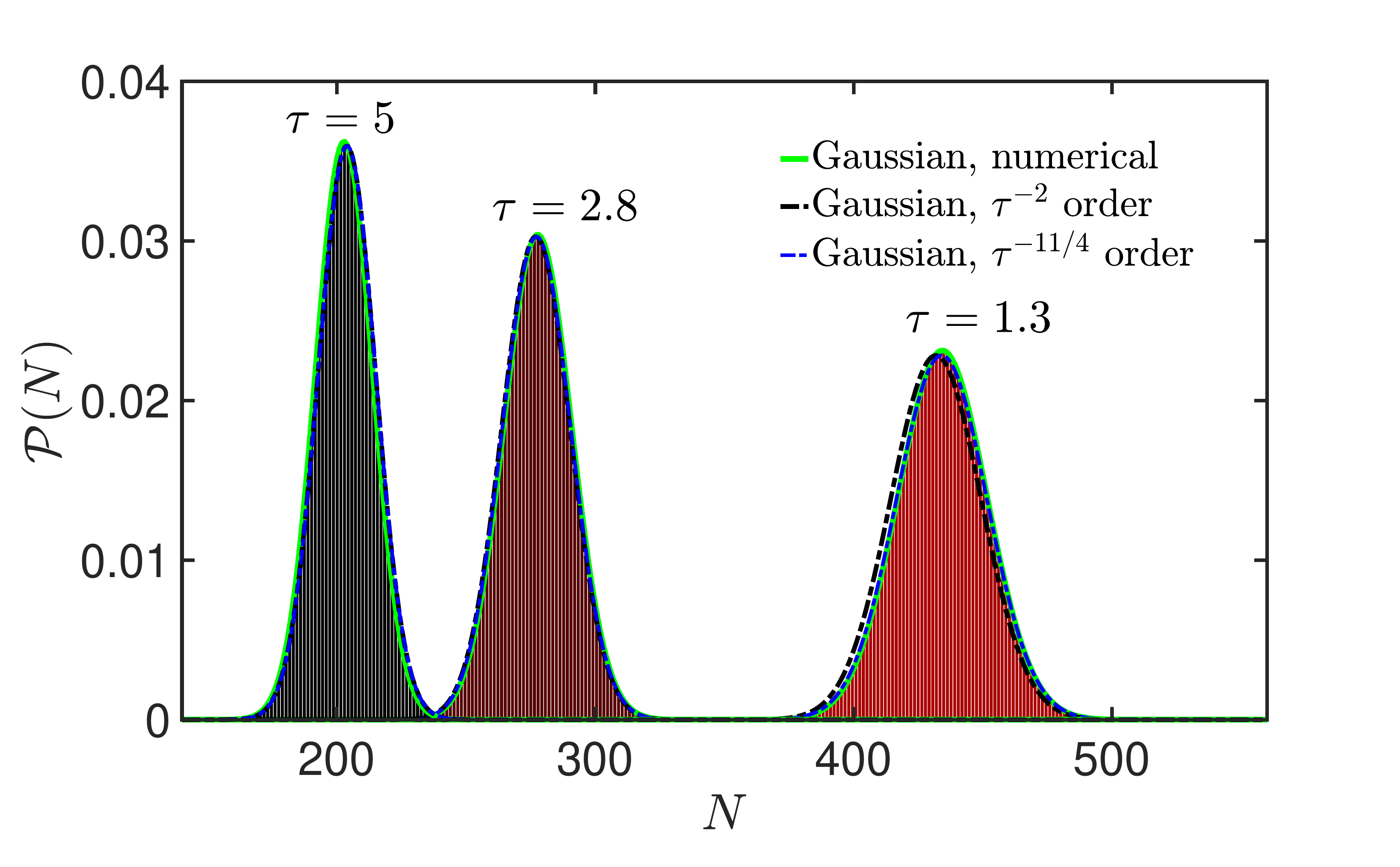}
    \caption{Histograms of the kink number distribution approaching the fast quench breakdown from the slow annealing limit. The shape of the distributions is approximately Gaussian, with corrections required to match the numerically exact histogram accurately.}
    \label{fig:Stat_slow}    
\end{figure}

\subsection{Cumulant generating function and cumulants in the sudden quench limit}

Let us start with the zeroth order cumulant generating function of the defect number distribution described by independent Bernoulli random variables with success probabilities given by  the first term in Eq.~\eqref{eq:p_k_fast}, $p_k\approx\cos^2\frac{k}{2}$,
%\begin{widetext}
\begin{equation}
    \log\tilde P(\theta;0) =\sum_{k>0}\log\left[1+\left(e^{2i\theta}-1\right)p_k\right]\approx \frac{L}{\pi}\int_0^\pi \mathrm dk \, \log \left[1+ \left(e^{2i\theta} - 1 \right) \cos^2\frac{k}{2} \right].
\end{equation}
The corresponding integral can be evaluated by expanding the logarithm in an analogous way as in Ref.~\cite{delCampo2018Universal}, $\log(1+x)=\sum_{p=1}^\infty (-1)^{p+1}x^p/p$,

\begin{eqnarray}\label{eq: logP_sudden_eps_f0}
    \log\tilde P(\theta;0) &\approx& \frac{L}{2\pi}\sum_{p=1}^\infty\,(-1)^{p+1}\frac{(e^{2i\theta}-1)^p}{p}\int_0^\pi\mathrm dk\,\cos^{2p} \frac{k}{2}=-\frac{L}{2}\sum_{p=1}^\infty\,(-1)^p\frac{(e^{2i\theta}-1)^p}{p}\,\frac{(2p)!}{4^p(p!)^2}\nonumber\\
    &=&-\frac{L}{2}\sum_{p=1}^\infty\,\frac{1}{p}\left(\frac{1-e^{2i\theta}}{4}\right)^p\binom{2p}{p}=-\frac{L}{2}\left[\log 4-2\log\left(1+e^{i\theta}\right)\right].
    \end{eqnarray}
%\end{widetext}
Here, we used the integral of $\int_0^\pi\mathrm dk\,\cos^{2p} \frac{k}{2}=\sqrt\pi\,\frac{\Gamma(p+1/2)}{\Gamma(p+1)}=\pi\frac{(2p)!}{2^{2p}(p!)^2}$ and that for integer values of $p$, $\Gamma(1+p)=p!$ and $\Gamma(p+1/2)=\frac{(2p)!}{4^pp!}\sqrt\pi $. Note that the criterion for the above series to converge, $\lvert\frac{e^{i\theta}-1}{4}\rvert<1$, is always satisfied. In addition, the following series representation was also used, $\sum_{p=0}^\infty\frac{x^p}{p}\binom{2p}{p}=\log 4-2\log\left(1+\sqrt{1-4x}\right)$.

From Eq.~\eqref{eq: logP_sudden_eps_f0} the cumulants can be generated by the relation
\begin{equation}
    \kappa_q=(-i\partial_\theta)^q \log\tilde P(\theta;\tau)\Big\vert_{\theta=0}.
\end{equation}
One can derive the exact relations for all the cumulants by looking at the first three and realizing that the derivatives are identical to those of the tangent function. This yields
\begin{align}
    &\kappa_1=-iL\partial_\theta\log(1+e^{i\theta})\Big\vert_{\theta=0}=\frac{L}{2},\\
    &\kappa_2=-iL\partial_\theta\frac{1}{1+e^{-i\theta}}\Big\vert_{\theta=0}=\frac{L}{4},\\
    &\kappa_3=-iL\partial_\theta\frac{1}{4\cos^2\frac{\theta}{2}}\Big\vert_{\theta=0}=-L\partial^2_\theta\tan\frac{\theta}{2}\Big\vert_{\theta=0}=0,
\end{align}
which explains the numerical observation in Fig.~\ref{fig:Ratios}, proving that $\kappa_2/\kappa_1=1/2$ and that $\kappa_3/\kappa_1$ identically vanishes. 
We note that the derivative of the cumulant generating function can be written as $-i\partial_\theta\log\tilde P(\theta;\tau)=L/2\tan\frac{\theta}{2}+L/2$. As a result, the $q$-th cumulant can be expressed as the $(q-1)$-th derivative of $L\tan\frac{\theta}{2}$ at $\theta=0$. One only needs to exploit the Taylor-series representation of the tangent to find
%\begin{widetext}
\begin{eqnarray}
\label{eq:cumulatns_tau=0eps_f=0}
    \kappa_{2q}&=&(-1)^q\frac{L}{2}\partial^{2q-1}_\theta \tan\frac{\theta}{2}\Big\vert_{\theta=0}
    =(-1)^q\frac{L}{2}\frac{(2q-1)!(-1)^{q-1}2^{2q}(2^{2q}-1)B_{2q}}{(2q)!2^{2q-1}}
    =-\frac{L}{2}\frac{2^{2q}-1}{q}B_{2q},\\
    \kappa_{2q+1}&=&0,\quad q=1,2,3,\dots
\end{eqnarray}
%\end{widetext}
where all odd cumulants are zero and $B_{2q}$ denotes the $2q$-th Bernoulli number.
The relative strength of the cumulants is best characterized by the successive ratios of them
\begin{equation}
    \kappa_{2q}/\kappa_{2(q-1)}=\frac{q-1}{q}\frac{2^{2q}-1}{2^{2q-2}-1}\frac{B_{2q}}{B_{2q-2}}
    \approx \frac{1}{q(\pi e)^2}\frac{q^{2q}}{(q-1)^{2q-2}}\approx 4\frac{q}{\pi^2},
\end{equation}
where we have exploited that for large enough $q$, $B_{2q}\approx4\sqrt{\pi q}\left(\frac{q}{\pi e}\right)^{2q}$ and that $q^{2q}/(q-1)^{2q-2}\approx (e\,q)^2$.

\subsection{Correction to the cumulant generating function and the cumulants near sudden quenches}
Based on the expansion of the transition probabilities in Eq.~\eqref{eq:p_k_fast}, that we denote by $p_k=\cos^2\frac{k}{2}+a_{1/2}\tau^{1/2}+a_1\tau+a_{3/2}\tau^{3/2}$ for compactness, one can deduce the corrections to the cumulant generating function for small values of $\tau$ as

%\begin{widetext}
\begin{eqnarray}\label{eq: LogP_tausmallaxpanded}
    \log\tilde P(\theta;\tau)&\approx& \frac{L}{2\pi}\int_0^\pi \mathrm dk \, \log \left[1+ \left(e^{2i\theta} - 1 \right) \left(\cos^2\frac{k}{2}+a_{1/2}\tau^{1/2}+a_1\tau+a_{3/2}\tau^{3/2}+a_2\tau^2\right) \right]\nonumber\\
    &\approx& \log\tilde P(\theta;0)+b_{1/2}(\theta)\tau^{1/2}+b_1(\theta)\tau+b_{3/2}(\theta)\tau^{3/2},
\end{eqnarray}
%\end{widetext}
where the coefficients become rather involved functions of $\theta$. The first two corrections are given by 
%terms acquire an analytically treatable and meaningful expression,
%\begin{widetext}
\begin{eqnarray}\label{eq:b_coefficients}
    b_{1/2}(\theta)&=&(e^{2i\theta}-1)\frac{L}{2\pi}\int_0^\pi\mathrm dk\,\frac{-\frac{\sqrt\pi}{2}\sin^2k}{1+(e^{2i\theta}-1)\cos^2\frac{k}{2}}=-\frac{\sqrt\pi L}{2}\frac{(e^{i\theta}-1)^2}{1-e^{-2i\theta}},\\
    b_1(\theta)&=&(e^{2i\theta}-1)\frac{L}{2\pi}\int_0^\pi\mathrm dk\,\frac{-\frac{\pi}{2}\sin^2k\cos k}{1+(e^{2i\theta}-1)\cos^2\frac{k}{2}}-\frac{(e^{2i\theta}-1)^2}{2}\frac{L}{2\pi}\int_0^\pi\mathrm dk\,\frac{\frac{\pi}{4}\sin^4k}{\left(1+(e^{2i\theta}-1)\cos^2\frac{k}{2}\right)^2}\nonumber\\
    &=&\frac{\pi L}{4}e^{2i\theta}\frac{-3+2e^{-i\theta}(1+e^{2i\theta})-\cos(2\theta)}{(e^{2i\theta}-1)^2}.
    \end{eqnarray}
%\end{widetext}
    At the following order, $O(\tau^{3/2})$, the corresponding expressions are rather lengthy.  We report them in App.~\ref{app: Cumulant_Generating_function}, where we further compare them to the numerically exact cumulant generating function.
% \begin{widetext}
     
% \end{widetext}

In the case of the first three cumulants, we provide the expansion up to order $O(\tau^2)$ to show their general structure. The first two orders can conveniently be obtained from the cumulant generated function, while the $\sim\tau^{3/2}$ and $\sim\tau^2$ orders are easier to be computed directly from the excitation probabilities given in Eq.~\eqref{eq:p_k_fast}
\begin{eqnarray}\label{eq: fast_corrections}
    \delta\kappa_2&&\approx-\frac{\pi}{16}\,L\tau+\frac{L}{48}\left[16+9\pi^2-6\pi\left(2+\log32\right)\right]\tau^2,\\
    \delta\kappa_3&&\approx\frac{\sqrt\pi L}{8}\tau^{1/2}
    -\frac{\sqrt\pi L}{64}\left(4+\pi-\log 64\right)\tau^{3/2}.
\end{eqnarray}
Note that all further cumulants involve non-universal behavior apart from the fact that the second cumulant contains even powers of $\sqrt\tau$. In contrast, only odd powers of $\sqrt\tau$ enter the third cumulant and the average in Eq.~\eqref{eq: average_tau_small}. 

Finally, we characterize the maxima of the ratios $\kappa_2/\kappa_1$ and $\kappa_3/\kappa_1$ as a function of $\tau$. Numerically, they are found to be located at $\tau^{1,2}_\mathrm{max}\approx0.63,\quad\tau^{3,1}_\mathrm{max}\approx0.76$ reaching values $\mathrm{max}_\tau\,\kappa_2/\kappa_1\approx0.76$ and $\mathrm{max}_\tau\,\kappa_3/\kappa_1\approx0.48$. As a note, these values can be approximately determined analytically using the corrections given in Eq.~\eqref{eq: fast_corrections}, consistent with the driving times in the fast quench regime. As detailed in App.~\ref{app: max_cumulant_ratios}, the maximal driving times captured within the $O(\tau^2)$  cumulant expansion results in remarkably precise matching with the numerical values, $\tilde\tau^{1,2}_\mathrm{max}\approx0.64$ and $\tilde\tau^{1,3}_\mathrm{max}\approx0.75$. The associated maximum values can be obtained by plugging back $\tilde\tau^{1,2}_\mathrm{max}$ and $\tilde\tau^{1,3}_\mathrm{max}$ into Eq.~\eqref{eq: fast_corrections} yielding $\mathrm{max}_\tau\,\kappa_2/\kappa_1\approx0.77$ and $\mathrm{max}_\tau\,\kappa_3/\kappa_1\approx0.52$.

\end{widetext}

\begin{figure}
    \includegraphics[width=1.\linewidth]{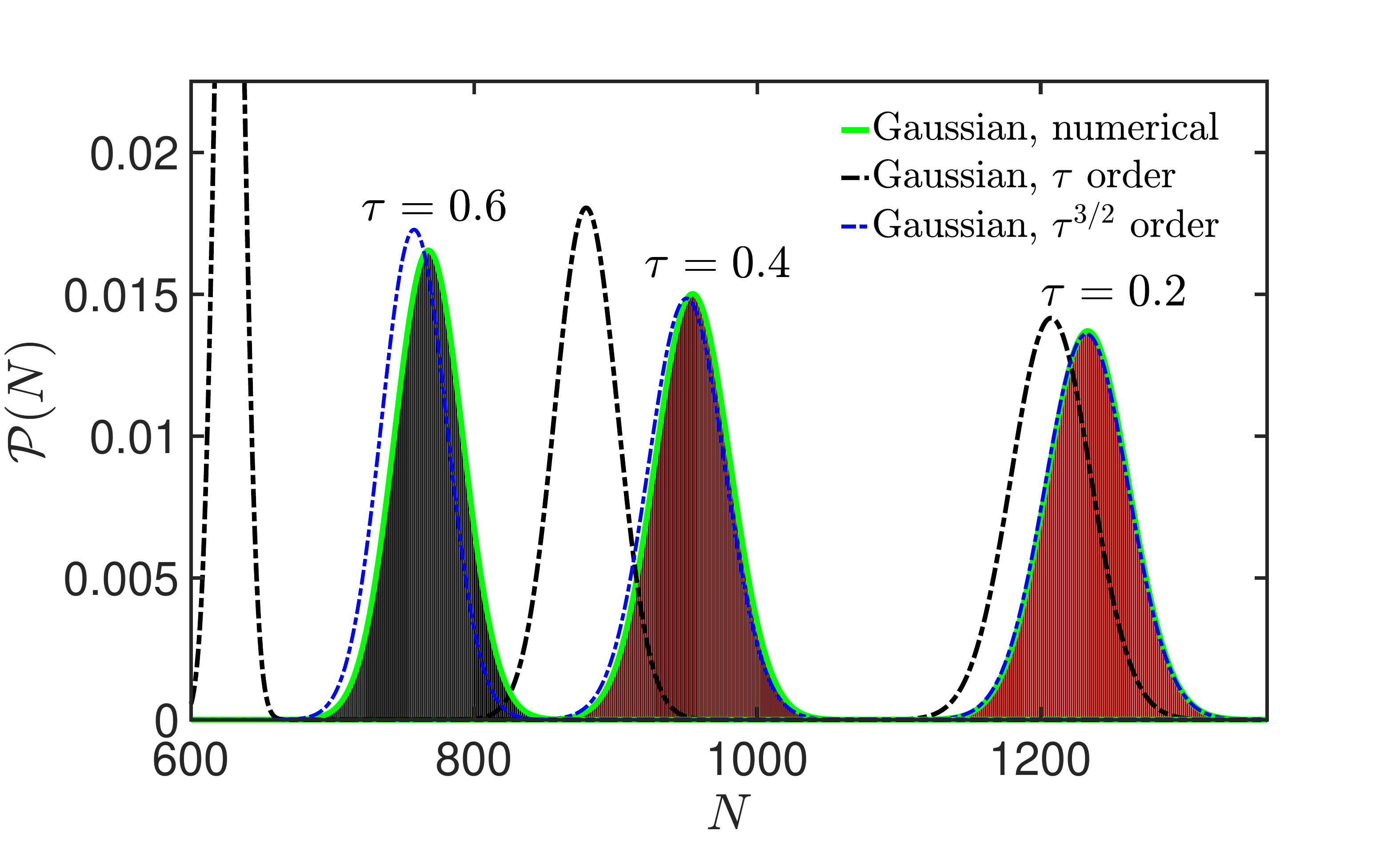}
    \caption{Histograms of the kink number distribution approaching the fast quench breakdown from the sudden quench limit for $L=4000$ and driving times $\tau=0.2,\,0.4,\,0.2$.}
    \label{fig:Stat_slow2}    
\end{figure}

\section{Summary and Outlook}

In the study of nonequilibrium quantum phenomena, the transverse-field Ising chain has played a key role historically. Thanks to its exact solvability even for processes driven at finite rate, its use was decisive in establishing the validity of the KZM in the quantum domain ~\cite{Zurek2005Dynamics,Dziarmaga2005Dynamics}, its generalization to inhomogeneous systems~\cite{Collura10,DziarmagaRams10,GomezRuiz19}, and its suppression by counterdiabatic driving~\cite{delcampo12,Saberi14}. Its analysis also made possible the discovery of universality beyond KZM~\cite{delCampo2018Universal,Balducci2023Large} and the KZM breakdown at fast quenches~\cite{Huabi2023FastQuench}. 

Using it, we have provided a thorough analysis of the quantum critical dynamics resulting from a driving protocol across a quantum critical point at an arbitrary rate. In doing so, we have put to the test the fast-quench scaling of the defect density recently introduced~\cite{Chesler2015,Huabi2023FastQuench,Xia23prd,Xia2024}, providing the first rigorous analytical evidence supporting it.
Further, we have reported a complete characterization of the defect number distribution encompassing both the slow and fast quench limits.
In particular, we characterized the cumulant generating function of excitations at arbitrary driving times and computed exact correction terms to the limiting cases of sudden and slow driving.

First, we considered the slow limit, where we derived a new analytical result for the excitation probabilities beyond the simplest Landau-Zener approximation. In particular, we showed that the LZ formula loses its validity above a given momentum scaling with the driving time as $k\gtrsim \tau^{-1/4}$. This becomes especially relevant close to the fast quench breakdown, where the simple LZ prediction loses its validity. In this regime, the first two cumulants, as well as the cumulant generating function, are reproduced up to excellent precision compared to the numerical results. 
Remarkably, the leading and next-to-leading order corrections are identical for all cumulants.
This implies the breakdown of the proportionality between the cumulants around the fast quench breakdown.

In the fast annealing regime, as a first step, we derived the exact form of the cumulant generating function in the sudden quench limit. Next, we computed the first three correction terms for the excitation probabilities, precisely reproducing the parabolic cylinder functions around the fast quench breakdown, $\tau\lesssim1$. With the help of these findings, we provided exact corrections to the cumulant generating function. In contrast with the slow annealing limit, the fast quench cumulant corrections exhibit non-universal behavior and vary with the order of the cumulant.

Finally, we further demonstrated the precision of our results by comparing the numerically obtained excitation statistics with the limiting Gaussian distribution with the analytical average and variance. In both cases of approaching from the sudden quench and the slow annealing limit, a precise agreement was found. An analysis of the cumulant rations across the complete range of driving rates indicates that the kink number statistics remain sub-Poissonian, with the value of the cumulant ratios varying nonmonotonically from the slow to the sudden quench limit. 

Our theoretical predictions are readily testable with current programmable quantum annealing devices and digital simulators, see, e.g., Refs.~\cite{Cui20,Keesling19,Bando20,King22}. We expect that our findings provide a guide for further studies, e.g., in exploring the role of multicritical points and critical lines~\cite{Divakaran09,Dziarmaga10,Suzuki2012Quantum}, nonlinear quenches~\cite{Diptiman08,Barankov08,GomezRuiz19,Grabarits24NAQO}, inhomogeneous systems~\cite{DKZ13}, systems with long-range interactions~\cite{Caneva08,Jaschke17,Dutta17,Balducci2023Large,Gherardini24}, disorder and a spin-glass character~\cite{Dziarmaga06,Caneva07}, shortcuts to adiabaticity~\cite{Hatomura24}, and complex topologies beyond one spatial dimension~\cite{Sengupta08,Schmitt22,King24,Ali2024}. The relevance of our study is further enhanced by the limited coherence times in quantum annealing devices and the available depths of quantum circuits using the gate model for digital quantum simulation and computation, which currently restrict the study of coherent dynamics to moderate and fast quenches.  Our results can also be applied to advance and generalize specific applications in other fields through the analysis of driven critical processes at arbitrary rates, i.e., in the quantum thermodynamics of quantum critical devices~\cite{Revathy20}, quantum critical metrology~\cite{Rams18}, and dynamical phase transitions~\cite{Zhang2024dqpt}. 
To conclude, we note that the quest for universality as a function of the quench depth for arbitrary driving rate remains an intriguing prospect.

\acknowledgments
This project was supported by the Luxembourg National Research Fund (FNR Grant Nos.\ 17132054 and 16434093). It has also received funding from the QuantERA II Joint Programme and co-funding from the European Union’s Horizon 2020 research and innovation programme.

\appendix
\onecolumngrid
\section{Probability density function and cumulants of kinks and kink pairs.}\label{app: PDF_CGF_kink_kink_pair}
In this appendix, we provide a summary of the probability density function and the cumulants of both the number of kinks and the number of kink pairs in the TFIM \cite{Cincio07,delCampo2018Universal,Cui20,Bando20}. As a consequence of the periodic boundary condition, kinks are always formed in pairs, whose number takes values $\tilde N=0,\,1,\,\dots,\,L/2$. This is the case as in a ring of $L$ sites at most $L$ domain walls can be formed corresponding to $L/2$ pairs. In the TFIM, a kink pair corresponds to the simultaneous excitations of the TLSs at momentum $k$ and $-k$, where, to fix the notation, we use expressions with $k>0$ over the positive momentum sector. Thus, the probability density function of the number of kink pairs is given by the sum of $L/2$ independent Bernoulli random variables with success probability $P(X_k=1)=p_k$ and $P(X_k=0)=1-p_k,\quad k=\frac{\pi}{L},\,\frac{3\pi}{L},\,\dots,\,\pi-\frac{\pi}{L}$, where $X_k$ denotes the random variable of the possible number $(0 \text{ or } 1)$ in the $k$-th TLS. This results in the following Poisson-binomial distribution,
\begin{eqnarray}
    P(\tilde N)=\sum_{X_1,\,X_2,\,\dots,\,X_{L/2}\in\left\{0,1\right\}^{L/2 
    }}\prod_{k>0}\,\left[\delta_{0,X_k}(1-p_k)+\delta_{1,X_k}p_k\right]\,\delta_{\tilde N,\sum_{k>0}X_k},
\end{eqnarray}
where the last Kronecker delta restricts the summation only to those cases where only $\tilde N$ number of $X_k$ equals one, and the rest is zero. 
As a result, the probability density function of the number of kinks is given by the sum of independent discrete random variables with success probability  $P(Y_k=2)=p_k$ and $P(Y_k=0)=1-p_k,\quad k=\frac{\pi}{L},\,\frac{3\pi}{L},\,\dots,\,\pi-\frac{\pi}{L}$. The corresponding kink number probability density function reads as
\begin{eqnarray}
    P(N)=\sum_{Y_1,\,Y_2,\,\dots,\,Y_{L/2}\in\left\{0,2\right\}^{L/2}}\prod_{k>0}\,\left[\delta_{0,Y_k}(1-p_k)+2\delta_{1,Y_k}p_k\right]\,\delta_{N,\sum_{k>0}Y_k},
\end{eqnarray}
where $N$ was introduced for the number of kinks.
In both cases, we omitted the $\tau$ dependence of the excitation probabilities for brevity.

The cumulants of the kink pair number provide deeper analytical insight. The cumulant generating function contains all the information about the cumulants. It is defined as the logarithm of the characteristic function of the kink pair number, $\tilde P_{\tilde N}(\theta)=\mathbb E\left[e^{i\theta\tilde N}\right]$,
\begin{eqnarray}
    \log\tilde P_{\tilde N}(\theta)=\log\mathbb E\left[e^{i\theta \tilde N}\right]=\log\left[\prod_{k>0}1+\left(e^{i\theta }-1\right)p_k\right]=\sum_{k>0}\log\left[1+\left(e^{i\theta }-1\right)p_k\right].
\end{eqnarray}
%The last expression allows one to conclude that the kink pair number distribution is a Poisson binomial distribution. 
The corresponding kink pair cumulants are generated via the relation $\kappa_q=(-i\partial_\theta)^q\log\tilde P_N(\theta)\Big\vert_{\theta=0}$. The first three cumulants equal
\begin{eqnarray}
    \tilde\kappa_1=\sum_{k>0}\,p_k,\quad\tilde\kappa_2=\sum_{k>0}\,p_k(1-p_k),\quad\tilde\kappa_3=\sum_{k>0}\,p_k(1-p_k)(1-2p_k).
\end{eqnarray}

As each kink pair involves $2$ kinks, the kink number cumulant generating function inherits a multiplicative factor, 
\begin{eqnarray}
    \log\tilde P_{N}(\theta)=\log\mathbb E\left[e^{i\theta N}\right]=\log\left[\prod_{k>0}1+\left(e^{i2\theta }-1\right)p_k\right]=\sum_{k>0}\log\left[1+\left(e^{i2\theta }-1\right)p_k\right]\equiv\log\tilde P_{\tilde N}(2\theta).
\end{eqnarray}
Thus, the $q$-th kink number cumulant  differs from the kink-pair number cumulants by a factor of $2^q$,
\begin{eqnarray}
    \kappa_q=(-i\partial_\theta)^q\log\tilde P_N(\theta)\Big\vert_{\theta=0}=(-i\partial_\theta)^q\log\tilde P_{\tilde N}(2\theta)\Big\vert_{\theta=0}=2^q(-i\partial_\theta)^q\log\tilde P_{\tilde N}(\theta)\Big\vert_{\theta=0}\equiv2^q\,\tilde\kappa_q.
\end{eqnarray}
For completeness, we also write out the first three cumulants of the kink number distribution
\begin{eqnarray}
    \kappa_1=2\sum_{k>0}\,p_k,\quad\kappa_2=4\sum_{k>0}\,p_k(1-p_k),\quad\kappa_3=8\sum_{k>0}\,p_k(1-p_k)(1-2p_k).
\end{eqnarray}

\section{Derivation of the exact solution of the Landau-Zener problem}
\label{app: LZ_exact}

In this appendix, we show the analytical steps to derive the exact solutions for the Schr\"odinger equation, Eq.~\eqref{eq:TLS_eq}.
We start by introducing the new variables
\begin{equation}
    s \equiv 2J \sin k \, (t + \tau \cos k) \qquad \text{ and } \qquad
    \phi_{k,i}(s) \equiv \psi_{k,i} (t),\quad i=1,2.
\end{equation}
Using them, the Schr\"odinger equation reads
\begin{equation}
    i \partial_s
    \begin{pmatrix}
        \phi_{k,1} \\
        \phi_{k,2}
    \end{pmatrix}
    = \left(-\frac{s}{\tau'} \tau^z + \tau^x \right)
    \begin{pmatrix}
        \phi_{k,1} \\
        \phi_{k,2}
    \end{pmatrix},
\end{equation}
with a new parameter given by
\begin{equation}
    \tau' \equiv 2J\tau \sin^2 k.
\end{equation}
We expand this system of equations for clarity as
\begin{eqnarray}
    \label{eq:EoM_system}
%    \begin{cases}
        i \partial_s \phi_{k,1} &=& -\frac{s}{\tau'} \phi_{k,1} + \phi_{k,2},\\
        i \partial_s \phi_{k,2} &=& \phi_{k,1} + \frac{s}{\tau'} \phi_{k,2}.
%    \end{cases}
\end{eqnarray}
We now reduce the system to one single second-order ODE:
\begin{eqnarray}\label{eq: s_ODE}
    i \partial_{ss} \phi_{k,1} &=& -\frac{1}{\tau'} \phi_{k,1} - \frac{s}{\tau'} \partial_s \phi_{k,1} + \partial_s \phi_{k,2} \nonumber\\
    &=& -\frac{1}{\tau'} \phi_{k,1} - \frac{s}{\tau'} \partial_s \phi_{k,1} - i \left[ \phi_{k,1} + \frac{s}{\tau'} \left( i \partial_s \phi_{k,1} + \frac{s}{\tau'} \phi_{k,1} \right)\right]\nonumber\\
    \label{eq:eq_psi1}
    &=& \left[ -i - \frac{1}{\tau'} - i \frac{s^2}{(\tau')^2}\right] \phi_{k,1}.
\end{eqnarray}
This equation can be reduced to the Weber differential equation
\begin{equation}\label{eq: Weber}
    \partial_{zz} W(z) + \left( \nu + \frac{1}{2} - \frac{z^2}{4} \right) W(z) = 0,
\end{equation}
whose general solution is expressed in terms of two of the four parabolic cylinder functions $D_{\nu}(z)$, $D_\nu(-z)$, $D_{-\nu-1}(iz)$ and $D_{-\nu-1}(-iz)$. To do so, let us set $z \equiv s/\alpha$, from which
\begin{equation}
    \partial_{zz} \phi_{k,1} = \left[ -\alpha^2 + \frac{i\alpha^2}{\tau'} + \frac{\alpha^4 z^2}{(\tau')^2}\right] \phi_{k,1}.
\end{equation}
It follows that
\begin{equation}
    \alpha = e^{i\pi/4} \sqrt{\frac{\tau'}{2}} \qquad \text{ and } \qquad
    \nu = \frac{i\tau'}{2}.
\end{equation}
To arrive at the canonical form of the Weber equation, one needs to introduce the new variable  $z=s\,e^{\pm i\pi/4}\sqrt{2/\tau^\prime}$. This choice ensures the correct form of the quadratic term, $-z^2/4$. Ensuring that the $1/2$ constant appears, with an overall minus sign, singles out the choice $z=s\,e^{-i\pi/4}\sqrt{2/\tau^\prime}$, such that
\begin{equation}
    \partial_{zz}\phi_{k,1}=-\left(\frac{-\tau^\prime}{2i}+\frac{1}{2}-\frac{z^2}{4}\right)\phi_{k,1}.
\end{equation}
This equation identifies $\nu$ and keeps the canonical form of the Weber equation.

Next, we choose two out of the four functions, $D_{\nu}(z)$, $D_\nu(-z)$, $D_{-\nu-1}(iz)$ and $D_{-\nu-1}(-iz)$, to express the solution to Eq.~\eqref{eq:eq_psi1}. To do so, we look at the expansions for $s\to -\infty$, and find that only $D_{-\nu-1}(-iz)$ vanishes. In particular~\cite{NIST-DLMF},
\begin{equation}
    \label{eq:asymp_Dnu_large_arg}
    D_\nu(z) \simeq e^{-z^2/4} z^\nu, \qquad \text{ for }|\arg z| < \frac{3\pi}{4}, \; |z|\to \infty.
\end{equation}
Note the important detail that due to the limit of $s\rightarrow-\infty$ the phase is shifted by $i\pi$, so that $\arg (-iz)=\pi/4$. To ensure the bound on the phase, $D_{-\nu-1}(iz)$ is excluded. The exponent thus remains purely imaginary, so the solution can only vanish via the polynomial term, $z^\nu$. This can only be realized when the exponent has a finite real part leading to the choice of $D_{-1-\nu}(-iz)$.

Since we need to enforce $\psi_{k,1}(-\infty)=0$, we conveniently set 
\begin{equation}
    \phi_{k,1}(s) = \mathcal{A} \, D_{-i\tau'/2-1}\left( e^{-3i\pi/4}\sqrt{\frac{2}{\tau'}} s \right),
\end{equation}
with some proportionality constant $\mathcal{A}$, that can be fixed by further imposing $|\phi_{k,2}(-\infty)| = 1$. To determine $\phi_{k,2}$, one may use Eq.~\eqref{eq:EoM_system}:
\begin{equation}
    \phi_{k,2}(s) = i \partial_s \phi_{k,1}(s) + \frac{s}{\tau'} \phi_{k,1}(s).
\end{equation}
Using the relation
\begin{equation}
    {z} D_\nu(z) = \frac{z}{2} D_\nu(z) - D_{\nu+1}(z),
\end{equation}
one finds
\begin{equation}
    \phi_{k,2}(s) = \mathcal{A} \sqrt{\frac{2}{\tau'}} e^{3i\pi/4} D_{-i\tau'/2}\left( e^{-3i\pi/4} \sqrt{\frac{2i}{\tau'}} s \right).
\end{equation}
Using the asymptotic expansion of the parabolic cylinder functions for large arguments [for the phases of the particular $\nu$ and $z$ under consideration, see Eq.~\eqref{eq:asymp_Dnu_large_arg}], one finds 
\begin{equation}
    \left\lvert \mathcal A\sqrt{\frac{2}{\tau^\prime}}e^{3i\pi/4}e^{-\frac{i s^2}{2\tau^\prime}}\sqrt{\frac{2i}{\tau^\prime}}^{\frac{-i\tau^\prime}{2}}s^{\frac{-i\tau^\prime}{2}}\right\rvert=
    \left\lvert  \mathcal A\sqrt{\frac{2}{\tau^\prime}}e^{-\frac{i s^2}{2\tau^\prime}}e^{\frac{-i\tau^\prime}{2}\left(\log(-s)+\frac{1}{2}\log(2/\tau^\prime)+\frac{3}{2}\pi\,i \right)}\,e^{\pi\tau^\prime/8}\right\rvert,
\end{equation}
from where only the real exponents survive, giving
\begin{equation}
    1 = |\mathcal{A}| \sqrt{\frac{2}{\tau'}} e^{\pi\tau' / 8},
\end{equation}
and finally, up to an overall phase,
\begin{eqnarray}
    \phi_{k,1}(s) &=& \sqrt{\frac{\tau'}{2}} e^{-\frac{\pi}{8} \tau'} D_{-i\tau'/2-1} \left( e^{-3i\pi/4} \sqrt{\frac{2}{\tau'}}s \right),\nonumber\\
    \phi_{k,2}(s) &=& e^{-\frac{\pi}{8} \tau' +3i\pi/4} D_{-i\tau'/2} \left( e^{-3i\pi/4} \sqrt{\frac{2}{\tau'}}s \right).
\end{eqnarray}
Substituting the definitions of $\tau^\prime$ and $s$, one obtains the final wave function components at $t=0$ as given in Eq.~\eqref{eq:psi}. 

According to the expansion $D_\nu(z)\approx e^{-z^2/4}z^\nu$, the exciation probability reads, in the leading order of $k$, 
\begin{eqnarray}
    p_k&\approx& \Bigg\lvert\sin\frac{k}{2} \sqrt\tau \sin k e^{-\frac{\pi}{4}\tau\sin^2k}e^{-i\tau\cos^2k}\left(2\,e^{-3i\pi/4}\sqrt\tau\cos k\right)^{-i\tau\sin^2k-1}\nonumber\\
    & &+
    e^{-\frac{\pi}{4}\tau\sin^2k+3i\pi/4}e^{-i\tau\cos^2k}\left(2\,e^{-3i\pi/4}\sqrt\tau\cos k\right)^{-i\tau\sin^2k}\cos\frac{k}
    {2}\Bigg\rvert^2.
\end{eqnarray}
In the limit considered here, $k\lesssim\tau^{-1/4}$ only the second term survives in the leading order and yields the Landau-Zener formula by further taking the leading order limit of $\cos\frac{k}{2}\approx1$,
\begin{equation}
    p_k\approx e^{-2\pi k^2\,\tau}.
\end{equation}

\section{Saddle point approximations in the uniform asymptotic expansion of $\psi_{k,1}$}\label{app:Saddle_expansion_1}

In this appendix, we derive the saddle point approximation~\cite{Bleistein86} for the first component of the time-evolved wave function at $t=0$. We make use of the integral representation of the parabolic cylinder function given in Eq.~\eqref{eq: integral_representation}
\begin{equation}\label{eq: integral_representation_app}
    D_{-ip\tau-1} \left(e^{-3i\pi/4}\sqrt{\tau}q \right) = \frac{e^{-i \tau q^2/4}}{\Gamma(1+ip\tau)} \int_0^\infty dt \, e^{-e^{-3i\pi/4}\sqrt{\tau}qt - \frac{1}{2}t^2} t^{ip\tau},
\end{equation}
 with the new variables defined as
\begin{equation}
    \label{eq:def_pq}
    p \equiv  \sin^2 k>0, \qquad
    q \equiv 2 \cos k.
\end{equation}
To perform the saddle point approximation,  we define $x \equiv t/\sqrt{\tau}$, bringing the integral into the Laplace's form,
\begin{equation}
    D_{-ip\tau-1} \left(e^{-3i\pi/4}\sqrt{\tau}q \right) = \frac{e^{-i \tau q^2/4}}{\Gamma(1+ip\tau)} \tau^{(1+ip\tau)/2} \int_0^\infty dx\, e^{\tau f(x)} ,
\end{equation}
having further defined
\begin{equation}
    f(x) \equiv -e^{-3i\pi/4}qx - \frac{1}{2}x^2 + ip \log x .
\end{equation}
The saddle point is found by imposing 
\begin{equation}
    f'(x) = 0 = -e^{-3i\pi/4}q - x +\frac{ip}{x} \qquad \implies \qquad
    x^*_\pm = e^{i\pi/4} (\cos k \pm 1).
\end{equation}
We also compute, for later convenience,
\begin{equation}
    f(x^*_\pm) = \frac{i}{2} (3\cos^2 k \pm 2 \cos k -1) + i\sin^2 k \log\left[ e^{i\pi/4} (\cos k \pm 1)\right],
\end{equation}
along with
\begin{equation}
    f''(x) = - 1 - \frac{ip}{x^2} \qquad \implies \qquad
    f''(x^*_\pm) = \frac{2}{\mp\cos k -1}.
\end{equation}
The saddle-point approximation entails
\begin{equation}
    \int dx \, e^{ \lambda f(x)} = \sum_{x^*} \sqrt{\frac{2\pi}{-\lambda f''(x^*)}} e^{\lambda f(x^*)} \left[ 1 + O\left( \frac{1}{\lambda} \right) \right],
\end{equation}
where the sum runs on the admissible saddles, i.e., the ones such that the integration path can be deformed to pass through them. In this case, only $x^*_+$ is admissible, as can be checked from a contour plot of $\Re f(x)$.
 For the gamma function in the denominator of Eq.~\eqref{eq: integral_representation_app}, we use the expansion of
\begin{equation}
    \label{eq:expansion_Gamma_imag}
    \Gamma(1+iy) = \sqrt{2\pi} \, (iy)^{iy+1/2} e^{-iy} \left[ 1 - \frac{i}{12y} + O\left(\frac{1}{y^2} \right)\right],
\end{equation}
that is valid for $y\to + \infty$.
The total expression reads
\begin{eqnarray}\label{eq:psi1_asymp0}
    &&\sqrt{\tau}\sin k\,e^{-\frac{\pi}{4}\tau\sin^2k}\,e^{-i\tau\cos^2k}\tau^{(1+i\tau\sin^2k)/2}\sqrt{\frac{\pi(\cos k+1)}{\tau}}\,e^{i\frac{\tau}{2}\left(3\cos^2k+2\cos k-1\right)}\nonumber\\
    &&\times e^{i\tau\,\sin^2k\log((\cos k+1))}e^{-\frac{\pi}{4}\tau\sin^2k}\,\frac{e^{i\tau\sin^2k}}{\sqrt{2\pi}}\,(i\tau\sin^2k)^{-i\tau\sin^2k-1/2}\Rightarrow\\
    \psi_{k,1}(0)&&=e^{-i\pi/4}\left(\tau\sin^4k\right)^{-\frac{i}{2}\tau\sin^2k}e^{-i\frac{\tau}{2}\left(\cos^2k-2\cos k-1\right)+i\tau\,\sin^2k\log(\cos k+1)}\cos k/2.
\end{eqnarray}

Next, we show the steps to arrive at the next-to-leading order expansion of the first components. This is given by the expression in terms of the function $f(x)$ for the admissible saddles~\cite{Schafer1967Higher}

\begin{equation}
    \label{eq:saddle-point}
    \int dx \, e^{ \lambda f(x)} = \sum_{x^*} \sqrt{\frac{2\pi}{-\lambda f''(x^*)}} e^{\lambda f(x^*)} \left[ 1 + \frac{3f''''(x^*)f''(x^*)-5(f'''(x^*))^3}{24\lambda (f''(x^*))^3} + O\left( \frac{1}{\lambda^2} \right) \right].
\end{equation}
The additional correction term is given by $ \frac{3f''''(x^*)f''(x^*)-5(f'''(x^*))^3}{24\lambda (f''(x^*))^3}=\frac{i}{2} (4+5\cos k) \tan^2 \frac{k}{2}$. This leads to the final result,  
\begin{eqnarray}\label{eq:psi1_saddle_2_appendix}
        \psi_{k,1}(0)&\approx& e^{-i\pi/4} (\tau \sin^4 k)^{-\frac{i}{2}  \tau \sin^2 k} \cos \frac{k}{2} \left(1 + i \frac{4+5\cos k}{48\tau} \tan^2 \frac{k}{2} + \frac{i}{12 \tau \sin^2 k}\right)\nonumber\\
    & &\times \exp \left[ -\frac{i}{2} \tau (\cos^2 k -2\cos k-1) +i \tau \sin^2 k \, \log (1+\cos k) \right].
\end{eqnarray}

\section{Saddle point approximation in the uniform asymptotic expansion of $\psi_{k,2}$}\label{app:Saddle_expansion_2}

Exploiting the integral representation for the second components given in Eq.~\eqref{eq:saddle_inegral_psi2}, one can write again
\begin{equation}
    D_{-ip\tau} \left( e^{-3i\pi/4} q \sqrt{\tau} \right) = \sqrt{\frac{1}{2\pi}} e^{\frac{i}{4} q^2 \tau} \int_0^\infty dt \, t^{-ip\tau} e^{-\frac{1}{2}t^2} \left( e^{e^{-i\pi/4}q\sqrt{\tau}t - \frac{\pi}{2} p\tau} + e^{-e^{-i\pi/4}q\sqrt{\tau}t + \frac{\pi}{2} p\tau} \right),
\end{equation}
with the same $p$ and $q$ defined in Eq.~\eqref{eq:def_pq}. The two integrals can now be evaluated with the saddle-point method. The first one reads
\begin{equation}
    \int_0^\infty dt \, t^{-ip\tau} e^{-\frac{1}{2}t^2 + izt} = \tau^{(1-ip\tau)/2} \int_0^\infty dx \, e^{\tau f(x)},
\end{equation}
with $x \equiv t/\sqrt{\tau}$ as before, and 
\begin{equation}
    f(x) = -ip \log x -\frac{1}{2}x^2 + e^{-i\pi/4}qx.
\end{equation}
Proceeding as done above, one finds two saddles, 
\begin{equation}
    x^*_\pm = e^{-i\pi/4} (\cos k \pm 1),
\end{equation}
yielding
\begin{equation}
    f(x^*_\pm) = -\frac{i}{2}  (3 \cos^2 k \pm 2 \cos k -1) - i \sin^2 k \, \log \left[ e^{-i\pi/4}  (\cos k \pm 1) \right]
\end{equation}
and
\begin{equation}
    f''(x^*_\pm) = \frac{2}{\mp \cos k -1}.
\end{equation}
The only admissible saddle is $x^*_+$, as the integration path cannot be deformed through $x^*_-$ without avoiding the singularity at infinity. Combining everything together, one obtains 
\begin{eqnarray}
    \psi_{k,2}(0) &\approx& e^{-\frac{\pi}{4}\tau\sin^2k+3i\pi/4}\sqrt{\frac{1}{2\pi}}e^{i\tau\cos^2k}\tau^{(1-i\tau\sin^2k)/2}\sqrt{\frac{\pi(1+\cos k)}{\tau}}e^{-\frac{\pi}{2}\tau\sin^2k}\nonumber\\
    & &\times e^{-\frac{i}{2}  (3 \cos^2 k + 2 \cos k -1) - i \sin^2 k \, \log \left[ e^{-i\pi/4} (\cos k + 1) \right]}\\
    &=&e^{3i\pi/4} \tau^{-\frac{i}{2}  \tau \sin^2 k} \cos \frac{k}{2}  e^{ -\frac{i}{2} \tau (\cos^2 k + 2\cos k -1) -i \tau \sin^2 k \log (1+\cos k)  -\pi \tau \sin^2 k}.
\end{eqnarray}

Next, we also derive the next-to-leading order expansion in a similar way as in Eq.~\eqref{eq:saddle-point} in App.~\ref{app:Saddle_expansion_1}. The correction term is
\begin{equation}    \frac{3f''''(x^*)f''(x^*)-5(f'''(x^*))^3}{(f''(x^*))^3} = \frac{i (5 \cos k -4)}{2} \cot^2 \frac{k}{2},
\end{equation}
Thus, the final result within the next-to-leading order approximation for the second component reads as
\begin{eqnarray}
    \label{eq:psi2_saddle_2_appendix}
    \psi_{k,2}(0)&\approx& e^{3i\pi/4} \tau^{-\frac{i}{2}  \tau \sin^2 k} \sin \frac{k}{2} \left( 1 + i\frac{5\cos k-4}{48 \tau} \cot^2 \frac{k}{2} \right) \nonumber\\
    & & \times e^{ -\frac{i}{2} \tau (\cos^2 k - 2\cos k -1) -i \tau \sin^2 k \log (1-\cos k)}.
\end{eqnarray}

In the final excitation probability in the limit of $k\gtrsim \tau^{-1/4}$, the leading order expressions cancel out as they have the same phase factor and the sine and cosine terms in the opposite way as in the final excited state amplitudes; see Eq.~\eqref{eq:psi1_saddle_2_appendix} and Eq.~\eqref{eq:psi2_saddle_2_appendix} for the first and second components, respectively. As a result, one only needs to consider the next-to-leading order terms. It can be verified that 
\begin{equation}
    p_k=\left(\frac{\sin k}{8\tau}\right)^2.    
\end{equation}

\begin{figure}
\centering
\begin{tabular}{c c}
    \includegraphics[width=.45\linewidth]
    {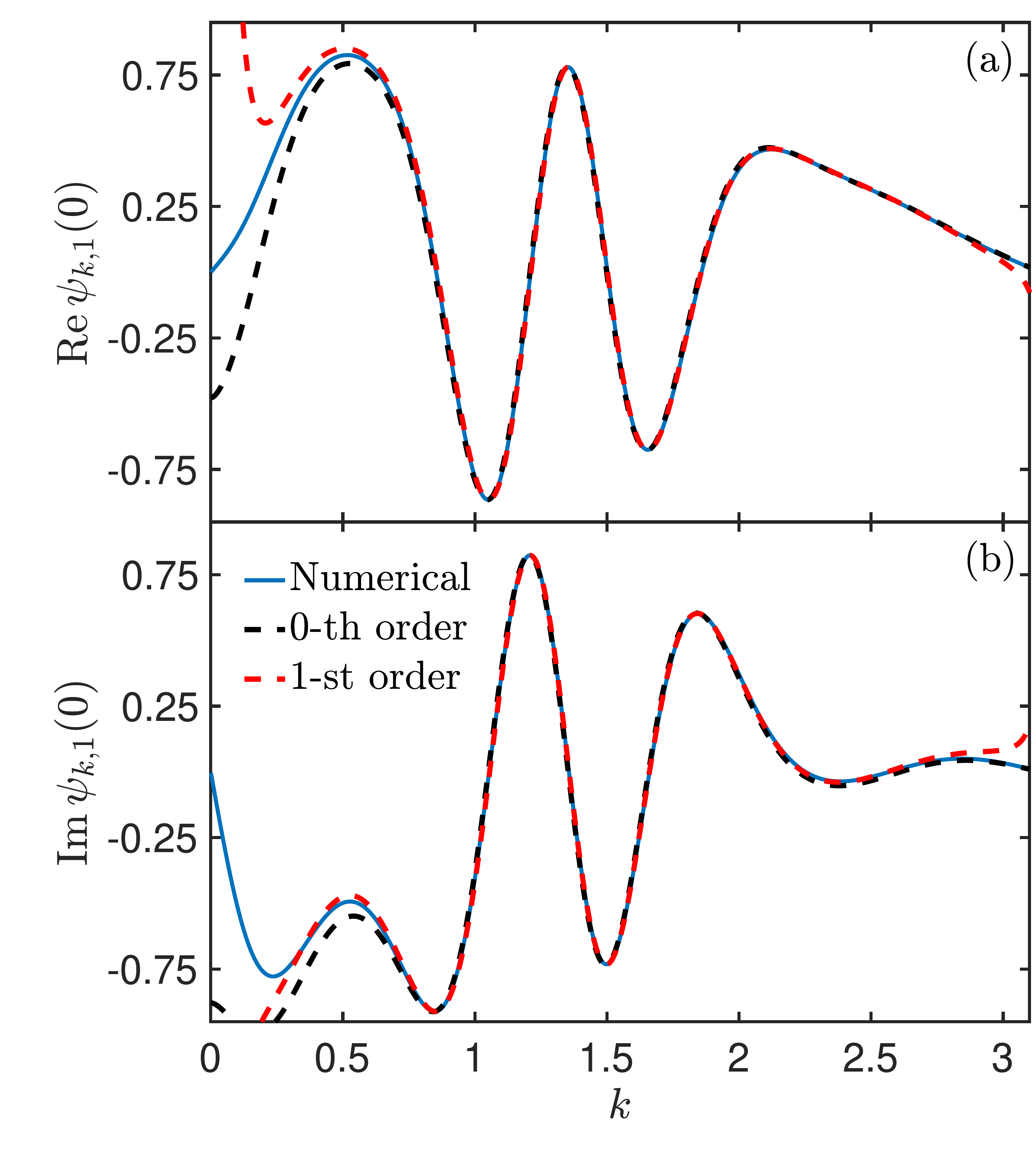}&
    \includegraphics[width=.45\linewidth]
    {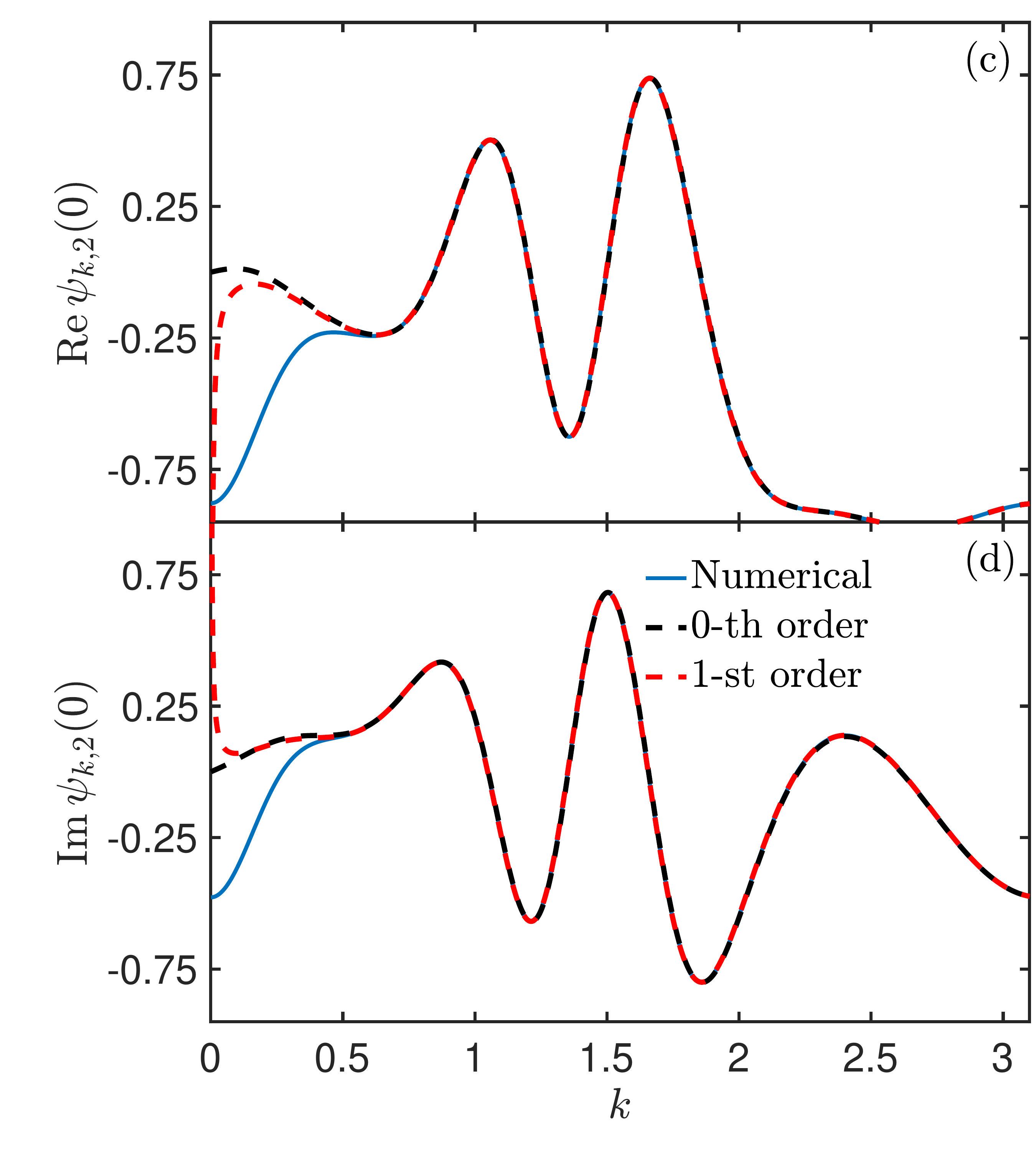}\\
\end{tabular} 
    \caption{Real and imaginary parts of the first and second component of the final time-evolved wave function. The solid line denotes the numerical results, while the dashed lines were obtained by the leading (black) and next-to-leading (red) order saddle point approximations  ($\tau=5$ and $L=4000$).
    (a) Real part of the first component. (b) Imaginary part of the first component. (c) Real part of the second component. (b) Imaginary part of the second component.
    A similar agreement holds for the real and imaginary parts of the second component.}
    \label{fig:Stat_slow3}     
\end{figure}

Finally, we also demonstrate how precise the saddle point approximations are by comparing directly the real and imaginary parts of the final time-evolved wave function to the numerical results. This can be seen in Fig.~\ref{fig:Stat_slow3}.

\section{Further numerical evidence for the cumulant generating function}\label{app: Cumulant_Generating_function}

In this appendix, we further test the precision of our approximations via the real and imaginary parts of the cumulant generating function for annealing times near the fast quench breakdown from both the sudden limit and the KZ scaling regime. First, we report the exact $O(\tau^{3/2})$ order correction given by
\begin{eqnarray}
    b_{3/2}(\theta)&=&(e^{2i\theta}-1)\frac{L}{2\pi}\int_0^\pi\mathrm dk\,\frac{-\frac{\sqrt{\pi}}{4} \sin^2 k  [4-(4+\pi-\log 4) \sin^2 k] }{1+(e^{2i\theta}-1)\cos^2\frac{k}{2}}\nonumber\\
    & &-2\frac{(e^{2i\theta}-1)^2}{2}\frac{L}{2\pi}\int_0^\pi\mathrm dk\,\frac{\frac{\pi^{3/2}}{4}\sin^4k\cos k}{\left(1+(e^{2i\theta}-1)\cos^2\frac{k}{2}\right)^2}
    -\frac{(e^{2i\theta}-1)^3}{3}\frac{L}{2\pi}\int_0^\pi\mathrm dk\,\frac{\frac{\pi^{3/2}}{8}\sin^6k}{\left(1+(e^{2i\theta}-1)\cos^2\frac{k}{2}\right)^3}\nonumber\\
    &=&\frac{1}{48(e^{2i\theta} - 1)^5} \bigg\{ -16 (e^{i\theta}) \bigg[ 112 (e^{2i\theta} - 1)^3 \pi + 3 (e^{2i\theta} - 1)^4 (-4 + 7\pi) \nonumber\\
    & &+ 16 (e^{2i\theta} - 1)^2 (12 + 10\pi - 3 \log(4))
    + 48 (4 + \pi - \log(4)) + 96 (e^{2i\theta} - 1)(4 + \pi - \log(4)) \bigg] \nonumber\\
& &+ (1 + e^{2i\theta}) \bigg[ (e^{2i\theta} - 1)^4 (-60 + 65\pi - 9 \log(4)) + 112 (e^{2i\theta} - 1)^2 (12 + 11\pi - 3 \log(4))\nonumber\\
& & + 384 (4 + \pi - \log(4)) 
+ 768 (e^{2i\theta} - 1)(4 + \pi - \log(4)) + 16 (e^{2i\theta} - 1)^3 (-12 + 53\pi + \log(64)) \bigg] \bigg\},
\end{eqnarray}
by which the cumulant generating function expanded up to order $O(\tau^{3/2})$ is given by 
\begin{eqnarray}
    \log\tilde P(\theta;\tau)\approx \log\tilde P(\theta;0)+b_{1/2}(\theta)\tau^{1/2}+b_{1}(\theta)\tau+b_{3/2}(\theta)\tau^{3/2},
\end{eqnarray}
with $b_{1/2}(\theta)$ and $b_1(\theta)$ given in Eq.~\eqref{eq:b_coefficients} and $\log\tilde P(\theta;0)$ in Eq.~\eqref{eq: logP_sudden_eps_f0}.

\begin{figure*}
\centering
\begin{tabular}{c  c}
    \includegraphics[width=.5\linewidth]{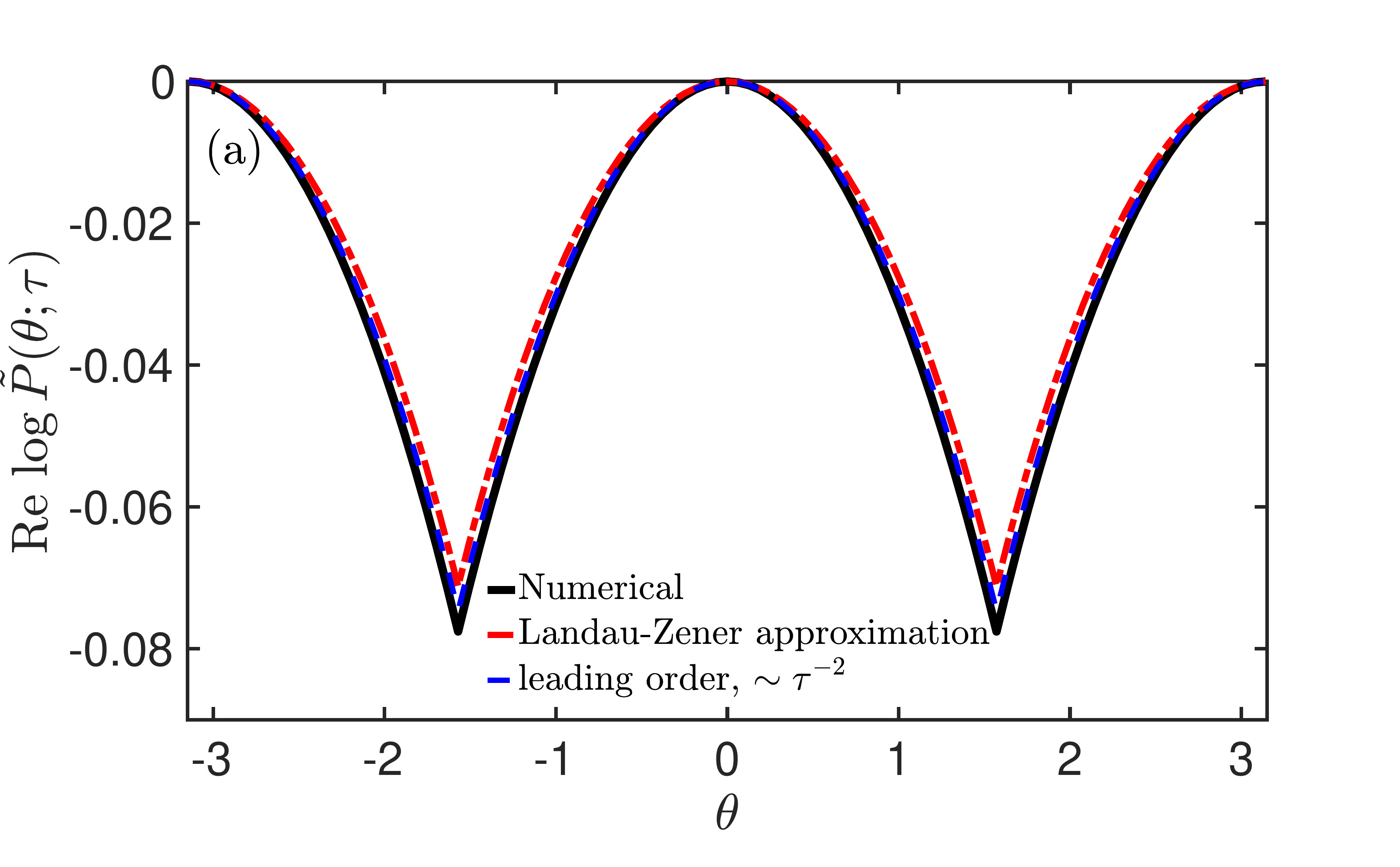}&
    \includegraphics[width=.5\linewidth]{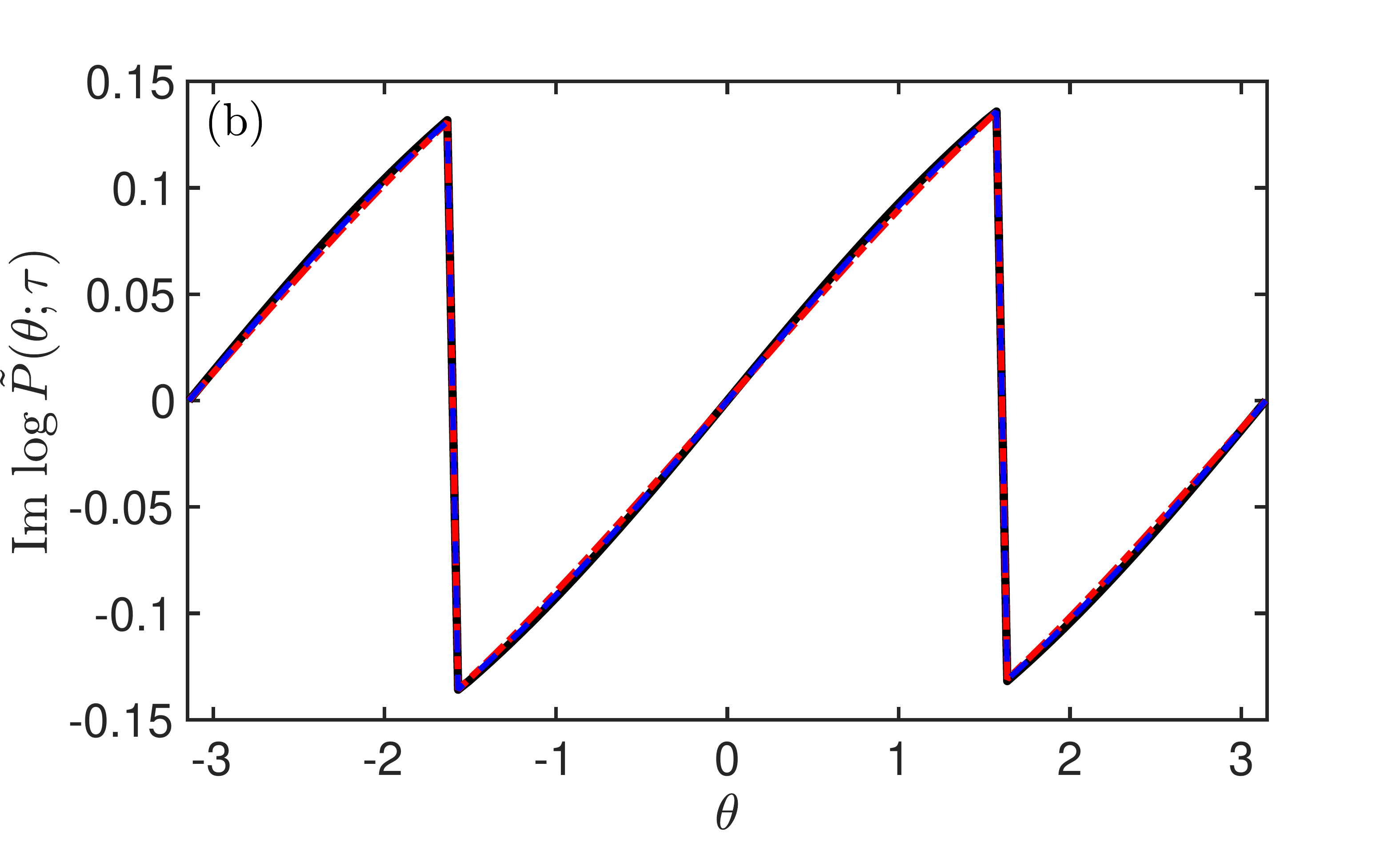}\\
    \end{tabular}
    \caption{(a) Real part of the cumulant generating function, with the leading order expansion reproducing up to high precision the numerical results, surpassing the precision of the Landau-Zener approximation ($\tau=1.5$ and $L=4000$). (b) The matching is even better for the imaginary part.}
    \label{fig:RelogP_tau_large}    
\end{figure*}

\begin{figure*}
\centering
\begin{tabular}{c  c}
    \includegraphics[width=.5\linewidth]{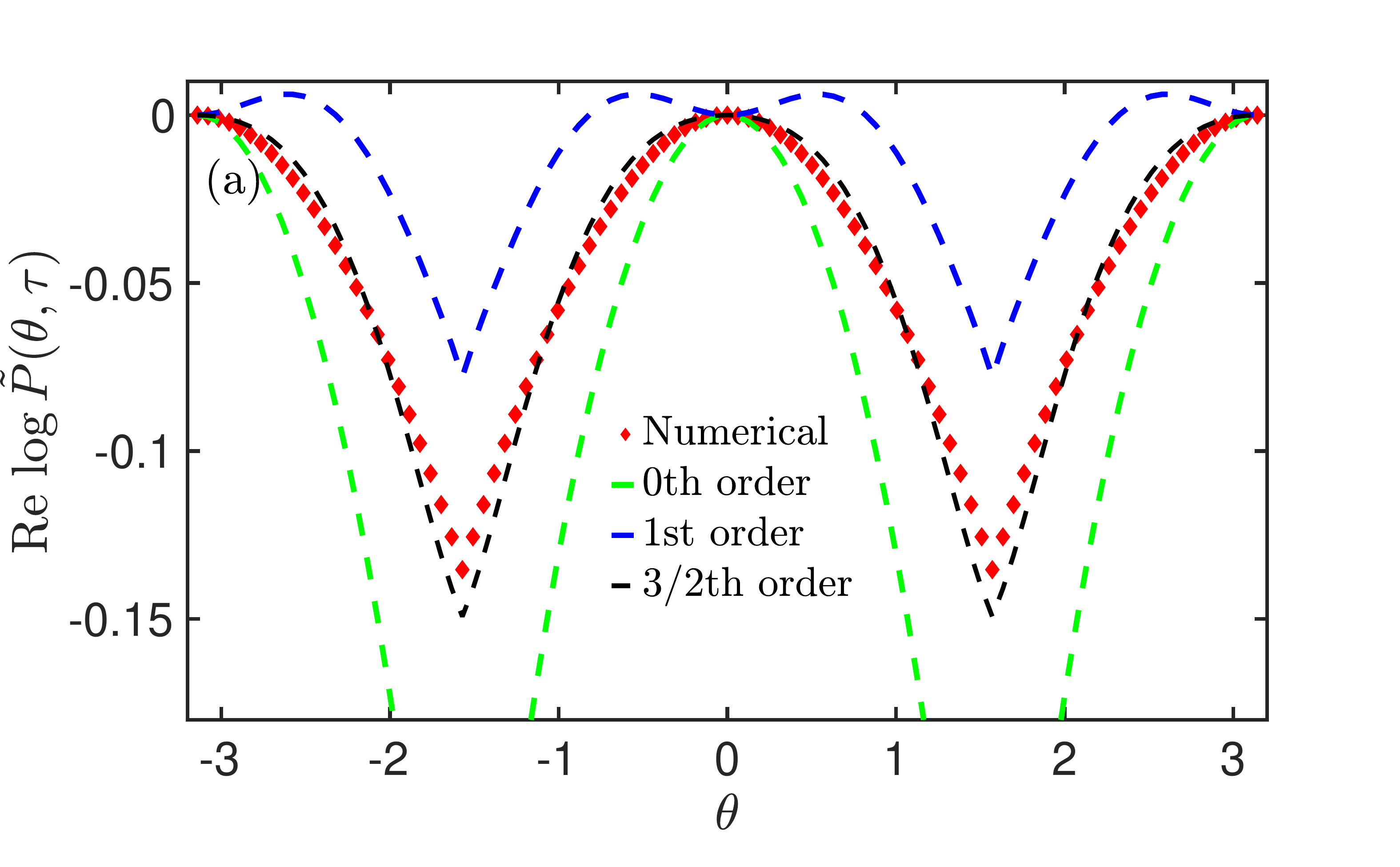}&
    \includegraphics[width=.5\linewidth]{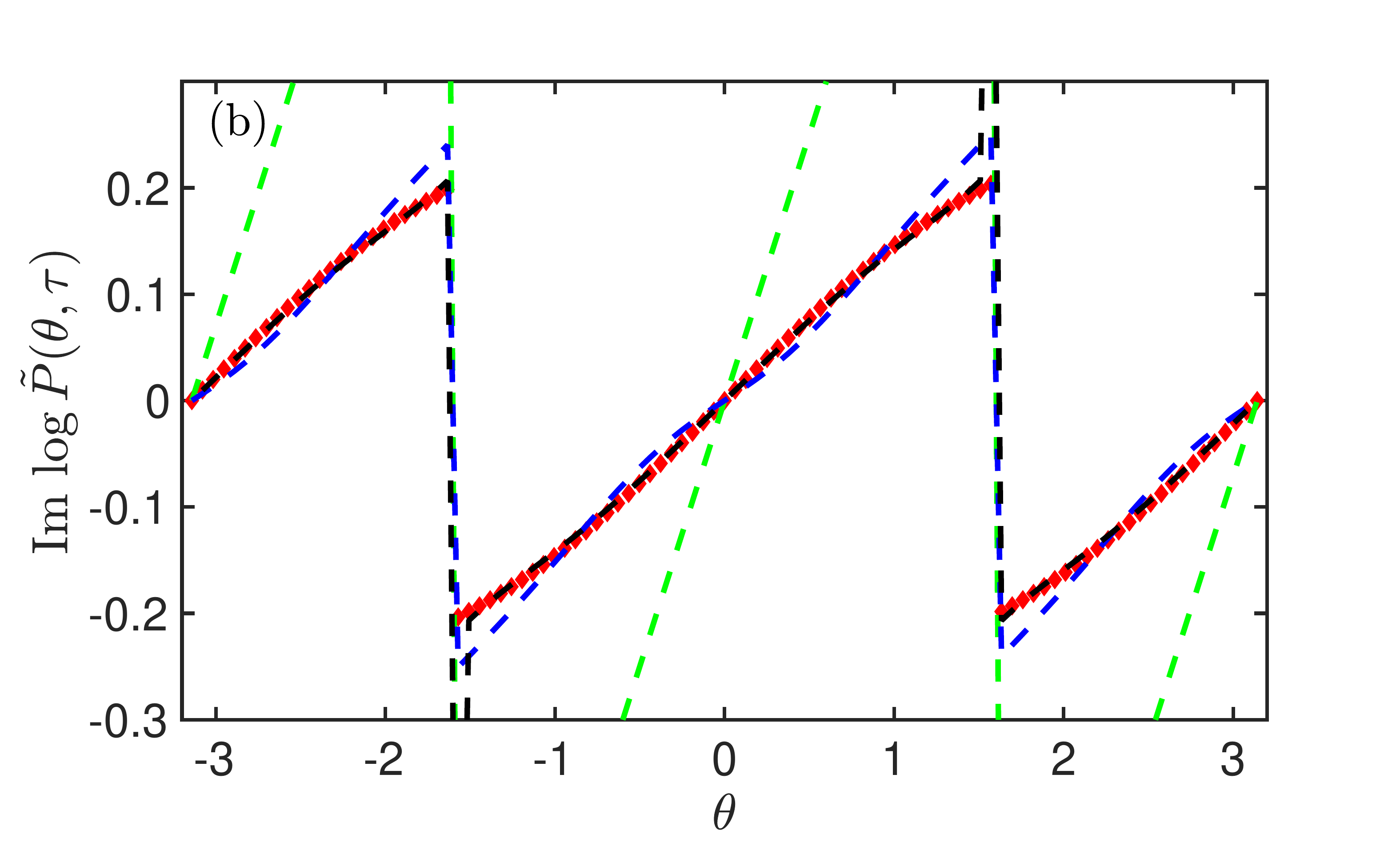}\\
    \end{tabular}
    \caption{(a) Real part of the cumulant generating function. The lower-order expansions show slight deviations. For a good matching, the $3/2$-th order is required to capture precisely the higher order cumulants as well ($\tau=0.8$ and $L=4000$). (b) A better matching for all orders is needed in the case of the imaginary part.}
    \label{fig:RelogP_tau_small}    
\end{figure*}

Figure~\ref{fig:RelogP_tau_large} shows that, even for $\tau=1.5$, the $\sim\tau^{-2}$ corrections provide much better fits for the cumulant generating function. Next,  we show in Fig.~\ref{fig:RelogP_tau_small} how 
 precise the expansions of orders $\sim\tau^{1/2},\,\tau,\,\tau^{3/2}$ can estimate the numerical results. Even for $\tau=0.8$, the contribution of order $\sim\tau^{3/2}$ precisely matches both the real and imaginary parts.

\section{Derivation of the driving times for the maximal cumulant ratios.}\label{app: max_cumulant_ratios}
In this section, we provide a brief summary of the steps to derive the values of $\tau^{1,2}_\mathrm{max}$ and $\tau^{1,3}_\mathrm{max}$ at which the cumulant ratios $\kappa_2/\kappa_1$ and $\kappa_3/\kappa_1$ take their maximum values. As it is clear from the numerical results, the ratio maxima are located considerably below the fast quench breakdown, a precise estimation is provided by considering the fast quench corrections given in Eq.~\eqref{eq: fast_corrections}.

In particular, the derivative of the variance ratio up to order $O(\tau^2)$ reads as
\begin{eqnarray}
    \frac{\mathrm d}{\mathrm d\tau}\left(\kappa_2/\kappa_1\right)&\approx&\frac{\mathrm d}{\mathrm d\tau}
    \left[\frac{\frac{1}{4}-\frac{\pi}{16}\tau+\frac{1}{48}\left[16+9\pi^2-6\pi(2\log(32))\right]\tau^2}{\frac{1}{2}-\frac{\sqrt\pi}{4}\tau^{1/2}-\frac{\sqrt\pi}{32}\left(4-3\pi+\log(64)\right)\tau^2}\right]\nonumber\\
    &=&-\frac{\left(\frac{1}{4} - \frac{\pi \tau}{16} + \frac{\tau^2}{48} \left(16 + 9\pi^2 - 6\pi (2 + \log 32)\right)\right) \left(-\frac{\sqrt{\pi}}{4 \sqrt{\tau}} - \frac{3}{32} \sqrt{\pi} \sqrt{\tau} \left(4 - 3\pi + \log 64\right)\right)}{\left(\frac{1}{2}- \frac{\sqrt{\pi} \sqrt{\tau}}{4} - \frac{\sqrt{\pi}}{32} \tau^{3/2} \left(4 - 3\pi + \log 64\right)\right)^2}\nonumber\\
&&+ \frac{-\frac{\pi}{16} + \frac{\tau}{24} \left(16 + 9\pi^2 - 6\pi (2 + \log 32)\right)}{\frac{1}{2} - \frac{\sqrt{\pi} \sqrt{\tau}}{4} - \frac{\sqrt{\pi}}{32} \tau^{3/2} \left(4 - 3\pi + \log 64\right)}=0.
\end{eqnarray}
Here, we did not use the $O(\tau^2)$ order expansion as it would have led to a third-order equation not providing any deeper analytical insight. Despite the complicated expression, it can readily be solved numerically, yielding $\tilde\tau^{1,2}_\mathrm{max}\approx0.64$ being remarkably close to the numerical value. Similar steps can be performed in the case of the ratio $\kappa_3/\kappa_1$,
\begin{eqnarray}
    \frac{\mathrm d}{\mathrm d\tau}\left(\kappa_3/\kappa_1\right)&\approx&\frac{\mathrm d}{\mathrm d\tau}
    \left[\frac{\frac{\sqrt\pi}{8}\sqrt\tau-\frac{\sqrt\pi}{32}\left[4+\pi-\log(64)\right]\tau^{3/2}}{\frac{1}{2}-\frac{\sqrt\pi}{4}\tau^{1/2}-\frac{\sqrt\pi}{64}\left(4-3\pi+\log(32)\right)\tau^2}\right]=0\Rightarrow\tilde \tau^{3,1}_\mathrm{max}\approx0.75,
\end{eqnarray}
agreeing with the numerical results up to high precision. In short, the finite-time corrections to the sudden and slow quench limits allow for interpolation to intermediate driving times, providing a complete description of arbitrary rates.

\twocolumngrid
\bibliography{references}

\begin{thebibliography}{72}%
\makeatletter
\providecommand \@ifxundefined [1]{%
 \@ifx{#1\undefined}
}%
\providecommand \@ifnum [1]{%
 \ifnum #1\expandafter \@firstoftwo
 \else \expandafter \@secondoftwo
 \fi
}%
\providecommand \@ifx [1]{%
 \ifx #1\expandafter \@firstoftwo
 \else \expandafter \@secondoftwo
 \fi
}%
\providecommand \natexlab [1]{#1}%
\providecommand \enquote  [1]{``#1''}%
\providecommand \bibnamefont  [1]{#1}%
\providecommand \bibfnamefont [1]{#1}%
\providecommand \citenamefont [1]{#1}%
\providecommand \href@noop [0]{\@secondoftwo}%
\providecommand \href [0]{\begingroup \@sanitize@url \@href}%
\providecommand \@href[1]{\@@startlink{#1}\@@href}%
\providecommand \@@href[1]{\endgroup#1\@@endlink}%
\providecommand \@sanitize@url [0]{\catcode `\\12\catcode `\$12\catcode
  `\&12\catcode `\#12\catcode `\^12\catcode `\_12\catcode `\%12\relax}%
\providecommand \@@startlink[1]{}%
\providecommand \@@endlink[0]{}%
\providecommand \url  [0]{\begingroup\@sanitize@url \@url }%
\providecommand \@url [1]{\endgroup\@href {#1}{\urlprefix }}%
\providecommand \urlprefix  [0]{URL }%
\providecommand \Eprint [0]{\href }%
\providecommand \doibase [0]{https://doi.org/}%
\providecommand \selectlanguage [0]{\@gobble}%
\providecommand \bibinfo  [0]{\@secondoftwo}%
\providecommand \bibfield  [0]{\@secondoftwo}%
\providecommand \translation [1]{[#1]}%
\providecommand \BibitemOpen [0]{}%
\providecommand \bibitemStop [0]{}%
\providecommand \bibitemNoStop [0]{.\EOS\space}%
\providecommand \EOS [0]{\spacefactor3000\relax}%
\providecommand \BibitemShut  [1]{\csname bibitem#1\endcsname}%
\let\auto@bib@innerbib\@empty
%</preamble>
\bibitem [{\citenamefont {Sachdev}(2011)}]{Sachdev2011Quantum}%
  \BibitemOpen
  \bibfield  {author} {\bibinfo {author} {\bibfnamefont {S.}~\bibnamefont
  {Sachdev}},\ }\href {https://doi.org/10.1017/CBO9780511973765} {\emph
  {\bibinfo {title} {{Quantum Phase Transitions}}}},\ \bibinfo {edition} {2nd}\
  ed.\ (\bibinfo  {publisher} {Cambridge University Press},\ \bibinfo {year}
  {2011})\BibitemShut {NoStop}%
\bibitem [{\citenamefont {Dziarmaga}(2010)}]{Dziarmaga10}%
  \BibitemOpen
  \bibfield  {author} {\bibinfo {author} {\bibfnamefont {J.}~\bibnamefont
  {Dziarmaga}},\ }\bibfield  {title} {\bibinfo {title} {{Dynamics of a quantum
  phase transition and relaxation to a steady state}},\ }\href
  {https://doi.org/10.1080/00018732.2010.514702} {\bibfield  {journal}
  {\bibinfo  {journal} {Adv. Phys.}\ }\textbf {\bibinfo {volume} {59}},\
  \bibinfo {pages} {1063} (\bibinfo {year} {2010})}\BibitemShut {NoStop}%
\bibitem [{\citenamefont {del Campo}\ and\ \citenamefont {Zurek}(2014)}]{DZ14}%
  \BibitemOpen
  \bibfield  {author} {\bibinfo {author} {\bibfnamefont {A.}~\bibnamefont {del
  Campo}}\ and\ \bibinfo {author} {\bibfnamefont {W.~H.}\ \bibnamefont
  {Zurek}},\ }\bibfield  {title} {\bibinfo {title} {{Universality of phase
  transition dynamics: Topological defects from symmetry breaking}},\ }\href
  {https://doi.org/10.1142/S0217751X1430018X} {\bibfield  {journal} {\bibinfo
  {journal} {Intl. J. Mod. Phys. A}\ }\textbf {\bibinfo {volume} {29}},\
  \bibinfo {pages} {1430018} (\bibinfo {year} {2014})}\BibitemShut {NoStop}%
\bibitem [{\citenamefont {Kibble}(1976)}]{Kibble76a}%
  \BibitemOpen
  \bibfield  {author} {\bibinfo {author} {\bibfnamefont {T.~W.~B.}\
  \bibnamefont {Kibble}},\ }\bibfield  {title} {\bibinfo {title} {{Topology of
  cosmic domains and strings}},\ }\href
  {https://doi.org/10.1088/0305-4470/9/8/029} {\bibfield  {journal} {\bibinfo
  {journal} {J. Phys. A: Math. Gen.}\ }\textbf {\bibinfo {volume} {9}},\
  \bibinfo {pages} {1387} (\bibinfo {year} {1976})}\BibitemShut {NoStop}%
\bibitem [{\citenamefont {Kibble}(1980)}]{Kibble76b}%
  \BibitemOpen
  \bibfield  {author} {\bibinfo {author} {\bibfnamefont {T.~W.~B.}\
  \bibnamefont {Kibble}},\ }\bibfield  {title} {\bibinfo {title} {{Some
  implications of a cosmological phase transition}},\ }\href
  {https://doi.org/10.1016/0370-1573(80)90091-5} {\bibfield  {journal}
  {\bibinfo  {journal} {Phys. Rep.}\ }\textbf {\bibinfo {volume} {67}},\
  \bibinfo {pages} {183} (\bibinfo {year} {1980})}\BibitemShut {NoStop}%
\bibitem [{\citenamefont {Zurek}(1985)}]{Zurek85}%
  \BibitemOpen
  \bibfield  {author} {\bibinfo {author} {\bibfnamefont {W.~H.}\ \bibnamefont
  {Zurek}},\ }\bibfield  {title} {\bibinfo {title} {{Cosmological experiments
  in superfluid helium?}},\ }\href {https://doi.org/10.1038/317505a0}
  {\bibfield  {journal} {\bibinfo  {journal} {Nature}\ }\textbf {\bibinfo
  {volume} {317}},\ \bibinfo {pages} {505} (\bibinfo {year}
  {1985})}\BibitemShut {NoStop}%
\bibitem [{\citenamefont {Zurek}(1996)}]{Zurek96c}%
  \BibitemOpen
  \bibfield  {author} {\bibinfo {author} {\bibfnamefont {W.~H.}\ \bibnamefont
  {Zurek}},\ }\bibfield  {title} {\bibinfo {title} {{Cosmological experiments
  in condensed matter systems}},\ }\href
  {https://doi.org/https://doi.org/10.1016/S0370-1573(96)00009-9} {\bibfield
  {journal} {\bibinfo  {journal} {Phys. Rep.}\ }\textbf {\bibinfo {volume}
  {276}},\ \bibinfo {pages} {177} (\bibinfo {year} {1996})}\BibitemShut
  {NoStop}%
\bibitem [{\citenamefont {del Campo}\ \emph {et~al.}(2010)\citenamefont {del
  Campo}, \citenamefont {De~Chiara}, \citenamefont {Morigi} \emph
  {et~al.}}]{delcampo10}%
  \BibitemOpen
  \bibfield  {author} {\bibinfo {author} {\bibfnamefont {A.}~\bibnamefont {del
  Campo}}, \bibinfo {author} {\bibfnamefont {G.}~\bibnamefont {De~Chiara}},
  \bibinfo {author} {\bibfnamefont {G.}~\bibnamefont {Morigi}}, \emph
  {et~al.},\ }\bibfield  {title} {\bibinfo {title} {{Structural Defects in Ion
  Chains by Quenching the External Potential: The Inhomogeneous Kibble-Zurek
  Mechanism}},\ }\href {https://doi.org/10.1103/PhysRevLett.105.075701}
  {\bibfield  {journal} {\bibinfo  {journal} {Phys. Rev. Lett.}\ }\textbf
  {\bibinfo {volume} {105}},\ \bibinfo {pages} {075701} (\bibinfo {year}
  {2010})}\BibitemShut {NoStop}%
\bibitem [{\citenamefont {Ulm}\ \emph {et~al.}(2013)\citenamefont {Ulm},
  \citenamefont {Ro{\ss}nagel}, \citenamefont {Jacob} \emph {et~al.}}]{Ulm13}%
  \BibitemOpen
  \bibfield  {author} {\bibinfo {author} {\bibfnamefont {S.}~\bibnamefont
  {Ulm}}, \bibinfo {author} {\bibfnamefont {J.}~\bibnamefont {Ro{\ss}nagel}},
  \bibinfo {author} {\bibfnamefont {G.}~\bibnamefont {Jacob}}, \emph {et~al.},\
  }\bibfield  {title} {\bibinfo {title} {{Observation of the Kibble--Zurek
  scaling law for defect formation in ion crystals}},\ }\href
  {https://doi.org/10.1038/ncomms3290} {\bibfield  {journal} {\bibinfo
  {journal} {Nat. Commun.}\ }\textbf {\bibinfo {volume} {4}},\ \bibinfo {pages}
  {2290} (\bibinfo {year} {2013})}\BibitemShut {NoStop}%
\bibitem [{\citenamefont {Pyka}\ \emph {et~al.}(2013)\citenamefont {Pyka},
  \citenamefont {Keller}, \citenamefont {Partner} \emph {et~al.}}]{Pyka13}%
  \BibitemOpen
  \bibfield  {author} {\bibinfo {author} {\bibfnamefont {K.}~\bibnamefont
  {Pyka}}, \bibinfo {author} {\bibfnamefont {J.}~\bibnamefont {Keller}},
  \bibinfo {author} {\bibfnamefont {H.~L.}\ \bibnamefont {Partner}}, \emph
  {et~al.},\ }\bibfield  {title} {\bibinfo {title} {{Topological defect
  formation and spontaneous symmetry breaking in ion Coulomb crystals}},\
  }\href {https://doi.org/10.1038/ncomms3291} {\bibfield  {journal} {\bibinfo
  {journal} {Nat. Commun.}\ }\textbf {\bibinfo {volume} {4}},\ \bibinfo {pages}
  {2291} (\bibinfo {year} {2013})}\BibitemShut {NoStop}%
\bibitem [{\citenamefont {Ejtemaee}\ and\ \citenamefont
  {Haljan}(2013)}]{Ejtemaee2013}%
  \BibitemOpen
  \bibfield  {author} {\bibinfo {author} {\bibfnamefont {S.}~\bibnamefont
  {Ejtemaee}}\ and\ \bibinfo {author} {\bibfnamefont {P.~C.}\ \bibnamefont
  {Haljan}},\ }\bibfield  {title} {\bibinfo {title} {Spontaneous nucleation and
  dynamics of kink defects in zigzag arrays of trapped ions},\ }\href
  {https://doi.org/10.1103/PhysRevA.87.051401} {\bibfield  {journal} {\bibinfo
  {journal} {Phys. Rev. A}\ }\textbf {\bibinfo {volume} {87}},\ \bibinfo
  {pages} {051401} (\bibinfo {year} {2013})}\BibitemShut {NoStop}%
\bibitem [{\citenamefont {Xu}\ \emph {et~al.}(2014)\citenamefont {Xu},
  \citenamefont {Han}, \citenamefont {Sun}, \citenamefont {Xu}, \citenamefont
  {Tang}, \citenamefont {Li},\ and\ \citenamefont {Guo}}]{Xu2014}%
  \BibitemOpen
  \bibfield  {author} {\bibinfo {author} {\bibfnamefont {X.-Y.}\ \bibnamefont
  {Xu}}, \bibinfo {author} {\bibfnamefont {Y.-J.}\ \bibnamefont {Han}},
  \bibinfo {author} {\bibfnamefont {K.}~\bibnamefont {Sun}}, \bibinfo {author}
  {\bibfnamefont {J.-S.}\ \bibnamefont {Xu}}, \bibinfo {author} {\bibfnamefont
  {J.-S.}\ \bibnamefont {Tang}}, \bibinfo {author} {\bibfnamefont {C.-F.}\
  \bibnamefont {Li}},\ and\ \bibinfo {author} {\bibfnamefont {G.-C.}\
  \bibnamefont {Guo}},\ }\bibfield  {title} {\bibinfo {title} {\textit{Quantum
  Simulation of Landau-Zener Model Dynamics Supporting the Kibble-Zurek
  Mechanism}},\ }\href {https://doi.org/10.1103/PhysRevLett.112.035701}
  {\bibfield  {journal} {\bibinfo  {journal} {Phys. Rev. Lett.}\ }\textbf
  {\bibinfo {volume} {112}},\ \bibinfo {pages} {035701} (\bibinfo {year}
  {2014})}\BibitemShut {NoStop}%
\bibitem [{\citenamefont {Cui}\ \emph {et~al.}(2016)\citenamefont {Cui},
  \citenamefont {Huang}, \citenamefont {Wang}, \citenamefont {Cao},
  \citenamefont {Wang}, \citenamefont {Lv}, \citenamefont {Luo}, \citenamefont
  {del Campo}, \citenamefont {Han}, \citenamefont {Li},\ and\ \citenamefont
  {Guo}}]{Cui16}%
  \BibitemOpen
  \bibfield  {author} {\bibinfo {author} {\bibfnamefont {J.-M.}\ \bibnamefont
  {Cui}}, \bibinfo {author} {\bibfnamefont {Y.-F.}\ \bibnamefont {Huang}},
  \bibinfo {author} {\bibfnamefont {Z.}~\bibnamefont {Wang}}, \bibinfo {author}
  {\bibfnamefont {D.-Y.}\ \bibnamefont {Cao}}, \bibinfo {author} {\bibfnamefont
  {J.}~\bibnamefont {Wang}}, \bibinfo {author} {\bibfnamefont {W.-M.}\
  \bibnamefont {Lv}}, \bibinfo {author} {\bibfnamefont {L.}~\bibnamefont
  {Luo}}, \bibinfo {author} {\bibfnamefont {A.}~\bibnamefont {del Campo}},
  \bibinfo {author} {\bibfnamefont {Y.-J.}\ \bibnamefont {Han}}, \bibinfo
  {author} {\bibfnamefont {C.-F.}\ \bibnamefont {Li}},\ and\ \bibinfo {author}
  {\bibfnamefont {G.-C.}\ \bibnamefont {Guo}},\ }\bibfield  {title} {\bibinfo
  {title} {Experimental trapped-ion quantum simulation of the kibble-zurek
  dynamics in momentum space},\ }\href {https://doi.org/10.1038/srep33381}
  {\bibfield  {journal} {\bibinfo  {journal} {Scientific Reports}\ }\textbf
  {\bibinfo {volume} {6}},\ \bibinfo {pages} {33381} (\bibinfo {year}
  {2016})}\BibitemShut {NoStop}%
\bibitem [{\citenamefont {Cui}\ \emph {et~al.}(2020)\citenamefont {Cui},
  \citenamefont {G{\'o}mez-Ruiz}, \citenamefont {Huang}, \citenamefont {Li},
  \citenamefont {Guo},\ and\ \citenamefont {del Campo}}]{Cui20}%
  \BibitemOpen
  \bibfield  {author} {\bibinfo {author} {\bibfnamefont {J.-M.}\ \bibnamefont
  {Cui}}, \bibinfo {author} {\bibfnamefont {F.~J.}\ \bibnamefont
  {G{\'o}mez-Ruiz}}, \bibinfo {author} {\bibfnamefont {Y.-F.}\ \bibnamefont
  {Huang}}, \bibinfo {author} {\bibfnamefont {C.-F.}\ \bibnamefont {Li}},
  \bibinfo {author} {\bibfnamefont {G.-C.}\ \bibnamefont {Guo}},\ and\ \bibinfo
  {author} {\bibfnamefont {A.}~\bibnamefont {del Campo}},\ }\bibfield  {title}
  {\bibinfo {title} {Experimentally testing quantum critical dynamics beyond
  the kibble--zurek mechanism},\ }\href
  {https://doi.org/10.1038/s42005-020-0306-6} {\bibfield  {journal} {\bibinfo
  {journal} {Communications Physics}\ }\textbf {\bibinfo {volume} {3}},\
  \bibinfo {pages} {44} (\bibinfo {year} {2020})}\BibitemShut {NoStop}%
\bibitem [{\citenamefont {Deutschl{\"a}nder}\ \emph {et~al.}(2015)\citenamefont
  {Deutschl{\"a}nder}, \citenamefont {Dillmann}, \citenamefont {Maret},\ and\
  \citenamefont {Keim}}]{Keim15}%
  \BibitemOpen
  \bibfield  {author} {\bibinfo {author} {\bibfnamefont {S.}~\bibnamefont
  {Deutschl{\"a}nder}}, \bibinfo {author} {\bibfnamefont {P.}~\bibnamefont
  {Dillmann}}, \bibinfo {author} {\bibfnamefont {G.}~\bibnamefont {Maret}},\
  and\ \bibinfo {author} {\bibfnamefont {P.}~\bibnamefont {Keim}},\ }\bibfield
  {title} {\bibinfo {title} {{K}ibble{\textendash}{Z}urek mechanism in
  colloidal monolayers},\ }\href {https://doi.org/10.1073/pnas.1500763112}
  {\bibfield  {journal} {\bibinfo  {journal} {Proc. Nat. Acad. Sciences}\
  }\textbf {\bibinfo {volume} {112}},\ \bibinfo {pages} {6925} (\bibinfo {year}
  {2015})}\BibitemShut {NoStop}%
\bibitem [{\citenamefont {Griffin}\ \emph {et~al.}(2012)\citenamefont
  {Griffin}, \citenamefont {Lilienblum}, \citenamefont {Delaney}, \citenamefont
  {Kumagai}, \citenamefont {Fiebig},\ and\ \citenamefont
  {Spaldin}}]{Griffin12}%
  \BibitemOpen
  \bibfield  {author} {\bibinfo {author} {\bibfnamefont {S.~M.}\ \bibnamefont
  {Griffin}}, \bibinfo {author} {\bibfnamefont {M.}~\bibnamefont {Lilienblum}},
  \bibinfo {author} {\bibfnamefont {K.~T.}\ \bibnamefont {Delaney}}, \bibinfo
  {author} {\bibfnamefont {Y.}~\bibnamefont {Kumagai}}, \bibinfo {author}
  {\bibfnamefont {M.}~\bibnamefont {Fiebig}},\ and\ \bibinfo {author}
  {\bibfnamefont {N.~A.}\ \bibnamefont {Spaldin}},\ }\bibfield  {title}
  {\bibinfo {title} {Scaling behavior and beyond equilibrium in the hexagonal
  manganites},\ }\href {https://doi.org/10.1103/PhysRevX.2.041022} {\bibfield
  {journal} {\bibinfo  {journal} {Phys. Rev. X}\ }\textbf {\bibinfo {volume}
  {2}},\ \bibinfo {pages} {041022} (\bibinfo {year} {2012})}\BibitemShut
  {NoStop}%
\bibitem [{\citenamefont {Lin}\ \emph {et~al.}(2014)\citenamefont {Lin},
  \citenamefont {Wang}, \citenamefont {Kamiya}, \citenamefont {Chern},
  \citenamefont {Fan}, \citenamefont {Fan}, \citenamefont {Casas},
  \citenamefont {Liu}, \citenamefont {Kiryukhin}, \citenamefont {Zurek},
  \citenamefont {Batista},\ and\ \citenamefont {Cheong}}]{Lin14}%
  \BibitemOpen
  \bibfield  {author} {\bibinfo {author} {\bibfnamefont {S.-Z.}\ \bibnamefont
  {Lin}}, \bibinfo {author} {\bibfnamefont {X.}~\bibnamefont {Wang}}, \bibinfo
  {author} {\bibfnamefont {Y.}~\bibnamefont {Kamiya}}, \bibinfo {author}
  {\bibfnamefont {G.-W.}\ \bibnamefont {Chern}}, \bibinfo {author}
  {\bibfnamefont {F.}~\bibnamefont {Fan}}, \bibinfo {author} {\bibfnamefont
  {D.}~\bibnamefont {Fan}}, \bibinfo {author} {\bibfnamefont {B.}~\bibnamefont
  {Casas}}, \bibinfo {author} {\bibfnamefont {Y.}~\bibnamefont {Liu}}, \bibinfo
  {author} {\bibfnamefont {V.}~\bibnamefont {Kiryukhin}}, \bibinfo {author}
  {\bibfnamefont {W.~H.}\ \bibnamefont {Zurek}}, \bibinfo {author}
  {\bibfnamefont {C.~D.}\ \bibnamefont {Batista}},\ and\ \bibinfo {author}
  {\bibfnamefont {S.-W.}\ \bibnamefont {Cheong}},\ }\bibfield  {title}
  {\bibinfo {title} {Topological defects as relics of emergent continuous
  symmetry and higgs condensation of disorder in ferroelectrics},\ }\href
  {https://doi.org/10.1038/nphys3142} {\bibfield  {journal} {\bibinfo
  {journal} {Nature Physics}\ }\textbf {\bibinfo {volume} {10}},\ \bibinfo
  {pages} {970} (\bibinfo {year} {2014})}\BibitemShut {NoStop}%
\bibitem [{\citenamefont {Du}\ \emph {et~al.}(2023)\citenamefont {Du},
  \citenamefont {Fang}, \citenamefont {Won}, \citenamefont {De}, \citenamefont
  {Huang}, \citenamefont {Xu}, \citenamefont {You}, \citenamefont
  {G{\'o}mez-Ruiz}, \citenamefont {del Campo},\ and\ \citenamefont
  {Cheong}}]{Du2023}%
  \BibitemOpen
  \bibfield  {author} {\bibinfo {author} {\bibfnamefont {K.}~\bibnamefont
  {Du}}, \bibinfo {author} {\bibfnamefont {X.}~\bibnamefont {Fang}}, \bibinfo
  {author} {\bibfnamefont {C.}~\bibnamefont {Won}}, \bibinfo {author}
  {\bibfnamefont {C.}~\bibnamefont {De}}, \bibinfo {author} {\bibfnamefont
  {F.-T.}\ \bibnamefont {Huang}}, \bibinfo {author} {\bibfnamefont
  {W.}~\bibnamefont {Xu}}, \bibinfo {author} {\bibfnamefont {H.}~\bibnamefont
  {You}}, \bibinfo {author} {\bibfnamefont {F.~J.}\ \bibnamefont
  {G{\'o}mez-Ruiz}}, \bibinfo {author} {\bibfnamefont {A.}~\bibnamefont {del
  Campo}},\ and\ \bibinfo {author} {\bibfnamefont {S.-W.}\ \bibnamefont
  {Cheong}},\ }\bibfield  {title} {\bibinfo {title} {Kibble--zurek mechanism of
  ising domains},\ }\href {https://doi.org/10.1038/s41567-023-02112-5}
  {\bibfield  {journal} {\bibinfo  {journal} {Nature Physics}\ }\textbf
  {\bibinfo {volume} {19}},\ \bibinfo {pages} {1495} (\bibinfo {year}
  {2023})}\BibitemShut {NoStop}%
\bibitem [{\citenamefont {Weiler}\ \emph {et~al.}(2008)\citenamefont {Weiler},
  \citenamefont {Neely}, \citenamefont {Scherer}, \citenamefont {Bradley},
  \citenamefont {Davis},\ and\ \citenamefont {Anderson}}]{Weiler08}%
  \BibitemOpen
  \bibfield  {author} {\bibinfo {author} {\bibfnamefont {C.~N.}\ \bibnamefont
  {Weiler}}, \bibinfo {author} {\bibfnamefont {T.~W.}\ \bibnamefont {Neely}},
  \bibinfo {author} {\bibfnamefont {D.~R.}\ \bibnamefont {Scherer}}, \bibinfo
  {author} {\bibfnamefont {A.~S.}\ \bibnamefont {Bradley}}, \bibinfo {author}
  {\bibfnamefont {M.~J.}\ \bibnamefont {Davis}},\ and\ \bibinfo {author}
  {\bibfnamefont {B.~P.}\ \bibnamefont {Anderson}},\ }\bibfield  {title}
  {\bibinfo {title} {Spontaneous vortices in the formation of {B}ose-{E}instein
  condensates},\ }\href {https://doi.org/10.1038/nature07334} {\bibfield
  {journal} {\bibinfo  {journal} {Nature}\ }\textbf {\bibinfo {volume} {455}},\
  \bibinfo {pages} {948} (\bibinfo {year} {2008})}\BibitemShut {NoStop}%
\bibitem [{\citenamefont {Lamporesi}\ \emph {et~al.}(2013)\citenamefont
  {Lamporesi}, \citenamefont {Donadello}, \citenamefont {Serafini},
  \citenamefont {Dalfovo},\ and\ \citenamefont {Ferrari}}]{Lamporesi2013}%
  \BibitemOpen
  \bibfield  {author} {\bibinfo {author} {\bibfnamefont {G.}~\bibnamefont
  {Lamporesi}}, \bibinfo {author} {\bibfnamefont {S.}~\bibnamefont
  {Donadello}}, \bibinfo {author} {\bibfnamefont {S.}~\bibnamefont {Serafini}},
  \bibinfo {author} {\bibfnamefont {F.}~\bibnamefont {Dalfovo}},\ and\ \bibinfo
  {author} {\bibfnamefont {G.}~\bibnamefont {Ferrari}},\ }\bibfield  {title}
  {\bibinfo {title} {Spontaneous creation of kibble--zurek solitons in a
  bose--einstein condensate},\ }\href {https://doi.org/10.1038/nphys2734}
  {\bibfield  {journal} {\bibinfo  {journal} {Nature Physics}\ }\textbf
  {\bibinfo {volume} {9}},\ \bibinfo {pages} {656} (\bibinfo {year}
  {2013})}\BibitemShut {NoStop}%
\bibitem [{\citenamefont {Navon}\ \emph {et~al.}(2015)\citenamefont {Navon},
  \citenamefont {Gaunt}, \citenamefont {Smith},\ and\ \citenamefont
  {Hadzibabic}}]{Navon2015}%
  \BibitemOpen
  \bibfield  {author} {\bibinfo {author} {\bibfnamefont {N.}~\bibnamefont
  {Navon}}, \bibinfo {author} {\bibfnamefont {A.~L.}\ \bibnamefont {Gaunt}},
  \bibinfo {author} {\bibfnamefont {R.~P.}\ \bibnamefont {Smith}},\ and\
  \bibinfo {author} {\bibfnamefont {Z.}~\bibnamefont {Hadzibabic}},\ }\bibfield
   {title} {\bibinfo {title} {Critical dynamics of spontaneous symmetry
  breaking in a homogeneous bose gas},\ }\href
  {https://doi.org/10.1126/science.1258676} {\bibfield  {journal} {\bibinfo
  {journal} {Science}\ }\textbf {\bibinfo {volume} {347}},\ \bibinfo {pages}
  {167} (\bibinfo {year} {2015})}\BibitemShut {NoStop}%
\bibitem [{\citenamefont {Ko}\ \emph {et~al.}(2019)\citenamefont {Ko},
  \citenamefont {Park},\ and\ \citenamefont {Shin}}]{Ko2019}%
  \BibitemOpen
  \bibfield  {author} {\bibinfo {author} {\bibfnamefont {B.}~\bibnamefont
  {Ko}}, \bibinfo {author} {\bibfnamefont {J.~W.}\ \bibnamefont {Park}},\ and\
  \bibinfo {author} {\bibfnamefont {Y.}~\bibnamefont {Shin}},\ }\bibfield
  {title} {\bibinfo {title} {Kibble--zurek universality in a strongly
  interacting fermi superfluid},\ }\href
  {https://doi.org/10.1038/s41567-019-0650-1} {\bibfield  {journal} {\bibinfo
  {journal} {Nature Physics}\ }\textbf {\bibinfo {volume} {15}},\ \bibinfo
  {pages} {1227} (\bibinfo {year} {2019})}\BibitemShut {NoStop}%
\bibitem [{\citenamefont {Goo}\ \emph {et~al.}(2021)\citenamefont {Goo},
  \citenamefont {Lim},\ and\ \citenamefont {Shin}}]{Goo21}%
  \BibitemOpen
  \bibfield  {author} {\bibinfo {author} {\bibfnamefont {J.}~\bibnamefont
  {Goo}}, \bibinfo {author} {\bibfnamefont {Y.}~\bibnamefont {Lim}},\ and\
  \bibinfo {author} {\bibfnamefont {Y.}~\bibnamefont {Shin}},\ }\bibfield
  {title} {\bibinfo {title} {Defect saturation in a rapidly quenched {B}ose
  gas},\ }\href {https://doi.org/10.1103/PhysRevLett.127.115701} {\bibfield
  {journal} {\bibinfo  {journal} {Phys. Rev. Lett.}\ }\textbf {\bibinfo
  {volume} {127}},\ \bibinfo {pages} {115701} (\bibinfo {year}
  {2021})}\BibitemShut {NoStop}%
\bibitem [{\citenamefont {Gardas}\ \emph {et~al.}(2018)\citenamefont {Gardas},
  \citenamefont {Dziarmaga}, \citenamefont {Zurek} \emph {et~al.}}]{Gardas18}%
  \BibitemOpen
  \bibfield  {author} {\bibinfo {author} {\bibfnamefont {B.}~\bibnamefont
  {Gardas}}, \bibinfo {author} {\bibfnamefont {J.}~\bibnamefont {Dziarmaga}},
  \bibinfo {author} {\bibfnamefont {W.~H.}\ \bibnamefont {Zurek}}, \emph
  {et~al.},\ }\bibfield  {title} {\bibinfo {title} {{Defects in Quantum
  Computers}},\ }\href {https://doi.org/10.1038/s41598-018-22763-2} {\bibfield
  {journal} {\bibinfo  {journal} {Sci. Rep.}\ }\textbf {\bibinfo {volume}
  {8}},\ \bibinfo {pages} {4539} (\bibinfo {year} {2018})}\BibitemShut
  {NoStop}%
\bibitem [{\citenamefont {Keesling}\ \emph {et~al.}(2019)\citenamefont
  {Keesling}, \citenamefont {Omran}, \citenamefont {Levine}, \citenamefont
  {Bernien}, \citenamefont {Pichler}, \citenamefont {Choi}, \citenamefont
  {Samajdar}, \citenamefont {Schwartz}, \citenamefont {Silvi}, \citenamefont
  {Sachdev}, \citenamefont {Zoller}, \citenamefont {Endres}, \citenamefont
  {Greiner}, \citenamefont {Vuleti{\'{c}}},\ and\ \citenamefont
  {Lukin}}]{Keesling19}%
  \BibitemOpen
  \bibfield  {author} {\bibinfo {author} {\bibfnamefont {A.}~\bibnamefont
  {Keesling}}, \bibinfo {author} {\bibfnamefont {A.}~\bibnamefont {Omran}},
  \bibinfo {author} {\bibfnamefont {H.}~\bibnamefont {Levine}}, \bibinfo
  {author} {\bibfnamefont {H.}~\bibnamefont {Bernien}}, \bibinfo {author}
  {\bibfnamefont {H.}~\bibnamefont {Pichler}}, \bibinfo {author} {\bibfnamefont
  {S.}~\bibnamefont {Choi}}, \bibinfo {author} {\bibfnamefont {R.}~\bibnamefont
  {Samajdar}}, \bibinfo {author} {\bibfnamefont {S.}~\bibnamefont {Schwartz}},
  \bibinfo {author} {\bibfnamefont {P.}~\bibnamefont {Silvi}}, \bibinfo
  {author} {\bibfnamefont {S.}~\bibnamefont {Sachdev}}, \bibinfo {author}
  {\bibfnamefont {P.}~\bibnamefont {Zoller}}, \bibinfo {author} {\bibfnamefont
  {M.}~\bibnamefont {Endres}}, \bibinfo {author} {\bibfnamefont
  {M.}~\bibnamefont {Greiner}}, \bibinfo {author} {\bibfnamefont
  {V.}~\bibnamefont {Vuleti{\'{c}}}},\ and\ \bibinfo {author} {\bibfnamefont
  {M.~D.}\ \bibnamefont {Lukin}},\ }\bibfield  {title} {\bibinfo {title}
  {Quantum kibble--zurek mechanism and critical dynamics on a programmable
  rydberg simulator},\ }\href {https://doi.org/10.1038/s41586-019-1070-1}
  {\bibfield  {journal} {\bibinfo  {journal} {Nature}\ }\textbf {\bibinfo
  {volume} {568}},\ \bibinfo {pages} {207} (\bibinfo {year}
  {2019})}\BibitemShut {NoStop}%
\bibitem [{\citenamefont {Weinberg}\ \emph {et~al.}(2020)\citenamefont
  {Weinberg}, \citenamefont {Tylutki}, \citenamefont {R\"onkk\"o} \emph
  {et~al.}}]{Weinberg20}%
  \BibitemOpen
  \bibfield  {author} {\bibinfo {author} {\bibfnamefont {P.}~\bibnamefont
  {Weinberg}}, \bibinfo {author} {\bibfnamefont {M.}~\bibnamefont {Tylutki}},
  \bibinfo {author} {\bibfnamefont {J.~M.}\ \bibnamefont {R\"onkk\"o}}, \emph
  {et~al.},\ }\bibfield  {title} {\bibinfo {title} {{Scaling and Diabatic
  Effects in Quantum Annealing with a D-Wave Device}},\ }\href
  {https://doi.org/10.1103/PhysRevLett.124.090502} {\bibfield  {journal}
  {\bibinfo  {journal} {Phys. Rev. Lett.}\ }\textbf {\bibinfo {volume} {124}},\
  \bibinfo {pages} {090502} (\bibinfo {year} {2020})}\BibitemShut {NoStop}%
\bibitem [{\citenamefont {Bando}\ \emph {et~al.}(2020)\citenamefont {Bando},
  \citenamefont {Susa}, \citenamefont {Oshiyama} \emph {et~al.}}]{Bando20}%
  \BibitemOpen
  \bibfield  {author} {\bibinfo {author} {\bibfnamefont {Y.}~\bibnamefont
  {Bando}}, \bibinfo {author} {\bibfnamefont {Y.}~\bibnamefont {Susa}},
  \bibinfo {author} {\bibfnamefont {H.}~\bibnamefont {Oshiyama}}, \emph
  {et~al.},\ }\bibfield  {title} {\bibinfo {title} {{Probing the universality
  of topological defect formation in a quantum annealer: Kibble-Zurek mechanism
  and beyond}},\ }\href {https://doi.org/10.1103/PhysRevResearch.2.033369}
  {\bibfield  {journal} {\bibinfo  {journal} {Phys. Rev. Research}\ }\textbf
  {\bibinfo {volume} {2}},\ \bibinfo {pages} {033369} (\bibinfo {year}
  {2020})}\BibitemShut {NoStop}%
\bibitem [{\citenamefont {King}\ \emph {et~al.}(2022)\citenamefont {King},
  \citenamefont {Suzuki}, \citenamefont {Raymond} \emph {et~al.}}]{King22}%
  \BibitemOpen
  \bibfield  {author} {\bibinfo {author} {\bibfnamefont {A.~D.}\ \bibnamefont
  {King}}, \bibinfo {author} {\bibfnamefont {S.}~\bibnamefont {Suzuki}},
  \bibinfo {author} {\bibfnamefont {J.}~\bibnamefont {Raymond}}, \emph
  {et~al.},\ }\bibfield  {title} {\bibinfo {title} {{Coherent quantum annealing
  in a programmable 2,000-qubit Ising chain}},\ }\href
  {https://doi.org/10.1038/s41567-022-01741-6} {\bibfield  {journal} {\bibinfo
  {journal} {Nat. Phys.}\ }\textbf {\bibinfo {volume} {18}},\ \bibinfo {pages}
  {1324} (\bibinfo {year} {2022})}\BibitemShut {NoStop}%
\bibitem [{\citenamefont {King}\ \emph {et~al.}(2024)\citenamefont {King},
  \citenamefont {Nocera}, \citenamefont {Rams}, \citenamefont {Dziarmaga},
  \citenamefont {Wiersema}, \citenamefont {Bernoudy}, \citenamefont {Raymond},
  \citenamefont {Kaushal}, \citenamefont {Heinsdorf}, \citenamefont {Harris},
  \citenamefont {Boothby}, \citenamefont {Altomare}, \citenamefont {Berkley},
  \citenamefont {Boschnak}, \citenamefont {Chern}, \citenamefont {Christiani},
  \citenamefont {Cibere}, \citenamefont {Connor}, \citenamefont {Dehn},
  \citenamefont {Deshpande}, \citenamefont {Ejtemaee}, \citenamefont {Farré},
  \citenamefont {Hamer}, \citenamefont {Hoskinson}, \citenamefont {Huang},
  \citenamefont {Johnson}, \citenamefont {Kortas}, \citenamefont {Ladizinsky},
  \citenamefont {Lai}, \citenamefont {Lanting}, \citenamefont {Li},
  \citenamefont {MacDonald}, \citenamefont {Marsden}, \citenamefont {McGeoch},
  \citenamefont {Molavi}, \citenamefont {Neufeld}, \citenamefont {Norouzpour},
  \citenamefont {Oh}, \citenamefont {Pasvolsky}, \citenamefont {Poitras},
  \citenamefont {Poulin-Lamarre}, \citenamefont {Prescott}, \citenamefont
  {Reis}, \citenamefont {Rich}, \citenamefont {Samani}, \citenamefont
  {Sheldan}, \citenamefont {Smirnov}, \citenamefont {Sterpka}, \citenamefont
  {Clavera}, \citenamefont {Tsai}, \citenamefont {Volkmann}, \citenamefont
  {Whiticar}, \citenamefont {Whittaker}, \citenamefont {Wilkinson},
  \citenamefont {Yao}, \citenamefont {Yi}, \citenamefont {Sandvik},
  \citenamefont {Alvarez}, \citenamefont {Melko}, \citenamefont {Carrasquilla},
  \citenamefont {Franz},\ and\ \citenamefont {Amin}}]{King24}%
  \BibitemOpen
  \bibfield  {author} {\bibinfo {author} {\bibfnamefont {A.~D.}\ \bibnamefont
  {King}}, \bibinfo {author} {\bibfnamefont {A.}~\bibnamefont {Nocera}},
  \bibinfo {author} {\bibfnamefont {M.~M.}\ \bibnamefont {Rams}}, \bibinfo
  {author} {\bibfnamefont {J.}~\bibnamefont {Dziarmaga}}, \bibinfo {author}
  {\bibfnamefont {R.}~\bibnamefont {Wiersema}}, \bibinfo {author}
  {\bibfnamefont {W.}~\bibnamefont {Bernoudy}}, \bibinfo {author}
  {\bibfnamefont {J.}~\bibnamefont {Raymond}}, \bibinfo {author} {\bibfnamefont
  {N.}~\bibnamefont {Kaushal}}, \bibinfo {author} {\bibfnamefont
  {N.}~\bibnamefont {Heinsdorf}}, \bibinfo {author} {\bibfnamefont
  {R.}~\bibnamefont {Harris}}, \bibinfo {author} {\bibfnamefont
  {K.}~\bibnamefont {Boothby}}, \bibinfo {author} {\bibfnamefont
  {F.}~\bibnamefont {Altomare}}, \bibinfo {author} {\bibfnamefont {A.~J.}\
  \bibnamefont {Berkley}}, \bibinfo {author} {\bibfnamefont {M.}~\bibnamefont
  {Boschnak}}, \bibinfo {author} {\bibfnamefont {K.}~\bibnamefont {Chern}},
  \bibinfo {author} {\bibfnamefont {H.}~\bibnamefont {Christiani}}, \bibinfo
  {author} {\bibfnamefont {S.}~\bibnamefont {Cibere}}, \bibinfo {author}
  {\bibfnamefont {J.}~\bibnamefont {Connor}}, \bibinfo {author} {\bibfnamefont
  {M.~H.}\ \bibnamefont {Dehn}}, \bibinfo {author} {\bibfnamefont
  {R.}~\bibnamefont {Deshpande}}, \bibinfo {author} {\bibfnamefont
  {S.}~\bibnamefont {Ejtemaee}}, \bibinfo {author} {\bibfnamefont
  {P.}~\bibnamefont {Farré}}, \bibinfo {author} {\bibfnamefont
  {K.}~\bibnamefont {Hamer}}, \bibinfo {author} {\bibfnamefont
  {E.}~\bibnamefont {Hoskinson}}, \bibinfo {author} {\bibfnamefont
  {S.}~\bibnamefont {Huang}}, \bibinfo {author} {\bibfnamefont {M.~W.}\
  \bibnamefont {Johnson}}, \bibinfo {author} {\bibfnamefont {S.}~\bibnamefont
  {Kortas}}, \bibinfo {author} {\bibfnamefont {E.}~\bibnamefont {Ladizinsky}},
  \bibinfo {author} {\bibfnamefont {T.}~\bibnamefont {Lai}}, \bibinfo {author}
  {\bibfnamefont {T.}~\bibnamefont {Lanting}}, \bibinfo {author} {\bibfnamefont
  {R.}~\bibnamefont {Li}}, \bibinfo {author} {\bibfnamefont {A.~J.~R.}\
  \bibnamefont {MacDonald}}, \bibinfo {author} {\bibfnamefont {G.}~\bibnamefont
  {Marsden}}, \bibinfo {author} {\bibfnamefont {C.~C.}\ \bibnamefont
  {McGeoch}}, \bibinfo {author} {\bibfnamefont {R.}~\bibnamefont {Molavi}},
  \bibinfo {author} {\bibfnamefont {R.}~\bibnamefont {Neufeld}}, \bibinfo
  {author} {\bibfnamefont {M.}~\bibnamefont {Norouzpour}}, \bibinfo {author}
  {\bibfnamefont {T.}~\bibnamefont {Oh}}, \bibinfo {author} {\bibfnamefont
  {J.}~\bibnamefont {Pasvolsky}}, \bibinfo {author} {\bibfnamefont
  {P.}~\bibnamefont {Poitras}}, \bibinfo {author} {\bibfnamefont
  {G.}~\bibnamefont {Poulin-Lamarre}}, \bibinfo {author} {\bibfnamefont
  {T.}~\bibnamefont {Prescott}}, \bibinfo {author} {\bibfnamefont
  {M.}~\bibnamefont {Reis}}, \bibinfo {author} {\bibfnamefont {C.}~\bibnamefont
  {Rich}}, \bibinfo {author} {\bibfnamefont {M.}~\bibnamefont {Samani}},
  \bibinfo {author} {\bibfnamefont {B.}~\bibnamefont {Sheldan}}, \bibinfo
  {author} {\bibfnamefont {A.}~\bibnamefont {Smirnov}}, \bibinfo {author}
  {\bibfnamefont {E.}~\bibnamefont {Sterpka}}, \bibinfo {author} {\bibfnamefont
  {B.~T.}\ \bibnamefont {Clavera}}, \bibinfo {author} {\bibfnamefont
  {N.}~\bibnamefont {Tsai}}, \bibinfo {author} {\bibfnamefont {M.}~\bibnamefont
  {Volkmann}}, \bibinfo {author} {\bibfnamefont {A.}~\bibnamefont {Whiticar}},
  \bibinfo {author} {\bibfnamefont {J.~D.}\ \bibnamefont {Whittaker}}, \bibinfo
  {author} {\bibfnamefont {W.}~\bibnamefont {Wilkinson}}, \bibinfo {author}
  {\bibfnamefont {J.}~\bibnamefont {Yao}}, \bibinfo {author} {\bibfnamefont
  {T.~J.}\ \bibnamefont {Yi}}, \bibinfo {author} {\bibfnamefont {A.~W.}\
  \bibnamefont {Sandvik}}, \bibinfo {author} {\bibfnamefont {G.}~\bibnamefont
  {Alvarez}}, \bibinfo {author} {\bibfnamefont {R.~G.}\ \bibnamefont {Melko}},
  \bibinfo {author} {\bibfnamefont {J.}~\bibnamefont {Carrasquilla}}, \bibinfo
  {author} {\bibfnamefont {M.}~\bibnamefont {Franz}},\ and\ \bibinfo {author}
  {\bibfnamefont {M.~H.}\ \bibnamefont {Amin}},\ }\href@noop {} {\bibinfo
  {title} {Computational supremacy in quantum simulation}} (\bibinfo {year}
  {2024}),\ \Eprint {https://arxiv.org/abs/2403.00910} {arXiv:2403.00910
  [quant-ph]} \BibitemShut {NoStop}%
\bibitem [{\citenamefont {Ali}\ \emph {et~al.}(2024)\citenamefont {Ali},
  \citenamefont {Xu}, \citenamefont {Bernoudy}, \citenamefont {Nocera},
  \citenamefont {King},\ and\ \citenamefont {Banerjee}}]{Ali2024}%
  \BibitemOpen
  \bibfield  {author} {\bibinfo {author} {\bibfnamefont {A.}~\bibnamefont
  {Ali}}, \bibinfo {author} {\bibfnamefont {H.}~\bibnamefont {Xu}}, \bibinfo
  {author} {\bibfnamefont {W.}~\bibnamefont {Bernoudy}}, \bibinfo {author}
  {\bibfnamefont {A.}~\bibnamefont {Nocera}}, \bibinfo {author} {\bibfnamefont
  {A.~D.}\ \bibnamefont {King}},\ and\ \bibinfo {author} {\bibfnamefont
  {A.}~\bibnamefont {Banerjee}},\ }\bibfield  {title} {\bibinfo {title}
  {Quantum quench dynamics of geometrically frustrated ising models},\ }\href
  {https://doi.org/10.1038/s41467-024-54701-4} {\bibfield  {journal} {\bibinfo
  {journal} {Nature Communications}\ }\textbf {\bibinfo {volume} {15}},\
  \bibinfo {pages} {10756} (\bibinfo {year} {2024})}\BibitemShut {NoStop}%
\bibitem [{\citenamefont {Chesler}\ \emph {et~al.}(2015)\citenamefont
  {Chesler}, \citenamefont {Garc\'{\i}a-Garc\'{\i}a},\ and\ \citenamefont
  {Liu}}]{Chesler2015}%
  \BibitemOpen
  \bibfield  {author} {\bibinfo {author} {\bibfnamefont {P.~M.}\ \bibnamefont
  {Chesler}}, \bibinfo {author} {\bibfnamefont {A.~M.}\ \bibnamefont
  {Garc\'{\i}a-Garc\'{\i}a}},\ and\ \bibinfo {author} {\bibfnamefont
  {H.}~\bibnamefont {Liu}},\ }\bibfield  {title} {\bibinfo {title} {Defect
  formation beyond kibble-zurek mechanism and holography},\ }\href
  {https://doi.org/10.1103/PhysRevX.5.021015} {\bibfield  {journal} {\bibinfo
  {journal} {Phys. Rev. X}\ }\textbf {\bibinfo {volume} {5}},\ \bibinfo {pages}
  {021015} (\bibinfo {year} {2015})}\BibitemShut {NoStop}%
\bibitem [{\citenamefont {Zeng}\ \emph {et~al.}(2023)\citenamefont {Zeng},
  \citenamefont {Xia},\ and\ \citenamefont {del Campo}}]{Huabi2023FastQuench}%
  \BibitemOpen
  \bibfield  {author} {\bibinfo {author} {\bibfnamefont {H.-B.}\ \bibnamefont
  {Zeng}}, \bibinfo {author} {\bibfnamefont {C.-Y.}\ \bibnamefont {Xia}},\ and\
  \bibinfo {author} {\bibfnamefont {A.}~\bibnamefont {del Campo}},\ }\bibfield
  {title} {\bibinfo {title} {Universal breakdown of kibble-zurek scaling in
  fast quenches across a phase transition},\ }\href
  {https://doi.org/10.1103/PhysRevLett.130.060402} {\bibfield  {journal}
  {\bibinfo  {journal} {Phys. Rev. Lett.}\ }\textbf {\bibinfo {volume} {130}},\
  \bibinfo {pages} {060402} (\bibinfo {year} {2023})}\BibitemShut {NoStop}%
\bibitem [{\citenamefont {Xia}\ \emph {et~al.}(2023)\citenamefont {Xia},
  \citenamefont {Zeng}, \citenamefont {Chen},\ and\ \citenamefont {del
  Campo}}]{Xia23prd}%
  \BibitemOpen
  \bibfield  {author} {\bibinfo {author} {\bibfnamefont {C.-Y.}\ \bibnamefont
  {Xia}}, \bibinfo {author} {\bibfnamefont {H.-B.}\ \bibnamefont {Zeng}},
  \bibinfo {author} {\bibfnamefont {C.-M.}\ \bibnamefont {Chen}},\ and\
  \bibinfo {author} {\bibfnamefont {A.}~\bibnamefont {del Campo}},\ }\bibfield
  {title} {\bibinfo {title} {Structural phase transition and its critical
  dynamics from holography},\ }\href
  {https://doi.org/10.1103/PhysRevD.108.026017} {\bibfield  {journal} {\bibinfo
   {journal} {Phys. Rev. D}\ }\textbf {\bibinfo {volume} {108}},\ \bibinfo
  {pages} {026017} (\bibinfo {year} {2023})}\BibitemShut {NoStop}%
\bibitem [{\citenamefont {Xia}\ \emph {et~al.}(2024)\citenamefont {Xia},
  \citenamefont {Zeng}, \citenamefont {Grabarits},\ and\ \citenamefont {del
  Campo}}]{Xia2024}%
  \BibitemOpen
  \bibfield  {author} {\bibinfo {author} {\bibfnamefont {C.-Y.}\ \bibnamefont
  {Xia}}, \bibinfo {author} {\bibfnamefont {H.-B.}\ \bibnamefont {Zeng}},
  \bibinfo {author} {\bibfnamefont {A.}~\bibnamefont {Grabarits}},\ and\
  \bibinfo {author} {\bibfnamefont {A.}~\bibnamefont {del Campo}},\ }\href
  {https://arxiv.org/abs/2406.09433} {\bibinfo {title} {Kibble-zurek mechanism
  and beyond: Lessons from a holographic superfluid disk}} (\bibinfo {year}
  {2024}),\ \Eprint {https://arxiv.org/abs/2406.09433} {arXiv:2406.09433
  [cond-mat.stat-mech]} \BibitemShut {NoStop}%
\bibitem [{\citenamefont {Zurek}\ \emph {et~al.}(2005)\citenamefont {Zurek},
  \citenamefont {Dorner},\ and\ \citenamefont {Zoller}}]{Zurek2005Dynamics}%
  \BibitemOpen
  \bibfield  {author} {\bibinfo {author} {\bibfnamefont {W.~H.}\ \bibnamefont
  {Zurek}}, \bibinfo {author} {\bibfnamefont {U.}~\bibnamefont {Dorner}},\ and\
  \bibinfo {author} {\bibfnamefont {P.}~\bibnamefont {Zoller}},\ }\bibfield
  {title} {\bibinfo {title} {{Dynamics of a Quantum Phase Transition}},\ }\href
  {https://doi.org/10.1103/PhysRevLett.95.105701} {\bibfield  {journal}
  {\bibinfo  {journal} {Phys. Rev. Lett.}\ }\textbf {\bibinfo {volume} {95}},\
  \bibinfo {pages} {105701} (\bibinfo {year} {2005})}\BibitemShut {NoStop}%
\bibitem [{\citenamefont {Dziarmaga}(2005)}]{Dziarmaga2005Dynamics}%
  \BibitemOpen
  \bibfield  {author} {\bibinfo {author} {\bibfnamefont {J.}~\bibnamefont
  {Dziarmaga}},\ }\bibfield  {title} {\bibinfo {title} {Dynamics of a quantum
  phase transition: Exact solution of the quantum ising model},\ }\href
  {https://doi.org/10.1103/PhysRevLett.95.245701} {\bibfield  {journal}
  {\bibinfo  {journal} {Phys. Rev. Lett.}\ }\textbf {\bibinfo {volume} {95}},\
  \bibinfo {pages} {245701} (\bibinfo {year} {2005})}\BibitemShut {NoStop}%
\bibitem [{\citenamefont {del Campo}(2018)}]{delCampo2018Universal}%
  \BibitemOpen
  \bibfield  {author} {\bibinfo {author} {\bibfnamefont {A.}~\bibnamefont {del
  Campo}},\ }\bibfield  {title} {\bibinfo {title} {{Universal Statistics of
  Topological Defects Formed in a Quantum Phase Transition}},\ }\href
  {https://doi.org/10.1103/PhysRevLett.121.200601} {\bibfield  {journal}
  {\bibinfo  {journal} {Phys. Rev. Lett.}\ }\textbf {\bibinfo {volume} {121}},\
  \bibinfo {pages} {200601} (\bibinfo {year} {2018})}\BibitemShut {NoStop}%
\bibitem [{\citenamefont {G\'omez-Ruiz}\ \emph {et~al.}(2020)\citenamefont
  {G\'omez-Ruiz}, \citenamefont {Mayo},\ and\ \citenamefont {del
  Campo}}]{GomezRuiz20}%
  \BibitemOpen
  \bibfield  {author} {\bibinfo {author} {\bibfnamefont {F.~J.}\ \bibnamefont
  {G\'omez-Ruiz}}, \bibinfo {author} {\bibfnamefont {J.~J.}\ \bibnamefont
  {Mayo}},\ and\ \bibinfo {author} {\bibfnamefont {A.}~\bibnamefont {del
  Campo}},\ }\bibfield  {title} {\bibinfo {title} {Full counting statistics of
  topological defects after crossing a phase transition},\ }\href
  {https://doi.org/10.1103/PhysRevLett.124.240602} {\bibfield  {journal}
  {\bibinfo  {journal} {Phys. Rev. Lett.}\ }\textbf {\bibinfo {volume} {124}},\
  \bibinfo {pages} {240602} (\bibinfo {year} {2020})}\BibitemShut {NoStop}%
\bibitem [{\citenamefont {Mayo}\ \emph {et~al.}(2021)\citenamefont {Mayo},
  \citenamefont {Fan}, \citenamefont {Chern},\ and\ \citenamefont {del
  Campo}}]{Mayo21}%
  \BibitemOpen
  \bibfield  {author} {\bibinfo {author} {\bibfnamefont {J.~J.}\ \bibnamefont
  {Mayo}}, \bibinfo {author} {\bibfnamefont {Z.}~\bibnamefont {Fan}}, \bibinfo
  {author} {\bibfnamefont {G.-W.}\ \bibnamefont {Chern}},\ and\ \bibinfo
  {author} {\bibfnamefont {A.}~\bibnamefont {del Campo}},\ }\bibfield  {title}
  {\bibinfo {title} {Distribution of kinks in an ising ferromagnet after
  annealing and the generalized kibble-zurek mechanism},\ }\href
  {https://doi.org/10.1103/PhysRevResearch.3.033150} {\bibfield  {journal}
  {\bibinfo  {journal} {Phys. Rev. Res.}\ }\textbf {\bibinfo {volume} {3}},\
  \bibinfo {pages} {033150} (\bibinfo {year} {2021})}\BibitemShut {NoStop}%
\bibitem [{\citenamefont {del Campo}\ \emph {et~al.}(2022)\citenamefont {del
  Campo}, \citenamefont {G\'omez-Ruiz},\ and\ \citenamefont
  {Zhang}}]{delcampo22}%
  \BibitemOpen
  \bibfield  {author} {\bibinfo {author} {\bibfnamefont {A.}~\bibnamefont {del
  Campo}}, \bibinfo {author} {\bibfnamefont {F.~J.}\ \bibnamefont
  {G\'omez-Ruiz}},\ and\ \bibinfo {author} {\bibfnamefont {H.-Q.}\ \bibnamefont
  {Zhang}},\ }\bibfield  {title} {\bibinfo {title} {Locality of spontaneous
  symmetry breaking and universal spacing distribution of topological defects
  formed across a phase transition},\ }\href
  {https://doi.org/10.1103/PhysRevB.106.L140101} {\bibfield  {journal}
  {\bibinfo  {journal} {Phys. Rev. B}\ }\textbf {\bibinfo {volume} {106}},\
  \bibinfo {pages} {L140101} (\bibinfo {year} {2022})}\BibitemShut {NoStop}%
\bibitem [{\citenamefont {Thudiyangal}\ and\ \citenamefont {del
  Campo}(2024)}]{Thudiyangal24}%
  \BibitemOpen
  \bibfield  {author} {\bibinfo {author} {\bibfnamefont {M.}~\bibnamefont
  {Thudiyangal}}\ and\ \bibinfo {author} {\bibfnamefont {A.}~\bibnamefont {del
  Campo}},\ }\bibfield  {title} {\bibinfo {title} {Universal vortex statistics
  and stochastic geometry of bose-einstein condensation},\ }\href
  {https://doi.org/10.1103/PhysRevResearch.6.033152} {\bibfield  {journal}
  {\bibinfo  {journal} {Phys. Rev. Res.}\ }\textbf {\bibinfo {volume} {6}},\
  \bibinfo {pages} {033152} (\bibinfo {year} {2024})}\BibitemShut {NoStop}%
\bibitem [{\citenamefont {Suzuki}\ \emph {et~al.}(2012)\citenamefont {Suzuki},
  \citenamefont {Inoue},\ and\ \citenamefont
  {Chakrabarti}}]{Suzuki2012Quantum}%
  \BibitemOpen
  \bibfield  {author} {\bibinfo {author} {\bibfnamefont {S.}~\bibnamefont
  {Suzuki}}, \bibinfo {author} {\bibfnamefont {J.}~\bibnamefont {Inoue}},\ and\
  \bibinfo {author} {\bibfnamefont {B.}~\bibnamefont {Chakrabarti}},\ }\href
  {https://doi.org/10.1007/978-3-642-33039-1} {\emph {\bibinfo {title}
  {{Quantum Ising Phases and Transitions in Transverse Ising Models}}}},\
  Lecture Notes in Physics\ (\bibinfo  {publisher} {Springer},\ \bibinfo {year}
  {2012})\BibitemShut {NoStop}%
\bibitem [{\citenamefont {del Campo}\ \emph {et~al.}(2012)\citenamefont {del
  Campo}, \citenamefont {Rams},\ and\ \citenamefont {Zurek}}]{delcampo12}%
  \BibitemOpen
  \bibfield  {author} {\bibinfo {author} {\bibfnamefont {A.}~\bibnamefont {del
  Campo}}, \bibinfo {author} {\bibfnamefont {M.~M.}\ \bibnamefont {Rams}},\
  and\ \bibinfo {author} {\bibfnamefont {W.~H.}\ \bibnamefont {Zurek}},\
  }\bibfield  {title} {\bibinfo {title} {{Assisted Finite-Rate Adiabatic
  Passage Across a Quantum Critical Point: Exact Solution for the Quantum Ising
  Model}},\ }\href {https://doi.org/10.1103/PhysRevLett.109.115703} {\bibfield
  {journal} {\bibinfo  {journal} {Phys. Rev. Lett.}\ }\textbf {\bibinfo
  {volume} {109}},\ \bibinfo {pages} {115703} (\bibinfo {year}
  {2012})}\BibitemShut {NoStop}%
\bibitem [{\citenamefont {Balducci}\ \emph {et~al.}(2023)\citenamefont
  {Balducci}, \citenamefont {Beau}, \citenamefont {Yang} \emph
  {et~al.}}]{Balducci2023Large}%
  \BibitemOpen
  \bibfield  {author} {\bibinfo {author} {\bibfnamefont {F.}~\bibnamefont
  {Balducci}}, \bibinfo {author} {\bibfnamefont {M.}~\bibnamefont {Beau}},
  \bibinfo {author} {\bibfnamefont {J.}~\bibnamefont {Yang}}, \emph {et~al.},\
  }\bibfield  {title} {\bibinfo {title} {{Large Deviations beyond the
  Kibble-Zurek Mechanism}},\ }\href
  {https://doi.org/10.1103/PhysRevLett.131.230401} {\bibfield  {journal}
  {\bibinfo  {journal} {Phys. Rev. Lett.}\ }\textbf {\bibinfo {volume} {131}},\
  \bibinfo {pages} {230401} (\bibinfo {year} {2023})}\BibitemShut {NoStop}%
\bibitem [{\citenamefont {Cincio}\ \emph {et~al.}(2007)\citenamefont {Cincio},
  \citenamefont {Dziarmaga}, \citenamefont {Rams},\ and\ \citenamefont
  {Zurek}}]{Cincio07}%
  \BibitemOpen
  \bibfield  {author} {\bibinfo {author} {\bibfnamefont {L.}~\bibnamefont
  {Cincio}}, \bibinfo {author} {\bibfnamefont {J.}~\bibnamefont {Dziarmaga}},
  \bibinfo {author} {\bibfnamefont {M.~M.}\ \bibnamefont {Rams}},\ and\
  \bibinfo {author} {\bibfnamefont {W.~H.}\ \bibnamefont {Zurek}},\ }\bibfield
  {title} {\bibinfo {title} {Entropy of entanglement and correlations induced
  by a quench: Dynamics of a quantum phase transition in the quantum ising
  model},\ }\href {https://doi.org/10.1103/PhysRevA.75.052321} {\bibfield
  {journal} {\bibinfo  {journal} {Phys. Rev. A}\ }\textbf {\bibinfo {volume}
  {75}},\ \bibinfo {pages} {052321} (\bibinfo {year} {2007})}\BibitemShut
  {NoStop}%
\bibitem [{\citenamefont {Polkovnikov}(2005)}]{Polkovnikov05}%
  \BibitemOpen
  \bibfield  {author} {\bibinfo {author} {\bibfnamefont {A.}~\bibnamefont
  {Polkovnikov}},\ }\bibfield  {title} {\bibinfo {title} {\textit{Universal
  adiabatic dynamics in the vicinity of a quantum critical point}},\ }\href
  {https://doi.org/10.1103/PhysRevB.72.161201} {\bibfield  {journal} {\bibinfo
  {journal} {Phys. Rev. B}\ }\textbf {\bibinfo {volume} {72}},\ \bibinfo
  {pages} {161201(R)} (\bibinfo {year} {2005})}\BibitemShut {NoStop}%
\bibitem [{\citenamefont {Damski}\ and\ \citenamefont
  {Zurek}(2006)}]{Damski2006}%
  \BibitemOpen
  \bibfield  {author} {\bibinfo {author} {\bibfnamefont {B.}~\bibnamefont
  {Damski}}\ and\ \bibinfo {author} {\bibfnamefont {W.~H.}\ \bibnamefont
  {Zurek}},\ }\bibfield  {title} {\bibinfo {title} {\textit{Adiabatic-impulse
  approximation for avoided level crossings: From phase-transition dynamics to
  Landau-Zener evolutions and back again}},\ }\href
  {https://doi.org/10.1103/PhysRevA.73.063405} {\bibfield  {journal} {\bibinfo
  {journal} {Phys. Rev. A}\ }\textbf {\bibinfo {volume} {73}},\ \bibinfo
  {pages} {063405} (\bibinfo {year} {2006})}\BibitemShut {NoStop}%
\bibitem [{NIS()}]{NIST-DLMF}%
  \BibitemOpen
  \href {http://dlmf.nist.gov/} {\bibinfo {title} {{NIST Digital Library of
  Mathematical Functions}}},\ \bibinfo {note} {release 1.1.9 of
  2023-03-15}\BibitemShut {NoStop}%
\bibitem [{\citenamefont {Damski}(2005)}]{Damski05}%
  \BibitemOpen
  \bibfield  {author} {\bibinfo {author} {\bibfnamefont {B.}~\bibnamefont
  {Damski}},\ }\bibfield  {title} {\bibinfo {title} {{The Simplest Quantum
  Model Supporting the Kibble-Zurek Mechanism of Topological Defect Production:
  Landau-Zener Transitions from a New Perspective}},\ }\href
  {https://doi.org/10.1103/PhysRevLett.95.035701} {\bibfield  {journal}
  {\bibinfo  {journal} {Phys. Rev. Lett.}\ }\textbf {\bibinfo {volume} {95}},\
  \bibinfo {pages} {035701} (\bibinfo {year} {2005})}\BibitemShut {NoStop}%
\bibitem [{\citenamefont {Bleistein}\ and\ \citenamefont
  {Handelsman}(1986)}]{Bleistein86}%
  \BibitemOpen
  \bibfield  {author} {\bibinfo {author} {\bibfnamefont {N.}~\bibnamefont
  {Bleistein}}\ and\ \bibinfo {author} {\bibfnamefont {R.}~\bibnamefont
  {Handelsman}},\ }\href@noop {} {\emph {\bibinfo {title} {Asymptotic
  Expansions of Integrals}}},\ Dover Books on Mathematics Series\ (\bibinfo
  {publisher} {Dover Publications},\ \bibinfo {year} {1986})\BibitemShut
  {NoStop}%
\bibitem [{\citenamefont {Collura}\ and\ \citenamefont
  {Karevski}(2010)}]{Collura10}%
  \BibitemOpen
  \bibfield  {author} {\bibinfo {author} {\bibfnamefont {M.}~\bibnamefont
  {Collura}}\ and\ \bibinfo {author} {\bibfnamefont {D.}~\bibnamefont
  {Karevski}},\ }\bibfield  {title} {\bibinfo {title} {Critical quench dynamics
  in confined systems},\ }\href
  {https://doi.org/10.1103/PhysRevLett.104.200601} {\bibfield  {journal}
  {\bibinfo  {journal} {Phys. Rev. Lett.}\ }\textbf {\bibinfo {volume} {104}},\
  \bibinfo {pages} {200601} (\bibinfo {year} {2010})}\BibitemShut {NoStop}%
\bibitem [{\citenamefont {Dziarmaga}\ and\ \citenamefont
  {Rams}(2010)}]{DziarmagaRams10}%
  \BibitemOpen
  \bibfield  {author} {\bibinfo {author} {\bibfnamefont {J.}~\bibnamefont
  {Dziarmaga}}\ and\ \bibinfo {author} {\bibfnamefont {M.~M.}\ \bibnamefont
  {Rams}},\ }\bibfield  {title} {\bibinfo {title} {Dynamics of an inhomogeneous
  quantum phase transition},\ }\href
  {https://doi.org/10.1088/1367-2630/12/5/055007} {\bibfield  {journal}
  {\bibinfo  {journal} {New Journal of Physics}\ }\textbf {\bibinfo {volume}
  {12}},\ \bibinfo {pages} {055007} (\bibinfo {year} {2010})}\BibitemShut
  {NoStop}%
\bibitem [{\citenamefont {G\'omez-Ruiz}\ and\ \citenamefont {del
  Campo}(2019)}]{GomezRuiz19}%
  \BibitemOpen
  \bibfield  {author} {\bibinfo {author} {\bibfnamefont {F.~J.}\ \bibnamefont
  {G\'omez-Ruiz}}\ and\ \bibinfo {author} {\bibfnamefont {A.}~\bibnamefont {del
  Campo}},\ }\bibfield  {title} {\bibinfo {title} {{Universal Dynamics of
  Inhomogeneous Quantum Phase Transitions: Suppressing Defect Formation}},\
  }\href {https://doi.org/10.1103/PhysRevLett.122.080604} {\bibfield  {journal}
  {\bibinfo  {journal} {Phys. Rev. Lett.}\ }\textbf {\bibinfo {volume} {122}},\
  \bibinfo {pages} {080604} (\bibinfo {year} {2019})}\BibitemShut {NoStop}%
\bibitem [{\citenamefont {Saberi}\ \emph {et~al.}(2014)\citenamefont {Saberi},
  \citenamefont {Opatrn\'y}, \citenamefont {M\o{}lmer},\ and\ \citenamefont
  {del Campo}}]{Saberi14}%
  \BibitemOpen
  \bibfield  {author} {\bibinfo {author} {\bibfnamefont {H.}~\bibnamefont
  {Saberi}}, \bibinfo {author} {\bibfnamefont {T.~c.~v.}\ \bibnamefont
  {Opatrn\'y}}, \bibinfo {author} {\bibfnamefont {K.}~\bibnamefont
  {M\o{}lmer}},\ and\ \bibinfo {author} {\bibfnamefont {A.}~\bibnamefont {del
  Campo}},\ }\bibfield  {title} {\bibinfo {title} {Adiabatic tracking of
  quantum many-body dynamics},\ }\href
  {https://doi.org/10.1103/PhysRevA.90.060301} {\bibfield  {journal} {\bibinfo
  {journal} {Phys. Rev. A}\ }\textbf {\bibinfo {volume} {90}},\ \bibinfo
  {pages} {060301} (\bibinfo {year} {2014})}\BibitemShut {NoStop}%
\bibitem [{\citenamefont {Divakaran}\ \emph {et~al.}(2009)\citenamefont
  {Divakaran}, \citenamefont {Mukherjee}, \citenamefont {Dutta},\ and\
  \citenamefont {Sen}}]{Divakaran09}%
  \BibitemOpen
  \bibfield  {author} {\bibinfo {author} {\bibfnamefont {U.}~\bibnamefont
  {Divakaran}}, \bibinfo {author} {\bibfnamefont {V.}~\bibnamefont
  {Mukherjee}}, \bibinfo {author} {\bibfnamefont {A.}~\bibnamefont {Dutta}},\
  and\ \bibinfo {author} {\bibfnamefont {D.}~\bibnamefont {Sen}},\ }\bibfield
  {title} {\bibinfo {title} {Defect production due to quenching through a
  multicritical point},\ }\href
  {https://doi.org/10.1088/1742-5468/2009/02/P02007} {\bibfield  {journal}
  {\bibinfo  {journal} {Journal of Statistical Mechanics: Theory and
  Experiment}\ }\textbf {\bibinfo {volume} {2009}},\ \bibinfo {pages} {P02007}
  (\bibinfo {year} {2009})}\BibitemShut {NoStop}%
\bibitem [{\citenamefont {Sen}\ \emph {et~al.}(2008)\citenamefont {Sen},
  \citenamefont {Sengupta},\ and\ \citenamefont {Mondal}}]{Diptiman08}%
  \BibitemOpen
  \bibfield  {author} {\bibinfo {author} {\bibfnamefont {D.}~\bibnamefont
  {Sen}}, \bibinfo {author} {\bibfnamefont {K.}~\bibnamefont {Sengupta}},\ and\
  \bibinfo {author} {\bibfnamefont {S.}~\bibnamefont {Mondal}},\ }\bibfield
  {title} {\bibinfo {title} {{Defect Production in Nonlinear Quench across a
  Quantum Critical Point}},\ }\href
  {https://doi.org/10.1103/PhysRevLett.101.016806} {\bibfield  {journal}
  {\bibinfo  {journal} {Phys. Rev. Lett.}\ }\textbf {\bibinfo {volume} {101}},\
  \bibinfo {pages} {016806} (\bibinfo {year} {2008})}\BibitemShut {NoStop}%
\bibitem [{\citenamefont {Barankov}\ and\ \citenamefont
  {Polkovnikov}(2008)}]{Barankov08}%
  \BibitemOpen
  \bibfield  {author} {\bibinfo {author} {\bibfnamefont {R.}~\bibnamefont
  {Barankov}}\ and\ \bibinfo {author} {\bibfnamefont {A.}~\bibnamefont
  {Polkovnikov}},\ }\bibfield  {title} {\bibinfo {title} {{Optimal Nonlinear
  Passage Through a Quantum Critical Point}},\ }\href
  {https://doi.org/10.1103/PhysRevLett.101.076801} {\bibfield  {journal}
  {\bibinfo  {journal} {Phys. Rev. Lett.}\ }\textbf {\bibinfo {volume} {101}},\
  \bibinfo {pages} {076801} (\bibinfo {year} {2008})}\BibitemShut {NoStop}%
\bibitem [{\citenamefont {Grabarits}\ \emph {et~al.}(2024)\citenamefont
  {Grabarits}, \citenamefont {Balducci}, \citenamefont {Sanders},\ and\
  \citenamefont {del Campo}}]{Grabarits24NAQO}%
  \BibitemOpen
  \bibfield  {author} {\bibinfo {author} {\bibfnamefont {A.}~\bibnamefont
  {Grabarits}}, \bibinfo {author} {\bibfnamefont {F.}~\bibnamefont {Balducci}},
  \bibinfo {author} {\bibfnamefont {B.~C.}\ \bibnamefont {Sanders}},\ and\
  \bibinfo {author} {\bibfnamefont {A.}~\bibnamefont {del Campo}},\ }\href
  {https://arxiv.org/abs/2407.09596} {\bibinfo {title} {Non-adiabatic quantum
  optimization for crossing quantum phase transitions}} (\bibinfo {year}
  {2024}),\ \Eprint {https://arxiv.org/abs/2407.09596} {arXiv:2407.09596
  [quant-ph]} \BibitemShut {NoStop}%
\bibitem [{\citenamefont {del Campo}\ \emph {et~al.}(2013)\citenamefont {del
  Campo}, \citenamefont {Kibble},\ and\ \citenamefont {Zurek}}]{DKZ13}%
  \BibitemOpen
  \bibfield  {author} {\bibinfo {author} {\bibfnamefont {A.}~\bibnamefont {del
  Campo}}, \bibinfo {author} {\bibfnamefont {T.~W.~B.}\ \bibnamefont
  {Kibble}},\ and\ \bibinfo {author} {\bibfnamefont {W.~H.}\ \bibnamefont
  {Zurek}},\ }\bibfield  {title} {\bibinfo {title} {{Causality and
  non-equilibrium second-order phase transitions in inhomogeneous systems}},\
  }\href {https://doi.org/10.1088/0953-8984/25/40/404210} {\bibfield  {journal}
  {\bibinfo  {journal} {J. Phys.: Condens. Matt.}\ }\textbf {\bibinfo {volume}
  {25}},\ \bibinfo {pages} {404210} (\bibinfo {year} {2013})}\BibitemShut
  {NoStop}%
\bibitem [{\citenamefont {Caneva}\ \emph {et~al.}(2008)\citenamefont {Caneva},
  \citenamefont {Fazio},\ and\ \citenamefont {Santoro}}]{Caneva08}%
  \BibitemOpen
  \bibfield  {author} {\bibinfo {author} {\bibfnamefont {T.}~\bibnamefont
  {Caneva}}, \bibinfo {author} {\bibfnamefont {R.}~\bibnamefont {Fazio}},\ and\
  \bibinfo {author} {\bibfnamefont {G.~E.}\ \bibnamefont {Santoro}},\
  }\bibfield  {title} {\bibinfo {title} {Adiabatic quantum dynamics of the
  lipkin-meshkov-glick model},\ }\href
  {https://doi.org/10.1103/PhysRevB.78.104426} {\bibfield  {journal} {\bibinfo
  {journal} {Phys. Rev. B}\ }\textbf {\bibinfo {volume} {78}},\ \bibinfo
  {pages} {104426} (\bibinfo {year} {2008})}\BibitemShut {NoStop}%
\bibitem [{\citenamefont {Jaschke}\ \emph {et~al.}(2017)\citenamefont
  {Jaschke}, \citenamefont {Maeda}, \citenamefont {Whalen}, \citenamefont
  {Wall},\ and\ \citenamefont {Carr}}]{Jaschke17}%
  \BibitemOpen
  \bibfield  {author} {\bibinfo {author} {\bibfnamefont {D.}~\bibnamefont
  {Jaschke}}, \bibinfo {author} {\bibfnamefont {K.}~\bibnamefont {Maeda}},
  \bibinfo {author} {\bibfnamefont {J.~D.}\ \bibnamefont {Whalen}}, \bibinfo
  {author} {\bibfnamefont {M.~L.}\ \bibnamefont {Wall}},\ and\ \bibinfo
  {author} {\bibfnamefont {L.~D.}\ \bibnamefont {Carr}},\ }\bibfield  {title}
  {\bibinfo {title} {Critical phenomena and kibble–zurek scaling in the
  long-range quantum ising chain},\ }\href
  {https://doi.org/10.1088/1367-2630/aa65bc} {\bibfield  {journal} {\bibinfo
  {journal} {New J. Phys.}\ }\textbf {\bibinfo {volume} {19}},\ \bibinfo
  {pages} {033032} (\bibinfo {year} {2017})}\BibitemShut {NoStop}%
\bibitem [{\citenamefont {Dutta}\ and\ \citenamefont {Dutta}(2017)}]{Dutta17}%
  \BibitemOpen
  \bibfield  {author} {\bibinfo {author} {\bibfnamefont {A.}~\bibnamefont
  {Dutta}}\ and\ \bibinfo {author} {\bibfnamefont {A.}~\bibnamefont {Dutta}},\
  }\bibfield  {title} {\bibinfo {title} {Probing the role of long-range
  interactions in the dynamics of a long-range kitaev chain},\ }\href
  {https://doi.org/10.1103/PhysRevB.96.125113} {\bibfield  {journal} {\bibinfo
  {journal} {Phys. Rev. B}\ }\textbf {\bibinfo {volume} {96}},\ \bibinfo
  {pages} {125113} (\bibinfo {year} {2017})}\BibitemShut {NoStop}%
\bibitem [{\citenamefont {Gherardini}\ \emph {et~al.}(2024)\citenamefont
  {Gherardini}, \citenamefont {Buffoni},\ and\ \citenamefont
  {Defenu}}]{Gherardini24}%
  \BibitemOpen
  \bibfield  {author} {\bibinfo {author} {\bibfnamefont {S.}~\bibnamefont
  {Gherardini}}, \bibinfo {author} {\bibfnamefont {L.}~\bibnamefont
  {Buffoni}},\ and\ \bibinfo {author} {\bibfnamefont {N.}~\bibnamefont
  {Defenu}},\ }\bibfield  {title} {\bibinfo {title} {Universal defects
  statistics with strong long-range interactions},\ }\href
  {https://doi.org/10.1103/PhysRevLett.133.113401} {\bibfield  {journal}
  {\bibinfo  {journal} {Phys. Rev. Lett.}\ }\textbf {\bibinfo {volume} {133}},\
  \bibinfo {pages} {113401} (\bibinfo {year} {2024})}\BibitemShut {NoStop}%
\bibitem [{\citenamefont {Dziarmaga}(2006)}]{Dziarmaga06}%
  \BibitemOpen
  \bibfield  {author} {\bibinfo {author} {\bibfnamefont {J.}~\bibnamefont
  {Dziarmaga}},\ }\bibfield  {title} {\bibinfo {title} {Dynamics of a quantum
  phase transition in the random ising model: Logarithmic dependence of the
  defect density on the transition rate},\ }\href
  {https://doi.org/10.1103/PhysRevB.74.064416} {\bibfield  {journal} {\bibinfo
  {journal} {Phys. Rev. B}\ }\textbf {\bibinfo {volume} {74}},\ \bibinfo
  {pages} {064416} (\bibinfo {year} {2006})}\BibitemShut {NoStop}%
\bibitem [{\citenamefont {Caneva}\ \emph {et~al.}(2007)\citenamefont {Caneva},
  \citenamefont {Fazio},\ and\ \citenamefont {Santoro}}]{Caneva07}%
  \BibitemOpen
  \bibfield  {author} {\bibinfo {author} {\bibfnamefont {T.}~\bibnamefont
  {Caneva}}, \bibinfo {author} {\bibfnamefont {R.}~\bibnamefont {Fazio}},\ and\
  \bibinfo {author} {\bibfnamefont {G.~E.}\ \bibnamefont {Santoro}},\
  }\bibfield  {title} {\bibinfo {title} {Adiabatic quantum dynamics of a random
  ising chain across its quantum critical point},\ }\href
  {https://doi.org/10.1103/PhysRevB.76.144427} {\bibfield  {journal} {\bibinfo
  {journal} {Phys. Rev. B}\ }\textbf {\bibinfo {volume} {76}},\ \bibinfo
  {pages} {144427} (\bibinfo {year} {2007})}\BibitemShut {NoStop}%
\bibitem [{\citenamefont {Hatomura}(2024)}]{Hatomura24}%
  \BibitemOpen
  \bibfield  {author} {\bibinfo {author} {\bibfnamefont {T.}~\bibnamefont
  {Hatomura}},\ }\bibfield  {title} {\bibinfo {title} {Shortcuts to
  adiabaticity: theoretical framework, relations between different methods, and
  versatile approximations},\ }\href {https://doi.org/10.1088/1361-6455/ad38f1}
  {\bibfield  {journal} {\bibinfo  {journal} {Journal of Physics B: Atomic,
  Molecular and Optical Physics}\ }\textbf {\bibinfo {volume} {57}},\ \bibinfo
  {pages} {102001} (\bibinfo {year} {2024})}\BibitemShut {NoStop}%
\bibitem [{\citenamefont {Sengupta}\ \emph {et~al.}(2008)\citenamefont
  {Sengupta}, \citenamefont {Sen},\ and\ \citenamefont {Mondal}}]{Sengupta08}%
  \BibitemOpen
  \bibfield  {author} {\bibinfo {author} {\bibfnamefont {K.}~\bibnamefont
  {Sengupta}}, \bibinfo {author} {\bibfnamefont {D.}~\bibnamefont {Sen}},\ and\
  \bibinfo {author} {\bibfnamefont {S.}~\bibnamefont {Mondal}},\ }\bibfield
  {title} {\bibinfo {title} {Exact results for quench dynamics and defect
  production in a two-dimensional model},\ }\href@noop {} {\bibfield  {journal}
  {\bibinfo  {journal} {Phys. Rev. Lett.}\ }\textbf {\bibinfo {volume} {100}},\
  \bibinfo {pages} {077204} (\bibinfo {year} {2008})}\BibitemShut {NoStop}%
\bibitem [{\citenamefont {Schmitt}\ \emph {et~al.}(2022)\citenamefont
  {Schmitt}, \citenamefont {Rams}, \citenamefont {Dziarmaga}, \citenamefont
  {Heyl},\ and\ \citenamefont {Zurek}}]{Schmitt22}%
  \BibitemOpen
  \bibfield  {author} {\bibinfo {author} {\bibfnamefont {M.}~\bibnamefont
  {Schmitt}}, \bibinfo {author} {\bibfnamefont {M.~M.}\ \bibnamefont {Rams}},
  \bibinfo {author} {\bibfnamefont {J.}~\bibnamefont {Dziarmaga}}, \bibinfo
  {author} {\bibfnamefont {M.}~\bibnamefont {Heyl}},\ and\ \bibinfo {author}
  {\bibfnamefont {W.~H.}\ \bibnamefont {Zurek}},\ }\bibfield  {title} {\bibinfo
  {title} {Quantum phase transition dynamics in the two-dimensional
  transverse-field ising model},\ }\href
  {https://doi.org/10.1126/sciadv.abl6850} {\bibfield  {journal} {\bibinfo
  {journal} {Science Advances}\ }\textbf {\bibinfo {volume} {8}},\ \bibinfo
  {pages} {eabl6850} (\bibinfo {year} {2022})}\BibitemShut {NoStop}%
\bibitem [{\citenamefont {B.~S}\ \emph {et~al.}(2020)\citenamefont {B.~S},
  \citenamefont {Mukherjee}, \citenamefont {Divakaran},\ and\ \citenamefont
  {del Campo}}]{Revathy20}%
  \BibitemOpen
  \bibfield  {author} {\bibinfo {author} {\bibfnamefont {R.}~\bibnamefont
  {B.~S}}, \bibinfo {author} {\bibfnamefont {V.}~\bibnamefont {Mukherjee}},
  \bibinfo {author} {\bibfnamefont {U.}~\bibnamefont {Divakaran}},\ and\
  \bibinfo {author} {\bibfnamefont {A.}~\bibnamefont {del Campo}},\ }\bibfield
  {title} {\bibinfo {title} {Universal finite-time thermodynamics of many-body
  quantum machines from kibble-zurek scaling},\ }\href
  {https://doi.org/10.1103/PhysRevResearch.2.043247} {\bibfield  {journal}
  {\bibinfo  {journal} {Phys. Rev. Res.}\ }\textbf {\bibinfo {volume} {2}},\
  \bibinfo {pages} {043247} (\bibinfo {year} {2020})}\BibitemShut {NoStop}%
\bibitem [{\citenamefont {Rams}\ \emph {et~al.}(2018)\citenamefont {Rams},
  \citenamefont {Sierant}, \citenamefont {Dutta}, \citenamefont {Horodecki},\
  and\ \citenamefont {Zakrzewski}}]{Rams18}%
  \BibitemOpen
  \bibfield  {author} {\bibinfo {author} {\bibfnamefont {M.~M.}\ \bibnamefont
  {Rams}}, \bibinfo {author} {\bibfnamefont {P.}~\bibnamefont {Sierant}},
  \bibinfo {author} {\bibfnamefont {O.}~\bibnamefont {Dutta}}, \bibinfo
  {author} {\bibfnamefont {P.}~\bibnamefont {Horodecki}},\ and\ \bibinfo
  {author} {\bibfnamefont {J.}~\bibnamefont {Zakrzewski}},\ }\bibfield  {title}
  {\bibinfo {title} {At the limits of criticality-based quantum metrology:
  Apparent super-heisenberg scaling revisited},\ }\href
  {https://doi.org/10.1103/PhysRevX.8.021022} {\bibfield  {journal} {\bibinfo
  {journal} {Phys. Rev. X}\ }\textbf {\bibinfo {volume} {8}},\ \bibinfo {pages}
  {021022} (\bibinfo {year} {2018})}\BibitemShut {NoStop}%
\bibitem [{\citenamefont {Zhang}\ \emph {et~al.}(2024)\citenamefont {Zhang},
  \citenamefont {Hu},\ and\ \citenamefont {Li}}]{Zhang2024dqpt}%
  \BibitemOpen
  \bibfield  {author} {\bibinfo {author} {\bibfnamefont {X.}~\bibnamefont
  {Zhang}}, \bibinfo {author} {\bibfnamefont {L.}~\bibnamefont {Hu}},\ and\
  \bibinfo {author} {\bibfnamefont {F.}~\bibnamefont {Li}},\ }\href
  {https://arxiv.org/abs/2409.13293} {\bibinfo {title} {Unique and universal
  scaling in dynamical quantum phase transitions}} (\bibinfo {year} {2024}),\
  \Eprint {https://arxiv.org/abs/2409.13293} {arXiv:2409.13293
  [cond-mat.stat-mech]} \BibitemShut {NoStop}%
\bibitem [{\citenamefont {Schafer}\ and\ \citenamefont
  {Kouyoumjian}(1967)}]{Schafer1967Higher}%
  \BibitemOpen
  \bibfield  {author} {\bibinfo {author} {\bibfnamefont {R.}~\bibnamefont
  {Schafer}}\ and\ \bibinfo {author} {\bibfnamefont {R.}~\bibnamefont
  {Kouyoumjian}},\ }\bibfield  {title} {\bibinfo {title} {Higher order terms in
  the saddle point approximation},\ }\href
  {https://doi.org/10.1109/PROC.1967.5863} {\bibfield  {journal} {\bibinfo
  {journal} {Proc. IEEE}\ }\textbf {\bibinfo {volume} {55}},\ \bibinfo {pages}
  {1496} (\bibinfo {year} {1967})}\BibitemShut {NoStop}%
\end{thebibliography}%

\end{document}